\documentclass[12pt]{article}
\usepackage{amsfonts,amsmath,amssymb,epsf,bbm,color,array,colortbl,xypic,longtable}
\usepackage{ifpdf}
\ifpdf
\usepackage{epstopdf,hyperref}
\else
\usepackage[hypertex]{hyperref}
\fi

\advance\textheight by \topskip
\usepackage[all]{xy}

\usepackage{amsthm} 
\newtheorem{Thm}{Theorem}

\newtheorem{Lem}[Thm]{Lemma}

\theoremstyle{definition}

\newtheorem{Def}[Thm]{Definition}
\numberwithin{equation}{section}
\numberwithin{Thm}{section}
                   
\newcommand\be            {\begin{equation}}
\newcommand\bea           {\begin{equation}\begin{array}l\displaystyle}
\newcommand\bearll        {\begin{array}{ll}\displaystyle}
\newcommand\ee            {\end{equation}}
\newcommand\eear          {\end{array}}
\newcommand\enl           {\\[1em]\displaystyle}
\newcommand\etb           {& \displaystyle}

\newcommand\labl[1]       {\label{#1}\ee}
\newcommand\nxt{\noindent\raisebox{.08em}{\rule{.44em}{.44em}}%
\hspace{.4em}}

\newcommand\Cdd            {\mathcal{C}^b}
\newcommand\Crr            {\mathcal{C}^r}

\newcommand\eev           {{}^\vee\hspace*{-1pt}}
\newcommand\eps           {\varepsilon}
\newcommand\Hom           {\mathrm{Hom}}
\newcommand\id            {{\rm id}}
\newcommand\Kdd            {\mathrm{K}_0^b}
\newcommand\Krr            {\mathrm{K}_0^r}

\newcommand\one           {{\bf1}}

\newcommand\Rep           {\mathrm{Rep}}
\newcommand\Repbb           {\mathrm{Rep}(\mathcal{W}_{2,3})^b}

\newcommand\Cb            {\mathbb{C}}

\newcommand\Zb            {\mathbb{Z}}

\newcommand\Bc            {\mathcal{B}}
\newcommand\Cc            {\mathcal{C}}
\newcommand\Hc            {\mathcal{H}}
\newcommand\Nc            {\mathcal{N}}

\newcommand\Qc            {\mathcal{Q}}
\renewcommand\Rc            {\mathcal{R}}
\newcommand\Vc            {\mathcal{V}}
\newcommand\Wc           {\mathcal{W}}

\begin{document}
\thispagestyle{empty}
\def\thefootnote{\fnsymbol{footnote}}
\begin{flushright}
KCL-MTH-09-04\\
arXiv:0905.0916
\end{flushright}
\vskip 2.5em
\begin{center}\LARGE
Fusion rules and boundary conditions \\ 
in the $c=0$ triplet model
\end{center}\vskip 2em
\begin{center}\large
  Matthias R. Gaberdiel%
  $^{a}$\footnote{Email: {\tt gaberdiel@itp.phys.ethz.ch}}, 
  Ingo Runkel%
  $^{b}$\footnote{Email: {\tt ingo.runkel@kcl.ac.uk}} \,%
  and
  Simon Wood%
  $^{a}$\footnote{Email: {\tt swood@itp.phys.ethz.ch}}
\end{center}
\begin{center}\it$^a$
Institute for Theoretical Physics, ETH Z\"urich \\
8093 Z\"urich, Switzerland
\end{center}
\begin{center}\it$^b$
  Department of Mathematics, King's College London \\
  Strand, London WC2R 2LS, United Kingdom  
\end{center}
\vskip 1em
\begin{center}
May 2009
\end{center}
\vskip 1em
\begin{abstract}
The logarithmic triplet model $\Wc_{2,3}$ at $c=0$ is studied. 
In particular, we determine the fusion
rules of the irreducible representations from first principles, and show 
that there exists a finite set of representations, including all irreducible
representations, that closes under fusion. With the help of these results
we then investigate the possible boundary conditions of the 
$\Wc_{2,3}$ theory.  
Unlike the familiar Cardy case where there is a consistent boundary condition
for every representation of the chiral algebra, we find that for 
$\Wc_{2,3}$ only a subset of representations gives rise to consistent boundary
conditions. These then have boundary spectra
with non-degenerate two-point correlators.
\end{abstract}

\setcounter{footnote}{0}
\def\thefootnote{\arabic{footnote}}

\newpage

\tableofcontents

\newpage

\section{Introduction and summary}

Logarithmic conformal field theories appear in the description of critical points 
in many interesting physical systems. Some examples are polymers, spin chains, 
percolation, and sand-pile models, see for example
\cite{Jeng:2006tg,Pearce:2006we,Read:2007qq,Ruelle:2007kg,Mathieu:2007pe,Rasmussen:2008ii,Ridout:2008cv,SaintAubin:2008,Nigro:2009si} 
for some recent papers. Logarithmic conformal field theories 
have also played an important role in recent attempts to understand chiral massive gravity
\cite{Grumiller:2008es,Maloney:2009ck}.
In some of these examples, in particular for critical systems with quenched disorder, 
for dilute self-avoiding polymers, and for percolation, as well as in the context of 
chiral gravity, the logarithmic conformal field theory has  central charge 
$c=0$,  see {\it e.g.}\ \cite{Gurarie:1999yx} for a discussion of $c=0$ theories. 
This has important consequences for the structure of the resulting theory.
Indeed logarithmic conformal field theories at $c=0$ behave rather differently 
from the examples that have been studied in detail so far, in particular 
from the $(1,p)$-series whose structure has now been largely understood 
\cite{Gaberdiel:1996np,Fuchs:2003yu,Carqueville:2005nu,Gaberdiel:2007jv,Adamovic:2007er}. 
\medskip

In this paper we want to study one particular $c=0$ logarithmic conformal 
field theory, namely the $\Wc_{2,3}$  triplet theory. This is the simplest example 
of a whole family of $\Wc_{p,q}$  triplet
theories that can be naturally associated 
to the minimal models \cite{Feigin:2006iv,Rasmussen:2008xi}. One peculiar 
feature of the $\Wc_{2,3}$ model 
(and of all $\Wc_{p,q}$ theories with 
$p,q\geq 2$) is that the vacuum representation is not irreducible. This is a 
generic property of $c=0$ logarithmic conformal field theories,\footnote{
For $c=0$ the descendant of the vacuum, $L_{-2}\Omega$, is a Virasoro highest weight 
vector since $L_1 L_{-2} \Omega = L_2 L_{-2} \Omega = 0$. Unless the stress tensor of 
the conformal field theory vanishes, the vacuum representation is reducible as 
a representation of the Virasoro algebra. It may of course still be irreducible 
as a representation of a larger chiral algebra (for example if one takes the 
product of two non-logarithmic theories with opposite central charge), but for 
$\Wc_{2,3}$, and in fact for all $\Wc_{p,q}$ with $p,q \ge 2$, the vacuum 
representation contains a non-trivial sub-represenatation of the entire $\Wc$-algebra.}
and we believe that it is responsible for the complicated and rather unfamiliar behaviour
we shall encounter.

Our first main result concerns the description of the $\Wc_{2,3}$ fusion rules 
of all indecomposable representations that appear as direct summands
in successive fusions of the irreducibles. 
Our analysis starts from the corresponding Virasoro fusion rules 
which we re-examine following \cite{Eberle:2006zn}. 
Using induced representations and associativity, we then determine the fusion rules of the 
irreducible $\Wc_{2,3}$-representations, as well as those of the resulting indecomposable 
representations. Our results agree
with \cite{Rasmussen:2008ii,Rasmussen:2008ez}, but go beyond them in that we
also determine the fusion rules of representations that are not accessible in their 
approach. Furthermore, we shall exhibit some of the unusual 
properties of these $\Wc$-representations and their fusion.  
For example, there is a subtle difference between `conjugate' and `dual' representations
that we shall explain in some detail (see Section~\ref{sec:contra-dual}), 
and the Grothendieck group (that appears naturally
in the construction of the boundary theory) does not possess the standard
ring structure, see Section~\ref{sec:Gr-group-vs-ring}. 
\medskip

The fusion rules are an important ingredient for the description of the 
possible boundary conditions. 
Boundary logarithmic conformal field theories have been investigated 
from several points of view, for example starting from an underlying 
lattice realisation 
\cite{Ruelle:2002jy,Piroux:2004vd,Piroux:2004bc,Izmailian:2005zz,Pearce:2006sz,Jeng:2006tg,%
Pearce:2006we,Ruelle:2007kg,Pearce:2008nn,Rasmussen:2008ii,Rasmussen:2008xi,%
Nigro:2009si}, 
from super group WZW models 
\cite{Creutzig:2007jy,Quella:2007sg,Creutzig:2008ek,Creutzig:2008ag,%
Mitev:2008yt,Creutzig:2008an}, 
or from logarithmic extensions of Virasoro minimal models 
\cite{Kogan:2000fa,Ishimoto:2001jv,Kawai:2001ur,Bredthauer:2002ct,Bredthauer:2002xb,%
Ishimoto:2004dk,Gaberdiel:2006pp,Creutzig:2006wk,Gaberdiel:2007jv,Gainutdinov:2007tc,%
Ridout:2008cv}.
The work of most direct relevance to our purposes is \cite{Rasmussen:2008ii}, where 
the fusion rules of the $\Wc_{2,3}$ model are analysed via the boundary 
theory (on a lattice) under the assumption that one can read off the fusion rules from the open 
string spectra as in
Cardy's analysis  \cite{Cardy:1989ir}. Indeed, in the usual (non-logarithmic rational) case,
there is a boundary condition for every 
representation of the chiral algebra, 
and the open string spectrum between two such boundary conditions agrees 
precisely with the fusion of the corresponding representations (or rather, the 
fusion where one of the two representation is replaced by its conjugate representation). 
For $\Wc_{2,3}$, on the other hand, not every
representation corresponds to a consistent boundary condition. 

More specifically, if  we try to construct a boundary theory where all representations of 
$\Wc_{2,3}$ correspond to boundary conditions,  it is possible to define 
an associative operator 
product expansion of boundary fields, but the two-point correlator of boundary 
fields will in general be degenerate.
The boundary conditions with non-degenerate two-point correlator 
correspond essentially to representations whose conjugate representation agrees 
with the dual representation (for more details see Section~\ref{sec:intro-W23-bnd} below). If 
$\Rc$ and $\mathcal{S}$ are two such representations, the open string 
spectrum between the corresponding boundary conditions is given by the 
fusion of $\Rc$ with the conjugate representation of $\mathcal{S}$, just as in 
Cardy's analysis of the non-logarithmic case. This is the second main result of our 
paper, and it reproduces precisely the lattice results of \cite{Rasmussen:2008ii} 
from an analysis intrinsic to conformal field theory.
\medskip

In non-logarithmic rational conformal field theories one 
can uniquely reconstruct the bulk theory from a consistent boundary theory 
\cite{Runkel:1998pm,Felder:1999mq,tft1,unique}, and 
every possible bulk theory (with the appropriate symmetry algebra) 
can be obtained in this way \cite{Kong:2008ci}. Furthermore,
two boundary theories give rise to isomorphic bulk theories if and only 
if the boundary theories are equivalent in the sense described in 
\cite{Kong:2007yv}. The boundary theory is typically simpler than the
bulk theory, and it is therefore often useful to start with the boundary theory
in order to construct the bulk theory. This is most pronounced in the  
charge-conjugation Cardy case \cite{Cardy:1989ir}, where there is a boundary 
condition whose open string spectrum consists just of the vacuum representation 
of the chiral algebra.

One may hope that the general idea --- to start from a boundary theory in order to 
construct the bulk theory that fits to it --- remains valid also in the logarithmic case, 
even if the detailed construction will start to deviate. For $\Wc_{1,p}$ models
this approach was used in \cite{Gaberdiel:2007jv} to obtain a modular invariant 
bulk partition function, which for $p=2$ reproduced the known local theory from 
\cite{Gaberdiel:1998ps}. This analysis was performed for the analogue of
the Cardy case, {\it i.e.}\ by starting with a boundary condition whose open string
spectrum consists just of the vacuum representation. 
However, for the $\Wc_{2,3}$ theory, such a boundary condition
does not exist since the corresponding boundary two-point correlators are degenerate. 
This suggests that the analogue of the charge-conjugation modular invariant for
the $\Wc_{2,3}$ model will be more involved than for the 
$\Wc_{1,p}$ series \cite{Gaberdiel:2007jv}.
Nonetheless, because a lattice realisation of the $\Wc_{2,3}$ theory is 
known \cite{Rasmussen:2008ii}, it seems plausible that a consistent 
bulk theory does in fact exist.
Furthermore, there is a fairly natural guess for how the construction of the
bulk theory could roughly work; this is briefly indicated in Section~\ref{sec:concl}.
\medskip

In the remainder of this introduction we give a detailed (but non-technical) 
overview of the results of the paper. 
Section~\ref{sec:W-rep} contains the detailed discussion of the 
$\Wc$-representations and the computation of their fusion products. 
In Section~\ref{sec:iHom-assoc-etc} we construct the boundary theory 
based on the abstract theory of internal Homs and dual objects in tensor 
categories, and Section~\ref{sec:concl} contains our conclusions.
In Appendix~\ref{app:more-rep} we list the characters of the $\Wc_{2,3}$-representations,
their embedding diagrams, and their fusion rules. 
We also spell out the dictionary between our notation
and that of \cite{Rasmussen:2008ii,Rasmussen:2008ez}, see Appendix~\ref{app:dict}. 
Finally Appendix~\ref{app:cat-stuff} contains some technicalities needed in 
Section~\ref{sec:iHom-assoc-etc}.

\subsection[$\Wc$-representations and fusion rules]{$\boldsymbol{\Wc}$-representations and fusion rules}\label{sec:intro-W-fusion}

Let us begin by reviewing the structure of the underlying Virasoro theory. 
Recall that the Virasoro minimal models have central charge 
\begin{align}
  c_{p,q}=1-6\frac{(p-q)^2}{pq} \ ,
\end{align}
where $p$ and $q$ are a pair of positive coprime integers. 
The vacuum representation is the irreducible representation based on
the highest weight state $\Omega$ with $h=0$. The corresponding Verma module 
has two independent null vectors: the null vector
${\cal N}_1 = L_{-1}\Omega$ of conformal dimension $h=1$
and a null vector ${\cal N}_2$ of conformal 
dimension $h=(p-1)\cdot (q-1)$. 
Setting ${\cal N}_1$ and ${\cal N}_2$ to zero we obtain the irreducible vacuum 
representation based on $\Omega$.
The  highest weight representations of the 
corresponding vertex operator algebra
are the representations of the 
Virasoro algebra for which the modes $V_n({\cal N}_1)$ and 
$V_n({\cal N}_2)$ act trivially. They have conformal weights
\be
  h_{r,s}=\frac{(ps-qr)^2-(p-q)^2}{4pq} \ ,
\labl{reps}
where $1\leq r \leq p-1$, $1\leq s \leq q-1$ and we have the identification
\be
  h_{r,s}=h_{p-r,q-s}\ .
\ee

We shall be mainly interested in the case $(p,q)=(2,3)$ for which $c_{2,3}=0$. 
In this case, the null vector ${\cal N}_2$ of the vacuum representation 
is just the vector ${\cal N}_2= L_{-2}\Omega$, and thus the irreducible
vacuum representation $\Vc(0)$ only consists of the vacuum state $\Omega$ itself. 
Furthermore, there is only one representation in \eqref{reps}, namely the vacuum 
representation $\Vc(0)$ itself. This is clearly a very trivial and boring
theory.
\medskip

The logarithmic theory we are interested in is obtained in a slightly different
fashion. Instead of taking the vertex operator algebra to be $\Vc(0)$, we 
consider the vertex operator algebra $\Vc$ that is obtained from the 
Verma module based on $\Omega$ by dividing out ${\cal N}_1=L_{-1}\Omega$,
but not $\Nc_2 \equiv T = L_{-2}\Omega$. 
This leads to a logarithmic conformal field
theory, but not to one that is rational. In order to make the theory rational
we then enlarge the chiral algebra by three fields of conformal dimension $15$.
The resulting vertex operator algebra will be denoted by $\Wc_{2,3}$ or just
$\Wc$, and it 
defines the so-called $\Wc_{2,3}$ model \cite{Feigin:2006iv}.
Its irreducible representations are described by the finite Kac table:
\be
  \begin{tabular}{c|ccc}
    & $s=1$ & $s=2$ & $s=3$ \\
    \hline
    $r=1$ & \colorbox[gray]{.8}{\(0,\,2,\,7\)}~&~ \colorbox[gray]{.8}{\(0,\,1,\,5\)}~
    &\(\frac{1}{3},\,\frac{10}{3}\)\\[.5em]
    $r=2$ & \(\frac58,\,\frac{33}{8}\)& \(\frac18,\,\frac{21}{8}\)& \(-\frac{1}{24},\,\frac{35}{24}\)
  \end{tabular}
\labl{eq:W23-irreps}
Here each entry $h$ is the conformal dimension of the highest
weight states of an irreducible representation, which we shall denote by $\Wc(h)$. 
There is only one representation corresponding to $h=0$, 
namely the one-dimensional vacuum representation $\Wc(0)$, spanned 
by the vacuum vector $\Omega$.

The representations $\Wc(h)$ for which the value $h$ is coloured grey
in \eqref{eq:W23-irreps} will not correspond to consistent boundary conditions, 
see Section~\ref{sec:intro-W23-bnd} below.

As is familiar from other logarithmic theories, the $13$ irreducible representations 
in \eqref{eq:W23-irreps} do 
not close among themselves under fusion. However, one can show that the fusion rules
close on some larger set, involving in addition $22$ indecomposable representations.
These will be described in more detail in Section~\ref{sec:W-rep}, and their 
characters will be given in Appendix~\ref{app:W23-chars}; the relation to the 
notation in \cite{Rasmussen:2008ii,Rasmussen:2008ez} is explained in 
Appendix~\ref{app:dict}.
\bea
  \colorbox[gray]{.8}{\rule[-2pt]{0pt}{1.1em}$\Wc$}~,~~
  \colorbox[gray]{.8}{\rule[-2pt]{0pt}{1.1em}$\Wc^*$}~,~~
  \colorbox[gray]{.8}{\rule[-2pt]{0pt}{1.1em}$\Qc$}~,~~
  \colorbox[gray]{.8}{\rule[-2pt]{0pt}{1.1em}$\Qc^*$}~,~~
  \Rc^{(2)}(0,2)_7~,~~
  \Rc^{(2)}(2,7)~,~~
  \Rc^{(2)}(0,1)_5~,~~
  \Rc^{(2)}(1,5)~,~~
\\[.3em]\displaystyle
  \Rc^{(2)}(0,2)_5~,~~
  \Rc^{(2)}(2,5)~,~~
  \Rc^{(2)}(0,1)_7~,~~
  \Rc^{(2)}(1,7)~,~~
  \Rc^{(2)}(\tfrac13,\tfrac13)~,~~
\\[.3em]\displaystyle
  \Rc^{(2)}(\tfrac13,\tfrac{10}{3})~,~~
  \Rc^{(2)}(\tfrac58,\tfrac58)~,~~
  \Rc^{(2)}(\tfrac58,\tfrac{21}{8})~,~~
  \Rc^{(2)}(\tfrac18,\tfrac18)~,~~
  \Rc^{(2)}(\tfrac18,\tfrac{33}{8})~,~~
\\[.3em]\displaystyle 
  \Rc^{(3)}(0,0,1,1)~,~~
  \Rc^{(3)}(0,0,2,2)~,~~
  \Rc^{(3)}(0,1,2,5)~,~~
  \Rc^{(3)}(0,1,2,7)
\eear\labl{eq:indec-W-rep}
Again, the representations whose names are coloured grey will not correspond to
consistent boundary conditions. We do not claim that \eqref{eq:W23-irreps} 
and \eqref{eq:indec-W-rep} are {\em all} indecomposable representations of 
$\Wc_{2,3}$. Indeed it is clear that they are not, as they are not closed under
taking quotients and subrepresentations. However, the representations 
\eqref{eq:W23-irreps} and \eqref{eq:indec-W-rep} 
form the minimal set of representations, containing the irreducible representations 
in \eqref{eq:W23-irreps}, that closes under fusion and taking conjugates.

Since the vertex operator algebra
$\Wc$ contains generating fields at the rather high conformal weight $h=15$, it is 
difficult to determine the commutation relations of this $\Wc$-algebra explicitly, and thus 
we do not know how to determine the fusion rules 
directly.\footnote{One may hope
that the description given in \cite{AM09} may allow one to overcome this limitation.}
However, we can infer 
the $\Wc$ fusion rules from the calculation of the fusion rules of the Virasoro vertex 
operator algebra $\Vc$, using induced representations. 
The $\Vc$ fusion rules, on the other hand, can be determined explicitly, using the 
techniques of \cite{Nahm:1994by,Gaberdiel:1996kx} (see Section~\ref{sec:W-rep}). 
In fact, this analysis has already been done some time ago by \cite{Eberle:2006zn}, but
it contained a small mistake which we have corrected here.
The resulting $\Vc$ fusion rules are associative
and commutative, and the same then also holds for the induced $\Wc$ fusion
rules. We list all fusion products for the representations in \eqref{eq:W23-irreps} 
and \eqref{eq:indec-W-rep} in Appendix \ref{app:fus}.
\medskip

The resulting fusion rules are much more complicated than for example those
of the well-understood logarithmic $(1,p)$ models. The source of 
this and many other difficulties is probably
the fact that the vertex operator algebra $\Wc$ is not irreducible. 
In fact, $\Wc$ does not agree with the irreducible 
representation $\Wc(0)$ based on $\Omega$, since in $\Wc$ 
the state $T=L_{-2}\Omega$ does not vanish, but generates the proper
subrepresentation $\Wc(2)\subset \Wc$. The structure of $\Wc$ is thus
described by the embedding diagram 
\begin{equation}
  \rule{0pt}{1.5em}
  \xymatrix@R=0em{
  \Omega \ar@{.>}[r] \ar@/^1.2em/[rr]&\times&T\ar@{->}[rr]|{\Wc(2)}&& \\
  h=0 & h=1 & h=2 
  }
\end{equation}
where the arrows describe the action of the $\Wc$-modes and `$\times$' refers to the null 
vector $\Nc_1$ which has been divided out. 
Alternatively, we can characterise $\Wc$ by the exact sequence
\be
  0 \longrightarrow \Wc(2)  \longrightarrow \Wc  \longrightarrow \Wc(0)  \longrightarrow 0  ~.
\labl{eq:W-seq}
In the following we shall summarise some of the rather peculiar features
of the resulting theory and its fusion rules.

\subsubsection{Conjugate and dual representations} \label{sec:contra-dual}

The conjugate representation $\Rc^*$ of a representation $\Rc$ 
is characterised by the property that the two-point conformal blocks
involving one state from $\Rc$ and one state from $\Rc^*$ 
define a non-degenerate bilinear form
on $\Rc\times \Rc^*$. 
Usually, the vertex operator algebra itself is self-conjugate since the vacuum 
state $\Omega$ is a self-conjugate state, and the vertex operator algebra
is irreducible. However, in the present case, the latter property does not hold,
and as a consequence $\Wc^*$ is not isomorphic to $\Wc$. In fact, 
$\Wc^*$ is characterised by the exact sequence 
\be\label{eq:Wbar-seq}
  0 \longrightarrow \Wc(0)  \longrightarrow \Wc^*  \longrightarrow \Wc(2)  \longrightarrow 0  ~,
\ee
and is therefore different from $\Wc$. It is generated from a 
state $t$ at conformal weight $h=2$
\begin{equation}
  \rule{0pt}{1.8em}
  \xymatrix@R=0em{
  \omega \ar@{.>}[r] &\times&\ar@/_1.2em/[ll]t\ar@{->}[rr]|{\Wc(2)}&&\\
   h=0 & h=1 & h=2 }
\end{equation}
but $t$ is not a highest weight state since $L_2 t=\omega$. On the other hand,
$\omega$ is annihilated by all $\Wc$ modes. The fact that $\Wc$ is not
self-conjugate means amongst other things, that $\Wc$ cannot appear by itself as the
open string spectrum of a boundary condition --- this will be explained in more detail
below. 
\smallskip

In non-logarithmic rational conformal field theories, the 
fusion $\Rc \otimes \Rc^*$ always contains the 
vertex operator algebra $\Wc$ itself. Thus it makes sense to call the
conjugate representation $\Rc^*$  also the `dual representation'. Furthermore, it
is then obvious that the fusion of $\Rc$ with $\Rc \otimes \Rc^*$
contains $\Rc$. These properties motivate an abstract categorical definition of duals
which we   review in section \ref{sec:dual-nondeg}. 
Two necessary conditions for the existence of a dual 
representation $\Rc^\vee$ are that there exist non-zero  intertwiners
\be
  b_\Rc : \Wc \rightarrow \Rc \otimes \Rc^\vee \quad \text{and} \quad 
  d_\Rc : \Rc^\vee \otimes \Rc \rightarrow \Wc \ ,
\labl{eq:bd-for-W}
and the image of $b_\Rc$ in $\Rc \otimes \Rc^\vee$ should not give zero when 
fused with either $\Rc$ or $\Rc^\vee$.
In the general logarithmic case, the conjugate 
representation  $\Rc^*$ does not automatically satisfy these properties, 
and thus the dual representation $\Rc^\vee$
may not agree with the conjugate representation $\Rc^*$
(or may not even exist at all).  For example, each of the 
irreducible representations $\Wc(h)$ is self-conjugate, $\Wc(h)^* = \Wc(h)$, 
but for $\Wc(0)$ and $\Wc(2)$ we have the fusions
\begin{equation}\label{1.8}
\Wc(0) \otimes \Wc(0) = \Wc(0) \ ,\qquad
\Wc(2) \otimes \Wc(2) = \Wc^*\ .
\end{equation}
Thus $\Wc(0)$ is not self-dual since $d_{\Wc(0)}$ is zero --- there is 
simply no non-zero intertwiner from $\Wc(0)$ to $\Wc$. 
Furthermore, because any intertwiner from $\Wc$ to $\Wc^*$ has to factor 
through $\Wc(0)$, the image of $b_{\Wc(2)}$ is contained in 
$\Wc(0) \subset \Wc(2) \otimes \Wc(2)$. But $\Wc(0) \otimes \Wc(2)=0$ 
and so $\Wc(2)$ is not self-dual either.  
In fact, neither $\Wc(0)$ nor $\Wc(2)$ have a dual representation at all.
The same also holds for $\Wc(1)$, $\Wc(5)$, and $\Wc(7)$. 
\medskip

We believe that the indecomposable representations listed in 
\eqref{eq:W23-irreps} and \eqref{eq:indec-W-rep} which are not
in grey boxes are all self-dual and self-conjugate, see Appendix \ref{app:embed}. 
In particular, for these representations the conjugates agree with the duals.

\subsubsection{Exactness of the fusion product}\label{sec:tens-exact}

Another strange feature of the $\Wc_{2,3}$ theory is that the fusion product is not
exact, {\it i.e.}\ that  fusion does not in general respect exact sequences. 
Indeed, if we consider the fusion of each entry of \eqref{eq:Wbar-seq} with $\Wc(0)$, 
using the fusion rules (\ref{1.8}) as well as
\begin{equation}\label{1.9}
  \Wc(0) \otimes \Wc^* = 0
  \quad \text{and} \quad
  \Wc(0) \otimes \Wc(h) = 0 \quad \text{for}~ h\neq 0 \  , 
\end{equation}
we get the sequence 
$0 \longrightarrow \Wc(0)  \longrightarrow 0  \longrightarrow 0  \longrightarrow 0$
which is clearly not exact. However, the fusion rules we have determined appear
to be right-exact, {\it i.e.}\ the last three entries of an exact sequence are mapped 
to an exact sequence under fusion.

\subsubsection{The Grothendieck group}\label{sec:Gr-group-vs-ring}

The Grothendieck group $\mathrm{K}_0 \equiv \mathrm{K}_0(\Rep(\Wc))$ 
of the category of representations of $\Wc$ is, roughly
speaking, the quotient set obtained by identifying two representations
if they have the same character\footnote{
This is true for $\Wc_{2,3}$, and whenever the characters 
of all irreducible representations are linearly independent. We recall the general definition 
in Section~\ref{sec:K0-and-K0d}.}. 
Let us denote the equivalence class
of a representation $\Rc$ by $[\Rc]$. The group operation is addition, defined via
the direct sum of representations,
\be
  [\Rc] + [\Rc'] = [\Rc \oplus \mathcal{R}']  \ .
\ee
For example, given the exact sequence (\ref{eq:Wbar-seq}),
we have 
\be
  [\Wc^*] = [\Wc(0) \oplus \Wc(2)] = [\Wc(0)] + [\Wc(2)]  
\ee  
since (\ref{eq:Wbar-seq}) implies that the characters obey
$\chi_{\Wc^*} = \chi_{\Wc(0)} +  \chi_{\Wc(2)}$. 
If $\Rc_1,\dots,\Rc_n$ are the
irreducible representations, one can convince oneself that the Grothendieck group
is the free abelian group generated by $[\Rc_1],\dots,[\Rc_n]$, 
{\it i.e.}\ $\mathrm{K}_0 = \Zb [\Rc_1] \oplus \cdots \oplus \Zb [\Rc_n]$.
In other words, the elements of $\mathrm{K}_0$ are all linear combinations of the $[\Wc(h)]$ 
with integer coefficients, where $h$ takes one of the 13 values from the Kac table 
in \eqref{eq:W23-irreps}.

For non-logarithmic rational conformal field theories,        
the Grothendieck group also has a product structure which is defined by
\be
  [\Rc] \cdot [\mathcal{S}] = [\Rc \otimes \mathcal{S}] \ .
\labl{eq:K0-prod}
The physical significance of this product is that the character associated
to $[\Rc] \cdot [\mathcal{S}^*]$ is precisely the character of the open string spectrum between
the Cardy boundary conditions $\Rc$ and $\mathcal{S}$ \cite{Cardy:1989ir}.
Furthermore, the structure constants of the multiplication \eqref{eq:K0-prod}
are determined by the Verlinde formula.
A similar structure also appears for the 
$\Wc_{1,p}$ models \cite{Fuchs:2003yu,Fuchs:2006nx,Gaberdiel:2007jv}.
\medskip

As we have mentioned before, the construction of the boundary conditions
is more subtle in the $\Wc_{2,3}$ model. This is reflected by the fact that the product
\eqref{eq:K0-prod} is actually not well-defined on $\mathrm{K}_0$. To see this we observe
that we can compute $[\Wc(0)] \cdot [\Wc^*]$ in two ways:
\be\begin{array}{ll}
  [\Wc(0)] \cdot [\Wc^*] \etb= [\Wc(0) \otimes \Wc^*] = 0~~\text{versus} \enl
  [\Wc(0)] \cdot [\Wc^*] 
  \etb= [\Wc(0)] \cdot ([\Wc(0)] + [\Wc(2)])
  \\[.2em]
  \etb= [\Wc(0) \otimes \Wc(0)] + [\Wc(0) \otimes \Wc(2)] = [\Wc(0)] ~.
\eear\labl{eq:no-induced-prod}
Thus for $\Wc_{2,3}$ the fusion of representations does not induce a product on the 
Grothendieck group. 
However, we can restrict ourselves to the subset of those representations
that correspond to consistent boundary conditions, and on this subset
it is in fact possible to define the product \eqref{eq:K0-prod} --- this
will be explained in more detail in Section~\ref{sec:cyl-pf}.

For completeness, we also mention that one can define the multiplication 
\eqref{eq:K0-prod}  on the quotient
\be
  \widetilde{\mathrm{K}}_0 = \mathrm{K}_0 \, / \, (\Zb \, [\Wc(0)])~.
\labl{eq:K0-quot}
This amounts to setting $[\Wc(0)]$ to zero. The resulting ring structure on the quotient 
$\widetilde{\mathrm{K}}_0$ coincides with the one described in \cite[Sect.\,6.3]{Feigin:2006iv} 
using quantum groups (we explain the relation to the notation of \cite{Feigin:2006iv} in 
Appendix~\ref{app:dict}).

\subsubsection[Properties of the category $\Rep(\Wc)$]{Properties of the category 
$\boldsymbol{\Rep(\Wc)}$}\label{sec:RepW-prop}

It is instructive to summarise the properties of the representation category
of the $\Wc_{2,3}$ vertex operator algebra, and compare them to those of
the usual non-logarithmic Virasoro minimal models  $\Vc_{p,q}$, and the 
logarithmic $\Wc_{1,p}$ models. In the following table,
$\Vc$ is the vertex operator algebra, 
the ticks `$\surd$' are results which have been proved, the ticks in brackets 
`$(\surd)$' are supported by evidence but not proved, and the negative results `
${\hspace{-3.5pt}-}$' are proved by counter-example.

\begin{center}
\begin{tabular}{rl|ccc}
& & $\Vc = \Vc_{p,q}$ & $\Vc = \Wc_{1,p}$ & $\Vc = \Wc_{2,3}$ \\
\hline
1) & $L_0$ diagonalisable on $\Vc$ & $\surd$ & $\surd$ & $\surd$ \\
2) & $\mathrm{End}(\Vc) = \Cb \id_{\Vc}$  & $\surd$ & $\surd$ & $\surd$ \\
3) & $\Vc$ irreducible  & $\surd$ & $\surd$ & $-$ \\
4) & $\Rep(\Vc)$ is a braided tensor category & $\surd$ & $\surd$ & $(\surd)$ \\
5) & $\Rep(\Vc)$ has duals & $\surd$ & $(\surd)$ & $-$ \\
6) & Tensor product is right-exact & $\surd$ & $(\surd)$ & $(\surd)$ \\
7) & Tensor product is exact & $\surd$ & $(\surd)$ & $-$ \\
8) & Tensor product induces a product on $\mathrm{K}_0$ & $\surd$ & $(\surd)$ & $-$ \\
9) & $\Vc$ has finite number of irreducibles & $\surd$ & $\surd$ & $(\surd)$ \\
10) & $\Rep(\Vc)$ is semi-simple & $\surd$ & $-$ &  $-$
\end{tabular}
\end{center}

\noindent A few comments are maybe in order: $\mathrm{End}(\Vc)$ in 2) 
describes the space  of  $\Vc$-intertwiners from $\Vc$ to itself, and in all three cases this 
just consists of multiplication by complex numbers. This is automatic if $\Vc$ is irreducible, 
but it is also true for $\Wc_{2,3}$ since the intertwiner is uniquely determined by 
its action on the cyclic vector $\Omega$, and because the $L_0$-eigenspace 
of eigenvalue $0$ is one-dimensional, it can only map $\Omega$ to a multiple of itself.
In category-speak this means that $\Wc_{2,3}$ is absolutely simple but not simple. 

\medskip

Let us give some references to the literature where the results in the above 
table can be found:
\\[.3em]
\nxt $\boldsymbol{\Vc_{p,q}:}$  1)--3) hold by construction, and for 9) and 10) see
\cite[Def.\ 2.3 \& Thm.\ 4.2]{Wang:1993}. The existence of a braiding\footnote{
In all three cases of the above table, the map $e^{-2\pi i L_0}$ should endow 
the representation category with a twist in the sense of \cite[Def.\,6.1]{Joyal:1993}. 
Alternatively, the twist can be introduced as a morphism derived from a functorial 
isomorphism from a representation to its double-dual as in \cite[Sect.\,2.2]{BaKi-book}. 
The latter formulation requires the existence of duals. As far as we can tell, in the vertex 
operator algebra literature the question of the existence of a twist and of duals has not 
been addressed separately, and so we have omitted the twist as a separate property 
from the table.}
and tensor product follow from  \cite[Thm.\,3.10]{Huang:1995}, and the duality 
morphisms are constructed in \cite[Thm.\,3.8]{Huang2005}, 
establishing 4) and 5). 
The existence of duals implies that the tensor product is exact 
(see \cite[Prop.\,2.1.8]{BaKi-book}), which in turn guarantees that the product on 
$\mathrm{K}_0$ is well-defined, so that 6)--8) hold as well.
\\[.3em] 
\nxt $\boldsymbol{\Wc_{1,p}:}$  1)--3) again hold by construction, see \cite{Kausch:1990vg} 
and the free field approach in \cite{Fuchs:2003yu}. 
A tensor product theory for vertex operator algebras with logarithmic intertwiners has 
been developed in \cite{Huang:2007ir}. By \cite[Prop.\,4.1]{Huang:2007mj}, the theory can 
be applied for vertex operator algebras which are $C_2$-cofinite and of positive energy. 
By \cite{Carqueville:2005nu,Adamovic:2007er} the $\Wc_{1,p}$-vertex operator algebras 
are of this type, and as a consequence also satisfy  9). It follows that $\Rep(\Wc_{1,p})$, 
defined as in \cite[Prop.\,4.3]{Huang:2007mj}, is a braided tensor category 
\cite[Thm.\,4.11]{Huang:2007mj}. This establishes point 4). That 10) does not hold can 
be seen for example from \cite[Sect.\,2.4]{Fuchs:2003yu} or \cite[Sect.\,4]{Adamovic:2007er}. 
Finally, 5) (and consequently 6)--8)) would follow from \cite[Conj.\,4.2]{Huang:2009}.
\\[.3em] 
\nxt $\boldsymbol{\Wc_{2,3}:}$ The explicit free field construction of 
\cite[Def.\,4.1\,\&\,Thm.\,4.2]{Feigin:2006iv} establishes 1) and 2). It also shows that
3) and 10) do not hold. Counterexamples to 
5), 7) and 8) were provided in Sections \ref{sec:contra-dual}, \ref{sec:tens-exact} and 
\ref{sec:Gr-group-vs-ring}, respectively. As far as we know, it has not been proven that
$\Wc_{2,3}$ is $C_2$-cofinite and that the tensor product theory of \cite{Huang:2007ir} can be 
applied to this model. However, the results of \cite{Rasmussen:2008ii,Rasmussen:2008ez} 
and the fusion rule computations of  Section \ref{sec:W-rep} strongly support 4).
The tests we have for 6) are far less stringent, but it is certainly a natural property to expect. 
Finally, in favour of 9) we observe that there are only a finite number of irreducible 
representations which can be obtained via the free field construction 
of \cite[Sect.\,4.3]{Feigin:2006iv}, and that no additional irreducible representations 
appear in our computation of fusion products.

\subsection{Boundary conditions and open string spectra}\label{sec:intro-W-bnd}

With this detailed understanding of the fusion rules, we
can now turn to describing the possible boundary conditions, their
open string spectra and the OPEs of the corresponding boundary fields. We shall only 
consider boundary conditions that preserve the $\Wc$ symmetry. 
Let us first explain more precisely what we mean by this --- since we want 
to construct a consistent boundary theory without having to specify the bulk theory first,
this is somewhat subtle.

\subsubsection[$\Wc$-symmetric boundary conditions]{$\boldsymbol{\Wc}$-symmetric boundary conditions}\label{sec:W-sym-bc}

In order to speak of a $\Wc$-symmetric boundary condition, we are implicitly assuming 
that the corresponding bulk theory has the symmetry 
$\Wc_\text{left} \otimes_\Cb \Wc_\text{right}$,  with 
$\Wc_\text{left} = \Wc_\text{right} = \Wc$. In other words, 
there is an inclusion of
$\Wc_\text{left} \otimes_\Cb \Wc_\text{right}$ into the space of states of the bulk theory
$\Hc_\text{bulk}$ which respects operator products. This then turns 
$\Hc_\text{bulk}$ into a representation of $\Wc_\text{left} \otimes_\Cb \Wc_\text{right}$.
For logarithmic conformal field theories, the inclusion of 
$\Wc_\text{left} \otimes_\Cb \Wc_\text{right}$ need not be a direct summand of 
$\Hc_\text{bulk}$; the $\Wc_{1,p}$ models provide an example of this 
\cite{Gaberdiel:1998ps,Gaberdiel:2007jv}.

Suppose we consider such a conformal field theory on the upper half plane with a
boundary condition on the real line labelled by $A$. For $A$ to be a 
$\Wc$-symmetric boundary condition we demand that on the real line the fields of the 
left- and right-moving copy of $\Wc$ are related by 
$W_\text{left}(x) = W_\text{right}(x)$, where $W \in \Wc$ and 
$W_\text{left} = W \otimes_\Cb \Omega$, 
$W_\text{right} = \Omega \otimes_\Cb W$ \cite{Cardy:1984bb}. 
This implies, in particular, 
that there is a map $\eta_A : \Wc \rightarrow \Hc_{A\rightarrow A}$ and
that the boundary fields  $\Hc_{A\rightarrow A}$ on $A$ form a representation of $\Wc$.
Similarly, the spaces $\Hc_{A\rightarrow B}$ of boundary changing fields between 
two $\Wc$-symmetric  boundary conditions $A$ and $B$ are $\Wc$-representations. 

Given two (not necessarily different) boundary conditions $A$ and $B$ we require that the two 
point correlators of boundary (changing) fields are 
non-degenerate. Otherwise, if, say, a field $\psi \in \Hc_{A \rightarrow B}$ had 
zero two-point correlator with all fields in $\Hc_{B\rightarrow A}$, then $\psi$ 
would vanish in all correlation functions and we should replace $\Hc_{A \rightarrow B}$ 
with its quotient by the kernel of the two-point correlator. We assume that this has been 
done, and so all two-point correlators are non-degenerate. The two-point correlators 
themselves are determined by the OPE of boundary fields and their one-point correlators. 
We describe the one-point correlator on a boundary with label $A$ by a 
$\Wc$-intertwiner $\eps_A : \Hc_{A\rightarrow A} \rightarrow \Wc^*$. The reason to 
take the image of $\eps_A$ to be $\Wc^*$ rather than $\Cb$ is that this can be more 
directly translated into a condition defined in the category $\Rep(\Wc)$. The 
interpretation is that each boundary field gives rise to a linear functional on 
$\Wc$ by placing the boundary field at $0$ and a field in $\Wc$ at $\infty$, using the 
embedding $\eta_A$. The one-point correlator itself is obtained by placing the vacuum 
$\Omega$ at $\infty$.

We shall also demand that $\eta_A : \Wc \rightarrow \Hc_{A\rightarrow A}$ is 
{\em injective}. For suppose $\Nc \subset \Wc$ is annihilated by $\eta_A$. 
The sewing constraint arising from the two-point correlator on the upper half plane 
\cite{Cardy:1991tv,Lewellen:1991tb} shows that a correlator on the upper half plane which 
involves at least one field from $\Nc$ has to vanish. It follows that $\Nc$ is an ideal in 
$\Wc$, and that $\Nc$-descendents in $\Hc_\text{bulk}$ act as zero. We are therefore
no longer describing a conformal field theory with $\Wc$-symmetry, but a 
conformal field theory with $\Wc/\Nc$-symmetry in the presence of a $\Wc/\Nc$-symmetric 
boundary condition.
In particular, the bulk theory reconstructed with the procedure of \cite{Gaberdiel:2007jv} may
then depend on the boundary condition we start from.
\medskip

Summarising the above discussion, a consistent $\Wc$-symmetric boundary
theory thus consists of the following data:
\begin{itemize} 
\itemsep -.1em
\item[-] a collection $\Bc = \{ A, B, \dots \}$ of labels for boundary conditions,
\item[-] for each pair of labels $A,B$ an open string spectrum $\Hc_{A\rightarrow B}$ 
which is a $\Wc$-representation,
\item[-] a boundary OPE $\Hc_{B\rightarrow C} \times \Hc_{A\rightarrow B} \rightarrow 
\Hc_{A\rightarrow C}$ compatible with the $\Wc$-symmetry,
\item[-] $\Wc$-intertwiners $\eta_A : \Wc \rightarrow \Hc_{A\rightarrow A}$,
\item[-] one-point correlators $\eps_A : \Hc_{A \rightarrow A} \rightarrow \Wc^*$ which are 
$\Wc$-intertwiners.
\end{itemize}
These data should satisfy the following sewing constraints and non-degeneracy conditions:
\begin{itemize}
\item[B1] The boundary OPE is associative.
\item[B2] The two-point correlator obtained by taking the OPE of two boundary fields and 
evaluating with $\eps_A$ is non-degenerate.
\item[B3] $\eta_A$ is injective and $\eta_A(\Omega)$ is the identity field on the $A$-boundary.
\end{itemize}
Conditions B1 and B2 are certainly necessary if we want a consistent theory whose states
are distinguishable in correlators. Condition B3 has a different status, because dropping 
it does not lead to inconsistencies of the boundary theory. We impose it in our analysis for
the reason outlined above.
 
For a non-logarithmic rational vertex operator algebra $\Vc$ there is a canonical 
boundary theory \cite{Cardy:1989ir,Runkel:1998pm,Behrend:1999bn,Felder:1999mq}: 
$\Bc$ consists of all $\Vc$-representations, $\Hc_{A\rightarrow B} = B \otimes A^*$ and 
the OPE can be defined using the duality intertwiners $d_A$ 
({\it cf.}\ Section~\ref{sec:dual-nondeg} below). The same is true for the logarithmic 
rational $\Wc_{1,p}$ models \cite{Gaberdiel:2006pp,Gaberdiel:2007jv}. 
For the $\Wc_{2,3}$ model this ansatz turns out to work as well, but with one crucial 
difference: we can no longer assign consistent boundary conditions to {\em all} 
$\Wc$-representations, but only to a subset, as we will illustrate now.

\subsubsection[A boundary theory for the $\Wc_{2,3}$ model]{A 
boundary theory for the $\boldsymbol{\Wc_{2,3}}$ model}\label{sec:intro-W23-bnd}

Given two representations $A$ and $B$, there is a general categorical construction,
called the `internal Hom' $[A,B]$ (see Section~\ref{sec:iHom-assoc} below), which is the natural 
candidate for the open string spectrum, $\Hc_{A \rightarrow B}=[A,B]$. The reason for this 
proposal is that the internal Hom construction provides us with an 
associative boundary OPE, {\it i.e.}\ that B1 is automatically satisfied. 

Actually, as also explained in Section~\ref{sec:iHom-assoc} below, we can always
express the internal Hom as 
\be
[A,B] = \big( A \otimes B^* \big)^{\!*}\ ,
\ee
where $A^*$ is the conjugate representation to $A$. 
In order to see that this proposal for the boundary spectrum is not so unnatural, 
consider the case $A=B=\Wc(2)$. Recall that $\Wc(2)^* \cong \Wc(2)$ 
and $\Wc(2) \otimes \Wc(2) = \Wc^*$ so that 
$\Hc_{\Wc(2) \rightarrow \Wc(2)} = [\Wc(2),\Wc(2)]=\Wc$. Had we taken 
$\Hc_{A \rightarrow B} = B \otimes A^*$ as in the non-logarithmic case,
the result for $\Hc_{\Wc(2) \rightarrow \Wc(2)}$ would have been $\Wc^*$ 
which is different from $\Wc$ and
does not allow for a unit $\eta : \Wc \rightarrow \Wc^*$ (because this would
factor through $\Wc(0)$ and $\Wc(0) \otimes \Wc^* = 0$).
To summarise:
\be
  \Bc_\text{first try} = \{ \, \text{all $\Wc$-representations} \, \}
  ~~ \text{satisfies B1} ~.
\labl{eq:B-bad-choice}
However, it turns out that this attempt violates B2 and B3. To see that B2 fails consider 
$A=B=\Wc(2)$, for which we have just seen that 
$\Hc_{\Wc(2) \rightarrow \Wc(2)} = \Wc$. Recall from Section~\ref{sec:contra-dual} 
that a boundary condition with self-spectrum $\Wc$ does not allow for a non-degenerate 
two-point correlator, irrespective of what we choose for $\eps$, simply because 
$\Wc \not\cong \Wc^*$. 
B3 fails for $A = B= \Wc(0)$ because in this case $\Hc_{\Wc(0) \rightarrow \Wc(0)} = \Wc(0)$, 
and while there is an intertwiner $\eta_{\Wc(0)} : \Wc \rightarrow \Wc(0)$, it is not injective.

The obvious method to circumvent these problems is to remove all boundary labels from 
\eqref{eq:B-bad-choice} for which B2 and B3
fail. A necessary condition for B3 to hold is that 
there exists an injective intertwiner $\Wc \rightarrow \Hc_{A \rightarrow A}$. We have just seen 
that this eliminates $\Wc(0)$, and in fact this is the only indecomposable representation ruled 
out by this criterion. A necessary condition for B2
to hold is that for any boundary label $A$ we 
have $\big( A \otimes A^* \big)^{\!*} \cong A \otimes A^*$. This eliminates the irreducible 
representations $\Wc(1)$, $\Wc(2)$, $\Wc(5)$, $\Wc(7)$, as well as the indecomposable 
representations $\Wc$, $\Wc^*$, $\Qc$, $\Qc^*$. Coming from the opposite direction, 
we will prove in Section~\ref{sec:dual-nondeg} that the following holds:
\be
\Bc = \Bigg\{ ~ \begin{minipage}{18em} all $\Wc$-representations $A$ for which\\
$A^*$ is a dual for $A$ such that $b_A$ is injective \end{minipage}  \Bigg\} \qquad
 \text{satisfies B1--B3} \ .
\labl{eq:B-good-choice}
Here $b_A$ is the duality morphism mentioned in \eqref{eq:bd-for-W}; it will serve to 
construct the unit $\eta_A$. We believe (but we have no proof) that of the 
35 indecomposable $\Wc_{2,3}$-representations we consider in this paper, namely those listed in 
\eqref{eq:W23-irreps} and \eqref{eq:indec-W-rep}, only the 
$26$ representations that are not written in a grey box are in $\Bc$. 

These $26$ indecomposable representations agree precisely with the boundary conditions 
considered in \cite{Rasmussen:2008ii}. There, the boundary conditions were found 
by analysing a lattice model on a strip, while we obtain the list by representation theoretic 
arguments intrinsic to conformal field theory. 
\smallskip

We will prove in Section~\ref{sec:dual-nondeg} that if $A,B \in \Bc$ then 
$\big( A \otimes B^* \big)^{\!*} \cong B \otimes A^*$. The open string spectra thus take the 
same form as in the non-logarithmic case,
\be
  \Hc_{A \rightarrow B} = B \otimes A^*
  \quad \text{for}~A,B \in \Bc \ .
\labl{eq:open-string}
The construction of the boundary theory is then completely analogous to 
non-logarithmic rational conformal field theories and the $\Wc_{1,p}$ models;
we provide the details in Section~\ref{sec:iHom-assoc-etc}. 

We also prove (see Theorem~\ref{thm:non-deg}) that the space $\Hc_{A \rightarrow B}$
is always non-zero, {\it i.e.}\ that there is a non-trivial spectrum
of open strings between any two boundary conditions in $\Bc$. This is not true if for
example the representation $\Wc(0)$ would be an allowed boundary condition. 
$\Wc(0)$ still satisfies B1 and B2 but not B3. Indeed, the spectrum of open strings between 
$\Wc(0)$ and $\Wc(2)$ would be
$\Hc_{\Wc(0) \rightarrow \Wc(2)} = \big( \Wc(0) \otimes \Wc(2)^* \big)^* = 0$ because 
$\Wc(0) \otimes \Wc(2)=0$.

\subsubsection{Cylinder partition functions}\label{sec:cyl-pf}

As was already alluded to in Section~\ref{sec:Gr-group-vs-ring}, 
the product structure of the Grothendieck group of a rational
non-logarithmic theory is closely related to the cylinder diagram between two 
Cardy boundary conditions $A$ and $B$
\be
  Z(q)_{A \rightarrow B} = \text{tr}_{\Hc_{A\rightarrow B}}\big( q^{L_0 - c/24} \big) \ ,
\labl{eq:ZAB-def}
where $q=\exp(2 \pi i \tau)$. 
Indeed, for Cardy boundary conditions, the open string spectrum is 
described by $B\otimes A^*$, and the character of $B\otimes A^*$ only
depends on the class $[B\otimes A^*]$ in the Grothendieck group. As we 
have seen above (\ref{eq:open-string}), for boundary conditions 
labelled by $A,B\in{\cal B}$, the
open string spectrum is still given by $B\otimes A^*$, and the 
character again only depends on $[B\otimes A^*]$. One may therefore expect
that there should be a consistent product on the subgroup of the Grothendieck
group that comes from the consistent boundary conditions. With this in mind we 
introduce the subgroup of $\mathrm{K}_0$ defined by
\be
  \Kdd = \big( \text{ subgroup generated by $[\Rc]$ for 
  all $\Rc \in \mathcal{B}$ as defined in \eqref{eq:B-good-choice} } \big) \ .
\labl{eq:K0dbar-def}
We shall give an explicit description of $\Kdd$ in 
Section~\ref{sec:W23-K0-calc} below, and we shall show in 
Section~\ref{sec:K0-and-K0d} that on $\Kdd$ the product 
$[\Rc] \cdot [\Rc'] := [\Rc \otimes \Rc']$ is indeed well-defined 
and associative. 

The fact that the product is well-defined now implies that two
boundary conditions $A,A' \in \Bc$ for which $[A]=[A']$
cannot be distinguished in any cylinder partition function,
\be
   A,A' \in \Bc \text{ and } [A]=[A']
   \quad 
   \Rightarrow
   \quad
   Z(q)_{A \rightarrow B} = Z(q)_{A' \rightarrow B}
   \text{ for all } B\in\Bc ~.
\labl{eq:ZAB=ZA'B}
Actually we will see in Section~\ref{sec:W23-K0-calc} that $A$ and $A'$ cannot be 
distinguished in cylinder partition functions (\ref{eq:ZAB=ZA'B}) 
even if $A$ and $A'$ only coincide 
in the quotient $\widetilde{\mathrm{K}}_0$ defined in \eqref{eq:K0-quot}. 
On the other hand, as opposed to $\Kdd$ one cannot read off the 
cylinder partition functions directly from the product in 
$\widetilde{\mathrm{K}}_0$. 
For example, if $A = 2 \Wc(\frac{5}{8})$ and $A' = {\cal R}^{(2)}(2,7)$, 
the partition functions $Z(q)_{A \rightarrow A}$ and $Z(q)_{A' \rightarrow A'}$ differ by 
$2\chi_{\Wc(0)}(q)$. This difference is visible in $\Kdd$ but not in 
$\widetilde{\mathrm{K}}_0$. It is therefore not clear to us whether 
$\widetilde{\mathrm{K}}_0$ has a direct physical interpretation.
\smallskip

It would be interesting to see if the product structure on $\Kdd$ can be 
described by a Verlinde-like formula; for $\widetilde{\mathrm{K}}_0$ such a formula 
was obtained in \cite{Semikhatov:2007qp}. 

\subsubsection{Boundary conditions and boundary states}\label{sec:bndcond-bndstate}

Finally, let us comment on the relation between boundary conditions and boundary states. 
Recall that the boundary states encode
the one-point functions of bulk fields 
on the disc. On the other hand, a boundary condition is in addition specified by the 
bulk-boundary OPE, as well as by
the OPE of the boundary fields amongst themselves. 
This then also specifies how the open string spectra decompose into
$\Wc$-representations.

Given the property \eqref{eq:ZAB=ZA'B} of cylinder partition functions, it seems likely that the
boundary conditions in logarithmic conformal field theories are in general not uniquely 
characterised  by their boundary states. This phenomenon is already visible
for the $\Wc_{1,p}$ models whose boundary theory was analysed in 
\cite{Gaberdiel:2006pp,Gaberdiel:2007jv}. There, 
boundary states (and in \cite{Gaberdiel:2006pp} even the entire boundary 
condition including OPEs)  were constructed for the irreducible representations.
The construction of the present paper shows that one can find a consistent boundary
theory in the sense of B1--B3
also for the other representations (including the 
indecomposable representations).\footnote{To connect these to a bulk theory,
one obviously still has to construct a consistent
bulk-boundary OPE for these additional boundary conditions, but
we believe that this is indeed possible.}
The open string spectra of these boundary conditions will still be
given by the fusion rules --- see \eqref{eq:open-string} above and 
Eq.\ (2.21) of  \cite{Gaberdiel:2006pp} --- and thus these 
boundary conditions will be different from the (superpositions of the) irreducible 
representations that make up the same character. On the other hand, given the 
analysis of  \cite{Gaberdiel:2006pp,Gaberdiel:2007jv} it is clear that there are no additional boundary states, 
and thus their boundary states must agree.

The same phenomenon is expected to arise for the $\Wc_{2,3}$ model, although
we have not yet constructed the corresponding bulk theory, and thus do not know
how many Ishibashi states the theory actually possesses. 
In fact the above considerations suggest that for the $\Wc_{2,3}$ model, there
are precisely $12$ different Ishibashi states since 
the lattice $\Kdd$ is spanned by 12 characters --- see
Section~\ref{sec:W23-K0-calc}.

\subsubsection{A boundary theory for other $\Wc$-symmetric models}

While we only consider the $\Wc_{2,3}$ model in detail in this paper, we believe
that much of the structure we have found generalises to other models (in particular,
but not exclusively, to the $\Wc_{p,q}$ models). The general analysis of 
Section~\ref{sec:W-sym-bc} should be applicable provided that the 
(interesting) representations of $\Wc$ form a braided tensor category. One should
then be able to find a boundary theory satisfying B1--B3
with boundary labels given by \eqref{eq:B-good-choice}, and where 
the open string spectra are of the form \eqref{eq:open-string}. This follows from 
three additional properties of $\Rep(\Wc)$  (namely that it is abelian and has the 
two properties stated in condition C in Section~\ref{sec:tensor}) together with 
Theorem~\ref{thm:non-deg}.
Formulas \eqref{eq:ZAB-def}--\eqref{eq:ZAB=ZA'B} for the cylinder partition functions 
are also valid in this case, as demonstrated in Section~\ref{sec:K0-and-K0d}.

\section{Representations and fusion rules}\label{sec:W-rep}

After this long summary, we shall now describe our results in more detail. We begin
by analysing the fusion rules of the $\Wc_{2,3}$ model. As we mentioned above,
we do not know how to attack this calculation directly, and we shall therefore first
revisit the fusion rules of the Virasoro theory.

\subsection{The Virasoro theory}

For the Virasoro theory the relevant 
vertex operator algebra $\Vc$ is obtained from the 
Virasoro Verma module based on $\Omega$, by dividing out ${\cal N}_1=L_{-1}\Omega$,
but not $T= L_{-2}\Omega$. The vector $T$ is then a highest weight
state, {\it i.e.}\ it is annihilated by $L_1$ and $L_2$, but it is not the cyclic
vector of $\Vc$. 

Actually, $T$ generates the irreducible representation
$\Vc(2)$ of the Virasoro algebra with highest weight $h=2$. This representation
is the quotient space of the Verma module based on $T$ by  two independent
null-vectors ${\cal N}_3$ and ${\cal N}_5$ at level three and five, respectively 
(see (\ref{N35}) below).
In the Verma module based on $\Omega$, both ${\cal N}_3$ and ${\cal N}_5$ are 
actually descendants of ${\cal N}_1 = L_{-1} \Omega$, and hence 
both ${\cal N}_3$ and ${\cal N}_5$ are set to zero in $\Vc$. It follows that $\Vc$
indeed contains $\Vc(2)$ as a subrepresentation. In terms of exact sequences, 
the structure of the vertex operator algebra is thus 
\be
 0\longrightarrow  \Vc(2)\longrightarrow \Vc\longrightarrow \Vc(0)\longrightarrow0 \ .
\ee
Alternatively, the structure of the vertex operator algebra is described by the left-most 
diagram in 
\be
\raisebox{6em}{
\xymatrix{
 \hspace{-3em}\Vc&&\hspace{-3em}\Vc^*
 &&\hspace{-3em}\mathcal{P}&&\hspace{-3em}\mathcal{P}^*\\
 \quad\circ\ T\ar@{>}[u]|(.4){\Vc(2)}&
 &\quad\bullet\ t\ar@{>}[u]|(.4){\Vc(2)}\ar@{.>}
[d]\ar@/_1.2em/[dd]&&\times&&\times&h=2\\
 \times&&\times&&\quad\circ\ \psi\ar@/^1.2em/@{>}[uu]|(0.75){\Vc(1)}\ar@{.>}[u]&
 &\quad\bullet\ \bar{\psi}\ar@/^1.2em/@{>}[uu]|(0.75){\Vc(1)}\ar[d]\ar@{.>}
[u]&h=1\\
 \quad\bullet\ \Omega\ar@{.>}[u]\ar@/^1.2em/[uu]&&\quad\circ\ 
\omega&&\quad\bullet\ 
 \Omega\ar[u]&&\quad \circ\ \omega&h=0 
}}
\labl{eq:lvls}
Here $\bullet$ denotes the cyclic vector that generates the entire representation, 
$\circ$ are images of $\bullet$, and $\times$ denotes null vectors (which have been 
set to zero). The vertex operator algebra $\Vc$ is then not irreducible, but still 
indecomposable. 

It is easy to see from the above structure that $\Vc$ is not self-conjugate. This is to 
say, the two-point correlators involving two states from $\Vc$ do not lead
to a non-degenerate bilinear form. Indeed, it is manifest that 
\be
\langle \phi(z) T(w) \rangle =0 \ , \qquad \hbox{for any $\phi\in \Vc$.}
\ee
In fact, the conjugate representation $\Vc^*$ of $\Vc$ 
is generated from a cyclic state $t$ at conformal weight two.
The state $t$ is quasiprimary ($L_1 t=0$), but it is not highest weight since 
$L_2 t = \omega$, where $\omega$ satisfies $L_n \omega =0$ for all $n$. In terms of
exact sequences $\Vc^*$ is characterised by 
\be
 0\longrightarrow  \Vc(0)\longrightarrow \Vc^* \longrightarrow \Vc(2)\longrightarrow0 \ ,
\ee
and the structure is sketched in the second diagram in (\ref{eq:lvls}) above. 
With this definition it is then easy to see that 
\be
\langle T(z) t(w) \rangle 
= (z-w)^{-2} \, \langle \Omega(z) \, \omega(w) \rangle \neq 0 \ .
\ee
This implies that the two point correlators involving one field from $\Vc$
and one field from $\Vc^*$ give rise to a non-degenerate bilinear form.
\smallskip

Both $\Vc$ and $\Vc^*$ are obtained from the Verma module based on 
$\Omega$ by taking ${\cal N}_2$ to be non-zero. Similarly, we can consider the 
Virasoro representation where we set ${\cal N}_2=0$, but not $\psi=L_{-1}\Omega$. 
This leads to the representations ${\cal P}$ and ${\cal P}^*$, see
(\ref{eq:lvls}) above. The representation ${\cal P}$ has a null-vector at
conformal weight two, namely
\be
\mathcal{N}_{2}=(L_{-2}-\frac32 L_{-1}^2)\Omega \ . 
\ee
This vector is annihilated by $L_1$, as one can easily verify. (Note that 
$L_1 L_{-2}\Omega = 3 L_{-1} \Omega \neq 0$.)
Again, the subrepresentation generated from
$\psi$ is the irreducible Virasoro representation $\Vc(1)$ with highest weight
$h=1$; its Verma module has independent null-vectors at levels four and six, but
these are automatically zero in ${\cal P}$ since they are descendants of ${\cal N}_2$.
We can also describe the structure of these representations more formally
in terms of exact sequences:
\be
 0 \longrightarrow \Vc(1) \longrightarrow \mathcal{P} \longrightarrow \Vc(0) \longrightarrow 0
 \quad , \quad
 0 \longrightarrow \Vc(0) \longrightarrow \mathcal{P}^* \longrightarrow \Vc(1) \longrightarrow 0 ~.
\labl{eq:P-Pbar-seq}

The vertex operator algebra $\Vc$ does not define a rational theory. 
However, it has a family of  `quasirational' representations \cite{Nahm:1994by} that will play 
an important role later when we enlarge $\Vc$ to a rational $\Wc$-algebra. These 
quasirational representations are labelled by entries in the extended Kac table, see 
Table~\ref{tab:kactable}.
\begin{table}[tp]\centering
  \begin{tabular}{ccc}
    &\(s\)\\\(r\)&
  \begin{tabular}{c|cccccccc}
    &1&2&3&4&5&6&7&8\\[2pt]
   \hline
    1&0&0&\(\frac{1}{3}\)&1&2&\(\frac{10}{3}\)&5&7\\[2pt]
    2&\(\frac58\)&\(\frac18\)&\(-\frac{1}{24}\)&\(\frac18\)&\(\frac58\)&\(\frac{35}{24}\)
     &\(\frac{21}{8}\)&\(\frac{33}{8}\)\\[2pt]
    3&2&1&\(\frac13\)&0&0&\(\frac13\)&1&2\\[2pt]
    4&\(\frac{33}{8}\)&\(\frac{21}{8}\)&\(\frac{35}{24}\)&\(\frac58\)&\(\frac18\)&\(-\frac{1}{24}\)&
    \(\frac{1}{8}\)&\(\frac{5}{8}\)\\[2pt]
    5&7&5&\(\frac{10}{3}\)&2&1&\(\frac{1}{3}\)&0&0\\[2pt]
  \end{tabular}
  &\(\cdots\)\\
  &\(\vdots\)
\end{tabular}
\caption{The extended Kac table for $c_{2,3}=0$
        showing the values $h_{r,s}$ determined by \eqref{reps}.
} \label{tab:kactable}
\end{table}

\subsection{The Virasoro fusion rules}

The fusion rules of the vertex operator algebra $\Vc$ were studied in 
\cite{Eberle:2006zn}. The simplest fusion products are those of the irreducible representation
$\Vc(0)$. Since for all $n$, $L_n \Omega=0$
in $\Vc(0)$, the fusion of $\Vc(0)$ with
any state that is in the image of a Virasoro generator, {\it i.e.}\ that can be written 
as a sum of states of the form $L_n\chi$, vanishes. In particular, this is the case for 
any state in the irreducible representation $\Vc(h)$ with $h\neq 0$. On the other
hand the product of $\Vc(0)$ with itself just gives $\Vc(0)$ again. Thus we conclude 
that\footnote{Throughout this paper we shall denote `fusion' by the symbol $\otimes$. 
In order to distinguish it from the tensor product over the complex
numbers, we shall denote the latter by  $\otimes_{\Cb}$. We shall also
reserve $\otimes$ for the $\Wc$ fusions to be considered below; fusion of
$\Vc$ representations will be denoted by $\otimes_\Vc$.}
\be  \label{eq:v0fus}
  \Vc(0)\otimes_\Vc\Vc(0) =\Vc(0) \quad , \qquad
  \Vc(0)\otimes_\Vc\Vc(h_{r,s})=0 ~\text{for}~ h_{r,s}\neq0\ .
\ee
The next simplest fusion rules are those that involve the representation $\Vc(2)$. 
It was claimed in \cite{Eberle:2006zn} that $\Vc(2) \otimes_\Vc \Vc(2) = \Vc$
but this is inconsistent with associativity. Indeed, if we assume associativity, then 
it follows 
\be\label{assop}
\Vc(0) \otimes_\Vc \Vc 
= \Vc(0) \otimes_\Vc \big(  \Vc(2) \otimes_\Vc \Vc(2)  \big)
=\big( \Vc(0) \otimes_\Vc \Vc(2)  \big) \otimes_\Vc \Vc(2)  
= 0 \otimes_\Vc \Vc(2)  
= 0 \ ,
\ee
but this is not possible since
$\Vc$ is the vertex operator algebra, and hence fusion with $\Vc$ must always
act as the identity, $\Vc(0) \otimes_\Vc \Vc = \Vc(0)$. 

In order to resolve this issue, we
re-analysed  the fusion 
$\Vc(2) \otimes_\Vc \Vc(2)$  using the algorithm of 
\cite{Nahm:1994by,Gaberdiel:1996kx} 
(that was also used in \cite{Eberle:2006zn}). In this approach the fusion of two 
representations ${\cal H}_1$ and ${\cal H}_2$ of the chiral 
algebra ${\cal A}$ is the product space
\begin{align}\label{eq:fusionspace}
  {\cal H}_1\otimes{\cal H}_2:=
  ({\cal H}_1\otimes_{\Cb}{\cal H}_2)/(\Delta_{z,w}-\tilde\Delta_{z,w}) \ ,
\end{align}
where we quotient ${\cal H}_1\otimes_{\Cb}{\cal H}_2$ by the subspace
generated by $(\Delta_{z,w}(S_n)-\tilde\Delta_{z,w}(S_n)) \chi$. Here
$\chi\in {\cal H}_1\otimes_{\Cb}{\cal H}_2$,
$S_n$ is an arbitrary element of the chiral algebra ${\cal A}$ and
$\Delta_{z,w}$ and $\tilde\Delta_{z,w}$ are the two comultiplication actions 
of \cite{Gaberdiel:1993td}. 
Furthermore, $z$ and $w$ are the two points in the complex plane 
where the representations ${\cal H}_1$ and ${\cal H}_2$ are inserted. The fusion product
${\cal H}_1\otimes{\cal H}_2$ then carries an action of the chiral algebra, given either
by $\Delta_{z,w}$ or by $\tilde\Delta_{z,w}$. 

In order to unravel the structure of this fusion product, one then considers a family of 
quotient spaces, the most important of which is the quotient of ${\cal H}$ by the
states that are in the image of the negative modes, 
\be
{\cal H}^{(0)} : = {\cal H} / {\cal A}_{< 0} {\cal H} \ ,
\ee
where ${\cal A}_{< 0} {\cal H}$ is the subspace spanned by the states of the form
$S_{-n} \chi$ with $n>0$. If ${\cal H}$ is an irreducible highest weight representation,
then ${\cal H}^{(0)}$ is spanned by the highest weight state.
However, one can also
determine ${\cal H}^{(0)}$ for the case of ${\cal H} ={\cal H}_1\otimes{\cal H}_2$.
Using the
above definition of the fusion product (\ref{eq:fusionspace}) the quotient space can 
be calculated algorithmically.
Let us illustrate the analysis for the case of 
$\Vc(2)\otimes_\Vc\Vc(2)$. 
We denote the highest weight vector of $\Vc(2)$ at conformal 
weight $h=2$ by $\mu$. 
As we have already mentioned before, $\mu$ has two independent 
null-vectors, namely
\begin{align}\label{N35}
    \mathcal{N}_3&=(L_{-3}-L_{-2}L_{-1}+\tfrac16 L_{-1}^3)\mu\ , \\
    \mathcal{N}_5&=(L_{-5}-\tfrac32 L_{-4} L_{-1}-\tfrac{16}{13} L_{-3} L_{-2}
    +\tfrac34 L_{-3} L_{-1}^2+\tfrac{16}{13} L_{-2}^2 L_{-1}-\tfrac{15}{26} L_{-2} L_{-1}^3
    +\tfrac{9}{208} L_{-1}^5)\mu \ . \nonumber
\end{align}
For the case of the Virasoro modes, the comultiplications are 
\begin{align}
\Delta_{1,0}(L_0)  &= L_{-1}\otimes_{\Cb}\mathbbm{1}+L_0
\otimes_{\Cb}\mathbbm{1}
  +\mathbbm{1}\otimes_{\Cb}L_0\ , \nonumber \\
\Delta_{1,0}(L_{-1})  &=  
L_{-1}\otimes_{\Cb}\mathbbm{1}+\mathbbm{1}\otimes_{\Cb}L_{-1}\ , 
\nonumber \\
\Delta_{1,0}(L_{-n}) &=\sum_{m=-1}^\infty\binom{n+m-1}{m+1}
  \, (-1)^{m+1}L_m\otimes_{\Cb}\mathbbm{1}
  +\mathbbm{1}\otimes_{\Cb}L_{-n}\ , \qquad n\geq 2 \nonumber \\
\tilde\Delta_{0,-1}(L_{-n})  &=L_{-n}\otimes_{\Cb}\mathbbm{1}+\sum_{m=-1}^\infty
  \binom{n+m-1}{m+1}
  \, (-1)^{n+1}\mathbbm{1}\otimes_{\Cb}L_{m}\ , \qquad n\geq 2 \ . \nonumber 
\end{align}
On the space $({\cal H}_1\otimes{\cal H}_2)^{(0)}$ also the action of 
$\tilde{\Delta}_{0,-1}(L_{-n})$ can be divided out since $\tilde{\Delta}_{0,-1}(L_{-n})$ only
differs by the action of negative modes from ${\Delta}_{1,0}(L_{-n})$.

First we use the null vector $\mathcal{N}_3$  at level $3$
to conclude that $(\Vc(2)\otimes_\Vc\Vc(2))^{(0)}$ must be contained in
\begin{align}
{\rm span}\{(L_{-1}^n\mu)\otimes_{\Cb}\mu\}\supset 
(\Vc(2)\otimes\Vc(2))^{(0)}\quad n=0,1,2\ .
\end{align}
Using $\mathcal{N}_5$ and the fact that $L_{-1}\mathcal{N}_3$ and 
$L_{-1}^2\mathcal{N}_3$ are also null in $\Vc(2)$ we find that
we have the relation
\begin{align}
  (L_{-1}^2\mu)\otimes_{\Cb}\mu\cong -7(L_{-1}\mu)\otimes_{\Cb}\mu
  -8\mu\otimes_{\Cb}\mu\ .
\end{align}
These are all the relations that can be extracted from the null vectors 
$\mathcal{N}_3$ and $\mathcal{N}_5$, and we therefore conclude that 
\begin{align}
  (\Vc(2)\otimes_\Vc\Vc(2))^{(0)}=
  {\rm span}\{\mu\otimes_{\Cb}\mu,\ (L_{-1}\mu)\otimes_{\Cb}\mu\} \ .
\end{align}
On this space the $L_0$ action is then given by 
\be\begin{array}{rl}
  \Delta_{1,0}(L_0)(\mu\otimes_{\Cb} \mu) \etb=
  (L_{-1}\mu)\otimes_{\Cb}\mu+4\mu\otimes_{\Cb}\mu\\
  \Delta_{1,0}(L_0)((L_{-1}\mu)\otimes_{\Cb}\mu)\etb=
  (L_{-1}^2\mu)\otimes_{\Cb}\mu+5(L_{-1}\mu)
  \otimes_{\Cb}\mu\\
  \etb\cong-2(L_{-1}\mu)\otimes_{\Cb}\mu-8\mu\otimes_{\Cb}\mu.
\eear\ee
Thus we can represent it by the matrix 
\begin{align}
  L_0=\left(
    \begin{array}{cc}
      4&-8\\1&-2
    \end{array}
    \right) \qquad \hbox{which is conjugate to} \qquad 
  \left(
    \begin{array}{cc}
      0&0\\0&2
    \end{array}
    \right)\ .
\end{align}
This shows that $\Vc(2)\otimes_\Vc\Vc(2)$ contains precisely two 
vectors (of conformal weight zero and two) that are not images under the action of the negative
Virasoro modes. In particular, the fusion product is therefore {\em not} equal to
$\Vc$ --- it is obvious from the definition that 
$\Vc^{(0)}= \Cb\,\Omega$. 
On the other hand, the result is consistent with\footnote{
  This has also been independently observed by J\o rgen Rasmussen. We thank him for
  communicating this to us.}
\be\label{correct}
\Vc(2)\otimes_\Vc\Vc(2)=\Vc^* \ .
\ee
We have actually checked (\ref{correct}) up to level $2$, by considering larger quotient
spaces as in \cite{Gaberdiel:1996kx}, and our results are perfectly consistent with 
(\ref{correct}). 
In particular, we have checked that the action of $L_2$ maps the state at
conformal weight two to the state at conformal weight zero.
Also note that with (\ref{correct}) instead of $\Vc(2) \otimes_\Vc \Vc(2) = \Vc$
the problem with (\ref{assop}) is resolved: now associativity implies that
\be
\Vc(0) \otimes_\Vc \Vc^* = 0  \ ,
\ee
and this is actually independently correct, since every state in $\Vc^*$ is in 
the image of a (possibly positive) Virasoro mode.

We have similarly re-analysed the fusions of $\Vc(1)$, and instead of 
the claim of \cite{Eberle:2006zn} we find 
\be
\Vc(1) \otimes_\Vc \Vc(2) =  {\cal P}^*
\quad , \quad
\Vc(1) \otimes_\Vc \Vc(1) =  \Vc^*\oplus\Vc(\tfrac13)\ .
\ee
On the other hand, we have no reason to believe that there are 
problems with the other fusion rules of \cite{Eberle:2006zn}, and 
we have in fact reproduced a number of them independently. 
We therefore believe that their results are otherwise
correct, and we have used a few of them in our analysis of the 
$\Wc$ fusion rules below.

It is also worth pointing out that some of the indecomposable 
representations that appear in the fusions are not just characterised
by their highest weight, but also by some additional parameters
\cite{Gaberdiel:1996kx,Rohsiepe:1996qj,Mathieu:2007pe,Kytola:2009ax}.
In particular, this is the
case for the presentations ${\cal R}^{(2)}(0,2)_5$ and 
${\cal R}^{(2)}(0,2)_7$ of \cite{Eberle:2006zn}, for which the 
relevant parameter is called $\beta_2$ and is listed in 
Table~2 of that paper. Incidentally, the two values for $\beta_2$ 
agree precisely with what was determined already in \cite{Gurarie:1999yx}
using slightly different methods, although the interpretation is now different: 
in \cite{Gurarie:1999yx}
it was thought this implied that only some 
subsector of representations could consistently exist in a given theory. 
In the present context (see also \cite{Ridout:2008cv}),
in particular in connection with our boundary
analysis below, we see that both representations appear in the same
theory, but never together in an open string spectrum $\Hc_{A\rightarrow B}$ for 
indecomposable $\Wc$-representations $A,B$.

\subsection{$\Wc$-representations}\label{sec:W-rep-from-Vir}

Up to now we have only considered the Virasoro theory. In order to make this theory
rational we have to extend it by adjoining 3 states at conformal weight $h=15$. 
We shall denote the resulting vertex operator algebra $\Wc$. As in the Virasoro
case discussed before, $\Wc$ is again not irreducible, and its structure is similar
to that of $\Vc$. 

The $\Wc$ theory is expected to have only finitely many irreducible
representations. They 
can be expressed in terms of infinite sums of quasirational Virasoro representations
\cite[Sect.\,3.5]{Feigin:2006iv}. 
The irreducible representations are characterised by the eigenvalue of 
$L_0$ on the ground state, and we shall denote them by 
$\Wc(h)$. They contain in particular the irreducible 
Virasoro representation $\Vc(h)$. In our case, $13$ irreducible $\Wc$-representations
appear, and their conformal weights are listed in the Kac table \eqref{eq:W23-irreps}.
In addition we have the $\Wc$-analogues of the $\Vc$-representations
$\Vc^*$, ${\cal P}$ and ${\cal P}^*$. We shall denote them by 
$\Wc^*$, $\Qc$ and $\Qc^*$, respectively. The exact sequences
characterising $\Wc$ and $\Wc^*$ have been given in \eqref{eq:W-seq} and 
\eqref{eq:Wbar-seq}. For $\Qc$ and $\Qc^*$ we
have in analogy with \eqref{eq:P-Pbar-seq}
\be
  0 \longrightarrow \Wc(1) \longrightarrow \Qc \longrightarrow \Wc(0) \longrightarrow 0
  \quad,\quad
  0 \longrightarrow \Wc(0) \longrightarrow \Qc^* \longrightarrow \Wc(1) \longrightarrow 0~.
\labl{eq:Q-seq}
Further logarithmic representations appear in the various fusion products; they were listed in 
\eqref{eq:indec-W-rep};
the notation is inspired from \cite{Eberle:2006zn} 
(but here denotes $\Wc$-representations, not $\Vc$-representations as in \cite{Eberle:2006zn}).

\subsubsection*{Lifting the Virasoro fusion to $\Wc$-fusion}

Unfortunately the high weight of the additional $\Wc$-fields makes 
it difficult to determine the full chiral algebra explicitly and thus to 
calculate the fusion directly, using the methods from above. However, we can
infer the fusion rules of at least certain $\Wc$-representations from
the corresponding Virasoro fusions using induced representations. These
are the $\Wc$-representations that are of the form
\be
{\cal H}^\Wc = \Wc \otimes_\Vc {\cal H}^\Vc \ ,
\ee 
where ${\cal H}^\Vc$ is a Virasoro representation, and $\Vc$ acts on $\Wc$ by restricting 
the $\Wc$-action to the Virasoro algebra; the fusion is with respect to the Virasoro algebra. 
The fusion (with respect to the vertex operator algebra $\Wc$ ) of two such representations is 
then 
\be
  \mathcal{H}_1^\Wc\otimes\mathcal{H}_2^\Wc
  = (\Wc\otimes_\Vc\mathcal{H}_1^\Vc)
  \otimes (\Wc\otimes_\Vc\mathcal{H}_2^\Vc)
  \cong \Wc \otimes_\Vc (\mathcal{H}_1^\Vc \otimes_\Vc\mathcal{H}_2^\Vc)\ .
\ee
This allows us to calculate a certain number of $\Wc$ fusion products, based
on our knowledge of the $\Vc$ fusion rules. Combining these results with
associativity, we have managed to determine all $\Wc$ fusion rules of all 
the representations we have mentioned above. Our analysis reproduces the 
results of \cite{Rasmussen:2008ii,Rasmussen:2008ez}, but it goes beyond their 
analysis since we can also determine the fusion rules of the irreducible 
representations $\Wc(1)$, $\Wc(2)$, $\Wc(5)$ and $\Wc(7)$, as well as of the
indecomposable representations $\Wc^*$ and $\Qc^*$ --- the fusion
rules involving  $\Wc$ and ${\cal Q}$ were recently conjectured in 
\cite{Rasmussen:2008ez}.
For example, we find that the fusion of $\Wc(0)$ is trivial
except for 
\begin{equation}
\Wc(0) \otimes \Wc(0) = \Wc(0) \ .
\end{equation}
The fusion with $\Wc$ acts as the identity on every representation. Furthermore, we find
\begin{equation}
\begin{array}{rclrcl}
\Wc(2) \otimes \Wc(2) & = & \Wc^* \qquad & \Wc^* \otimes \Wc^* & = & \Wc^* \\
\Wc(2) \otimes \Wc^* & = & \Wc^* \qquad & \Wc^* \otimes {\cal Q} & = & \Qc^*  \\
\Wc(2) \otimes {\cal Q} & = & \Wc(1) \qquad  \qquad & \Wc^* \otimes \Qc^* & = & \Qc^* \\
\Wc(2) \otimes \Qc^* & = & \Qc^*  \qquad & 
        \Wc^* \otimes \Wc(1) & = & \Qc^* \\
\Wc(2) \otimes \Wc(1) & = & \Qc^* \ ,\qquad & & & 
\end{array}
\end{equation}
and
\begin{equation}
\begin{array}{rclrcl}
{\cal Q} \otimes {\cal Q} & = & \Wc \oplus \Wc(\frac{1}{3}) \qquad \qquad 
   & \Qc^* \otimes \Qc^* & = & \Wc^* \oplus \Wc(\frac{1}{3}) \\
{\cal Q} \otimes \Qc^* & = & \Wc^* \oplus \Wc(\frac{1}{3}) \qquad \qquad 
   & \Qc^* \otimes \Wc(1) & = & \Wc^* \oplus \Wc(\frac{1}{3}) \\
{\cal Q} \otimes \Wc(1) & = & \Wc(\frac{1}{3}) \oplus \Wc(2) \qquad \qquad 
   & \Wc(1) \otimes \Wc(1) & = &  \Wc^* \oplus \Wc(\frac{1}{3}) \ .
\end{array}
\end{equation}
The complete list of fusion products is given in Appendix~\ref{app:fus}.

\subsection{Multiplication on the Grothendieck group}\label{sec:W23-K0-calc}

Finally we turn to the question to which extent the fusion of representations 
induces a product on the space of characters. More formally, we want to study 
the Grothendieck group $\mathrm{K}_0 = \mathrm{K}_0(\Rep(\Wc_{2,3}))$ of the
tensor category $\Rep(\Wc_{2,3})$.
We already saw in Section~\ref{sec:Gr-group-vs-ring} that it is inconsistent to define the 
product \eqref{eq:K0-prod} on all of $\mathrm{K}_0$. However, as we will recall in 
Section~\ref{sec:K0-and-K0d}, if a representation $\mathcal{M}$ has a dual representation 
then we get a well-defined map $\mathrm{K}_0 \rightarrow \mathrm{K}_0$ given by
\be
  [\Rc] \mapsto [\mathcal{M} \otimes \Rc]~.
\labl{eq:R-with-dual-K0-map}
In words this means that if two representations $\Rc$ and $\Rc'$ have the same 
character, then so have $\mathcal{M} \otimes \Rc$ and $\mathcal{M} \otimes \mathcal{R}'$. 
We denote by $\Krr$ the subgroup of $\mathrm{K}_0$ generated by 
$[\Rc]$ for all $\Rc$ which have a dual representation.\footnote{
Here $r$ stands for `rigid', see Definition~\ref{def:Cr} below.}
It is slightly larger than the 
subgroup $\Kdd$ introduced in \eqref{eq:K0dbar-def}.

Note that if a representation has a dual representation, the dual need not be 
the conjugate representation. For example one finds that the 
$\Wc$-algebra always has $\Wc$ as a dual (see Lemma~\ref{lem:Cd-tensorcat} below), 
even though in the $\Wc_{2,3}$ model the conjugate representation $\Wc^*$ is not 
isomorphic to $\Wc$. Similarly, since $\Wc$ appears as a direct summand in the fusion 
$\Qc \otimes \Qc$ we expect $\Qc$ to be self-dual. However, the representation $\Qc$ is 
again not isomorphic to its conjugate $\Qc^*$. In any case, we have already found 
two elements of $\Krr$, namely
\be
  [\Wc]\ , \,\, [\Qc] \in \Krr \ .
\labl{eq:K0d-part1}
As pointed out in Section~\ref{sec:contra-dual}, we expect the representations 
from \eqref{eq:W23-irreps} and \eqref{eq:indec-W-rep} which are not in grey boxes to have 
duals. These include
\be
  [\Wc(h)] \in \Krr
  ~~\text{for}~~ h \in \{ \tfrac{1}{3}, \tfrac{10}{3}, \tfrac{5}{8}, \tfrac{33}{8}, 
  \tfrac{1}{8}, \tfrac{21}{8}, \tfrac{-1}{24}, \tfrac{35}{24} \}  \ ,
\labl{eq:K0d-part2}
and
\be
  [\Rc^{(2)}(0,2)_7]\ , \,\, [\Rc^{(2)}(2,7)] \ , \,\,
  [\Rc^{(2)}(1,5)] \, \in \, \Krr \ .
\labl{eq:K0d-part3}
The other representations in $\eqref{eq:indec-W-rep}$ that are not in grey boxes 
define elements in $\mathrm{K}_0$  that can be expressed as integer linear combinations
of the generators \eqref{eq:K0d-part1}--\eqref{eq:K0d-part3}.
Furthermore,
by comparing characters, it is not difficult to show that the 13 elements of $\Krr$ 
given in \eqref{eq:K0d-part1}--\eqref{eq:K0d-part3} are linearly independent.

As a consistency check of the claim that the representations in 
\eqref{eq:K0d-part1}--\eqref{eq:K0d-part3} have duals we have verified that the 
map \eqref{eq:R-with-dual-K0-map} is indeed independent of the choice of
representative for $[\Rc]$, provided we choose $\mathcal{M}$ 
from \eqref{eq:K0d-part1}--\eqref{eq:K0d-part3}. We have also verified 
that the remaining representations cannot have duals. For example, the 
characters in Appendix~\ref{app:W23-chars} show that 
$[\Rc^{(2)}(0,2)_7] = [\Wc(0) \oplus 2\Wc(2)\oplus 2\Wc(7)]$, but if we take 
$\mathcal{M} = \Wc^*$, $\Qc^*$, $\Wc(0)$, $\Wc(1)$, $\Wc(2)$, $\Wc(5)$ or $\Wc(7)$ 
\be\label{nodual}
  \big[\,\mathcal{M} \otimes \Rc^{(2)}(0,2)_7\,\big] ~\neq~ \big[\,\mathcal{M} \otimes 
  \big(\Wc(0) \oplus 2\Wc(2)\oplus 2\Wc(7)\big)\,\big]  \ ,
\ee
and thus none of these ${\cal M}$ can have duals.
We will prove in Lemma~\ref{lem:Cd-tensorcat} that if two representations $\Rc$ and $\Rc'$ 
have duals, then so does their fusion product $\Rc \otimes \mathcal{R}'$. This shows that 
the multiplication $[\Rc] \cdot [\Rc'] := [\Rc \otimes \mathcal{R}']$ provides 
a well-defined associative product on $\Krr$. This can
also be verified explicitly, using the table of fusion products in Appendix~\ref{app:fus}.
\smallskip

It would be natural to work in a basis consisting of the irreducible representations,
but this is not quite possible. To see this note that
$[\Rc^{(2)}(0,2)_7] - [\Rc^{(2)}(2,7)] = [\Wc(0)]$, so that\footnote{Note that the fact
that $[\Wc(0)]\in \Krr$ does not imply that $\Wc(0)$ has a dual representation;
as we have seen in (\ref{nodual}) it does not.}
$[\Wc(0)] \in \Krr$. Because 
$[\Wc] = [\Wc(0)] + [\Wc(2)]$, then also
$[\Wc(2)] \in \Krr$. In the same way, 
$[\Qc] \in \Krr$ implies 
$[\Wc(1)] \in \Krr$. Finally, $[\Rc^{(2)}(2,7)]-2[\Wc(2)]=2[\Wc(7)]$ and
$[\Rc^{(2)}(1,5)]-2[\Wc(1)]=2[\Wc(5)]$. Altogether we see that
\be
   \Krr = \text{span}_{\Zb}\Big(  [\Wc(h)] \big| 
   h = 0,2,1,\tfrac{1}{3}, \tfrac{10}{3}, \tfrac{5}{8}, \tfrac{33}{8}, \tfrac{1}{8}, 
   \tfrac{21}{8}, \tfrac{-1}{24}, \tfrac{35}{24} \Big) \oplus 2 \Zb [\Wc(7)]  \oplus 2 \Zb [\Wc(5)] \ .  
\labl{eq:K0d-second-basis}
In particular, $[\Wc(5)]$ and $[\Wc(7)]$ are {\em not in}  $\Krr$, but
only $2[\Wc(5)]$ and $2[\Wc(7)]$.

For completeness let us also work out some of
the structure constants in the basis given
by  (\ref{eq:K0d-second-basis}). Note that by construction, these structure constants
will be integers, but they need not be non-negative. Indeed, the
product in this basis cannot just be calculated by taking 
$[\Rc] \cdot [\mathcal{R}'] := [\Rc \otimes \mathcal{R}']$ --- this formula is only
true if both $\Rc$ and $\mathcal{R}'$ have duals. Thus in order to calculate 
the structure constants in the basis  (\ref{eq:K0d-second-basis}), we have to rewrite
the generators in terms of \eqref{eq:K0d-part1}--\eqref{eq:K0d-part3}, and then use
the product formulae for these. For example.
$[\Wc(0)] \cdot [\Wc(0)]$ is {\em not} given by $[\Wc(0) \otimes \Wc(0)] = [\Wc(0)]$, 
because $\Wc(0)$ does not have a dual. Instead we have to write 
$[\Wc(0)] = [\Rc^{(2)}(0,2)_7] - [\Rc^{(2)}(2,7)]$ and compute
\bea
  [\Wc(0)] \cdot [\Wc(0)] = 
  [\Rc^{(2)}(0,2)_7\otimes \Rc^{(2)}(0,2)_7] 
  - [\Rc^{(2)}(2,7) \otimes \Rc^{(2)}(0,2)_7]
\\[.3em] \hspace{9em}
  - [\Rc^{(2)}(0,2)_7 \otimes \Rc^{(2)}(2,7)]
  + [\Rc^{(2)}(2,7) \otimes \Rc^{(2)}(2,7)]
  = 0 \ .
\eear\ee
This also explains why the problem encountered in \eqref{eq:no-induced-prod} when 
trying to define a multiplication on $\mathrm{K}_0$ does not occur for $\Krr$: 
while it is true that $[\Wc^*] = [\Wc(0)]+[\Wc(2)] \in \Krr$, the product 
$[\Wc^*] \cdot [\Wc(0)]$ is not given by $[\Wc^* \otimes \Wc(0)]$, because 
neither $\Wc^*$ nor $\Wc(0)$ have duals.
Instead we have to express 
$[\Wc^*] = [\Wc(0)]+[\Wc(2)]$ in terms of representations which do have duals. 
For example we can use $[\Wc^*] = [\Wc]$, and express 
$[\Wc(0)]$ as above. The result is $[\Wc^*] \cdot [\Wc(0)] = [\Wc(0)]$. 
Similarly, the action of $[\Wc(0)]$ in the basis \eqref{eq:K0d-second-basis} is 
fixed by
\be\label{expro}
  [\Wc(0)] \cdot [\Wc(h)] = \begin{cases}
  [\Wc(0)] &\hbox{if $h = 1,2$} \\
  -[\Wc(0)] &\hbox{if $h = 5,7$} \\
  0 & \text{else.}
  \end{cases}
\ee
In particular, this implies that $\Zb [\Wc(0)]$ is an ideal in $\Krr$. We can
thus consider the quotient space $\widetilde{\mathrm{K}}_0$, see \eqref{eq:K0-quot},
with the quotient map $\pi : \Krr \rightarrow \widetilde{\mathrm{K}}_0$ 
given by $\pi([\Rc]) = [\Rc] + \Zb[\Wc(0)]$. Because of the factors of two in 
\eqref{eq:K0d-second-basis} the map $\pi$ is not surjective, but one can check that 
there is a (unique) multiplication on $\widetilde{\mathrm{K}}_0$ such that $\pi$ is a 
ring homomorphism (this is not obvious because the factors of two in 
\eqref{eq:K0d-second-basis}  might lead to non-integer structure constants for 
$\widetilde{\mathrm{K}}_0$). We have verified that the product on 
$\widetilde{\mathrm{K}}_0$ obtained in this way agrees with the one in
\cite[Sect.\,6.3]{Feigin:2006iv} that was constructed using quantum groups (see
Appendix~\ref{app:dict} for how to translate the notations).

The representations \eqref{eq:B-good-choice} which actually correspond to boundary 
conditions generate the subgroup $\Kdd$ of $\Krr$. 
Compared to $\Krr$, the group $\Kdd$
is missing the generators $[\Wc]$ and $[\Qc]$,
but we have to add $[\Rc^{(2)}(2,5)]$ which could previously be expressed as 
$[\Rc^{(2)}(1,5)]+2[\Wc]-2[\Qc]$. Writing the basis in a similar fashion as 
\eqref{eq:K0d-second-basis} gives
\bea
   \Kdd = \text{span}_{\Zb}\Big(  [\Wc(h)] \big| h = 0,\tfrac{1}{3}, \tfrac{10}{3}, \tfrac{5}{8}, \tfrac{33}{8}, \tfrac{1}{8}, \tfrac{21}{8}, \tfrac{-1}{24}, \tfrac{35}{24} \Big) \\[.5em]\displaystyle
   \hspace{4em}
   \oplus ~ 2 \Zb \big( [\Wc(2)]{+}[\Wc(7)] ) 
   \, \oplus \, 2 \Zb \big( [\Wc(1)]{+}[\Wc(5)] ) 
   \, \oplus \, 2 \Zb \big( [\Wc(2)]{+}[\Wc(5)] ) ~. 
\eear\labl{eq:K0dd-basis}
In particular, the lattice $\Kdd$ has only 12 basis vectors. We will prove in 
Theorem~\ref{thm:Cdd-prop} that also $\Kdd$ is closed under multiplication. 
To see this explicitly, we observe that 
\eqref{eq:K0dd-basis} is the kernel of the map $[\Rc] \mapsto [\Wc(0)] \cdot [\Rc]$, 
which goes from $\Krr$ to itself. This description of $\Kdd$ 
implies that any two representations $\Rc$ and $\Rc'$ with the property that 
$[\Rc]-[\Rc'] = n [\Wc(0)]$ for some $n$ will have the same cylinder partition function relative 
to any $[\mathcal{S}] \in \Kdd$. Indeed, we have in $\Kdd$
\be
 [\Rc] \cdot [\mathcal{S}^*]
 = [\Rc'] \cdot [\mathcal{S}^*] + ([\Rc]{-}[\Rc']) \cdot  [\mathcal{S}^*]
 = [\Rc'] \cdot [\mathcal{S}^*] + n [\Wc(0)]  \cdot  [\mathcal{S}^*]
 = [\Rc'] \cdot [\mathcal{S}^*] \ ,
\ee
where we have used that $n [\Wc(0)]  \cdot  [\mathcal{S}^*]= 0$ in 
$\Kdd$.

\section{Internal Homs and associativity}\label{sec:iHom-assoc-etc}

With this detailed understanding of the fusion rules we are now in a position to
construct a boundary theory for the $\Wc_{2,3}$ model. 
We shall switch gears, and formulate our construction in a more categorical fashion. 
First we shall explain informally why category theory is the appropriate language
(see Sections~\ref{sec:tensor} and \ref{sec:assoc-ope}). Then we shall
introduce the relevant notions that will be important to us, in particular that of an 
internal Hom (see Section~\ref{sec:iHom-assoc}), and that of dual objects (see 
Section~\ref{sec:dual-nondeg}). There we also establish the result announced in 
the introduction, namely that for the list of boundary labels \eqref{eq:B-good-choice} we 
can find a boundary theory satisfying conditions B1--B3.

\subsection{Tensor categories}\label{sec:tensor}

From the results in Section~\ref{sec:W-rep} we see that it is reasonable to assume that 
the $\Wc_{2,3}$-representa\-tions we consider form a tensor category. This is a non-trivial 
assumption (see the discussion in Section~\ref{sec:RepW-prop}), as a tensor category 
contains quite a lot of structure.

A tensor category is a tuple $\Cc \equiv (\Cc, \otimes, \one, \alpha, \lambda,\rho)$, where 
$\Cc$ is a category, $\otimes : \Cc \times \Cc \rightarrow \Cc$ is the tensor product bifunctor, 
$\one \in \Cc$ is the tensor unit, 
$\alpha_{U,V,W} : U \otimes (V \otimes W) \overset{\sim}\rightarrow (U\otimes V)\otimes W$ 
is the associator, $\lambda_U : \one \otimes U \rightarrow U$ is the left unit isomorphism, 
and $\rho_U : U \otimes \one \rightarrow U$ is the right unit isomorphism. These data are 
subject to conditions, in particular $\alpha$ satisfies the pentagon axiom and 
$\lambda$, $\rho$, $\alpha$ obey the triangle axiom. For more details on tensor categories
 the reader could consult \cite{BaKi-book,MacLane-book}.

For a tensor category $\Rep(\Wc)$ arising as the representations of a suitable vertex operator 
algebra $\Wc$ we expect three additional features. First of all, $\Rep(\Wc)$ should be 
abelian, so that in particular we can speak about kernels and quotients. Second, to each 
representation $\Rc$ we can assign a conjugate 
representation\footnote{In the vertex operator algebra literature this representation is usually referred to as 
the `contragredient' representation.} $\Rc^*$ such that 
$\Rc^{**} \cong \Rc$. In fact, we have a contravariant functor $(-)^*$ from 
$\Rep(\Wc)$ to itself whose square is naturally equivalent to the identity 
functor \cite[Notation\,2.36]{Huang:2007ir}. 
Finally, for two representations $\Rc$ and $\mathcal{S}$ there is an isomorphism between the 
spaces of intertwiners 
$\Hom(\Rc,\mathcal{S}^*)$ and $\Hom(\Rc \otimes \mathcal{S}, \Wc^*)$. To see this note 
that $\Hom(\Rc,\mathcal{S}^*)$ is by definition isomorphic to the space of conformal 
two-point blocks on the complex plane with insertions of $\Rc$ at 0 and 
$\mathcal{S}$ at $\infty$ with standard local coordinates, and 
$\Hom(\Rc \otimes \mathcal{S}, \Wc^*)$ is isomorphic to the space of conformal 
three point blocks with insertions of $\Rc$ at $0$, $\mathcal{S}$ at a point 
$z \in \Cb^\times$ and $\Wc$ at $\infty$, all treated as `in-going' punctures.
Since an insertion of the vertex operator 
algebra itself does not affect the dimension of the space of conformal blocks, these two 
spaces are isomorphic\footnote{ 
This also follows from the study of logarithmic intertwiners \cite{Huang:2007ir}.
Indeed, $\Hom(\Rc,\mathcal{S}^*) \cong \Hom(\Rc \otimes \Wc,\mathcal{S}^*)$. The latter 
space is by construction the space of intertwiners from $\Rc \times \Wc$ to $\mathcal{S}^*$.
By \cite[Prop.\,3.46]{Huang:2007ir} this space is naturally isomorphic to the space of 
intertwiners from $\Rc \times \mathcal{S}^{**}$ to $\Wc^*$. Using that 
$\mathcal{S}^{**} \cong \mathcal{S}$, this shows that we have a natural isomorphism 
$\Hom(\Rc,\mathcal{S}^*) \cong \Hom(\Rc \otimes \mathcal{S}, \Wc^*)$.
We thank Yi-Zhi Huang for a discussion of this point.}.

The last two properties motivate the following condition:
\begin{quote}
  {\bf Condition C:} The tensor category $\Cc$ is equipped with a contravariant 
  involutive  functor $(-)^* : \Cc \rightarrow \Cc$ and isomorphisms 
  $\pi_{U,V} : \Hom(U,V^*) \rightarrow \Hom(U \otimes V, \one^{\!*})$ which are 
  natural in $U$ and $V$.
\end{quote}
We denote the natural isomorphism from the
identity functor on $\Cc$ to the square of $(-)^*$ by
\be 
   \delta_U : U \rightarrow U^{**}\ .
\labl{eq:delta-def}

\subsection{Associative and non-degenerate boundary OPE}\label{sec:assoc-ope}

Next we want to describe in more detail the conditions B1--B3
of Section~\ref{sec:W-sym-bc}
that any consistent boundary theory should satisfy. Some of these conditions are nothing but
the usual requirement that the sewing constraints \cite{Lewellen:1991tb} must be satisfied. We 
shall only spell them out in some detail to make the 
categorical conditions below look less mysterious. Let us start with the OPE. 
Consider two boundary fields $\psi \in \Hc_{A\rightarrow B}$ and 
$\psi' \in \Hc_{B\rightarrow C}$,
\be
\begin{xy} 
(0,4)*{}="1";%
(15,4)*{\!\!\times\!\!}="2";%
(30,4)*{\!\!\times\!\!}="3";%
(45,4)*{}="4";%
{\ar@{-} "1";"2"}?(.5)*\dir{>} ?*!/_.7em/{A};%
{\ar@{-} "2";"3"}?(.5)*\dir{>} ?*!/_.7em/{B};%
{\ar@{-} "3";"4"}?(.5)*\dir{>} ?*!/_.7em/{C};%
(15,0)*{\psi(0)}="2";%
(30,0)*{\psi'(x)}="3";%
\end{xy}
\ee
The line in this picture is the boundary of the upper half plane, {\it i.e.}\ the real axis with 
standard orientation.
The OPE of two such fields is described by a bilinear map\footnote{
This is not quite right as the $\Wc$-representations are defined as direct sums of 
generalised $L_0$-eigenspaces, and the map $V_{C,B,A}$ in general has contributions 
in an infinite number of generalised $L_0$-eigenspaces. Instead one should use 
formal power series, including formal logarithms. We refer to \cite[Sect.\,3]{Huang:2007ir} 
for more details.}
\be
  V_{C,B,A}(-,x) : \Hc_{B\rightarrow C} \times \Hc_{A\rightarrow B} \rightarrow \Hc_{A\rightarrow C}
\labl{eq:bnd-ope}
taking $(\psi',\psi)$ to $V_{C,B,A}(\psi',x)\psi$. The map $V_{C,B,A}(-,x)$ has to be 
compatible with the $\Wc$-symmetry in the sense of intertwining operators 
(see {\it e.g.}\ \cite[Sect.\,3]{Huang:2007ir}). Condition B1 demands the OPE to be 
associative: Consider three boundary fields as follows,
\be
\begin{xy} 
(0,4)*{}="1";%
(15,4)*{\!\!\times\!\!}="2";%
(30,4)*{\!\!\times\!\!}="3";%
(45,4)*{\!\!\times\!\!}="4";%
(60,4)*{}="5";%
{\ar@{-} "1";"2"}?(.5)*\dir{>} ?*!/_.7em/{A};%
{\ar@{-} "2";"3"}?(.5)*\dir{>} ?*!/_.7em/{B};%
{\ar@{-} "3";"4"}?(.5)*\dir{>} ?*!/_.7em/{C};%
{\ar@{-} "4";"5"}?(.5)*\dir{>} ?*!/_.7em/{D};%
(15,0)*{\psi_1(0)};%
(30,0)*{\psi_2(y)};%
(45,0)*{\psi_3(x)};%
\end{xy}
\ee
Associativity means that it does not matter if we first take the OPE of $\psi_2$ with 
$\psi_1$ or that of $\psi_3$ with $\psi_2$. Written out in terms of the bilinear maps 
$V$, this condition reads
\be
  V_{D,C,A}(\psi_3,x)
  V_{C,B,A}(\psi_2,y) \psi_1
  =  V_{D,B,A}\big(V_{D,C,B}(\psi_3,x-y)\psi_2,y\big) \psi_1 \ .
\labl{eq:ope-assoc}
Condition B2 states that the bilinear pairing
\be
  (\psi',\psi) \longmapsto
  \big\langle \eps_A(V_{A,B,A}(\psi',x)\psi) \,,\, \Omega \rangle
\labl{eq:ope-nondeg}
on $\Hc_{B \rightarrow A} \times \Hc_{A \rightarrow B}$ is non-degenerate. 
Recall that $\eps_A$ is an intertwiner from $\Hc_{A \rightarrow A}$ to $\Wc^*$, 
so that one can evaluate the result on the vacuum vector $\Omega \in \Wc$.
Finally, B3 requires in particular
that $\eta_A(\Omega)$ is the identity field on $A$, {\it i.e.}\ that 
for $\psi \in \Hc_{A\rightarrow B}$,
\be
  V_{B,B,A}(\eta_B(\Omega),x) \psi = \psi
  \quad \text{and} \quad
  \lim_{x \rightarrow 0} V_{B,A,A}(\psi,x) \eta_A(\Omega) = \psi \ . 
\labl{eq:ope-unit}
\medskip

We now reformulate \eqref{eq:bnd-ope}, \eqref{eq:ope-assoc}, \eqref{eq:ope-unit}, 
and \eqref{eq:ope-nondeg} in a way which makes sense in an arbitrary tensor category 
satisfying condition C: There are morphisms
\be
  m_{C,B,A} \in \Hom(\Hc_{B\rightarrow C} \otimes \Hc_{A\rightarrow B} , \Hc_{A\rightarrow C}) ~,~~
  \eta_A : \one \rightarrow \Hc_{A \rightarrow A} ~,~~
  \eps_A : \Hc_{A \rightarrow A} \rightarrow \one^{\!*} \ ,
\ee
such that
\be\begin{array}{ll}
  \text{(associativity)} \etb
  m_{D,C,A} \circ ( \id_{\Hc_{C\rightarrow D}} \otimes m_{C,B,A}) \\ \etb \qquad
  = m_{D,B,A} \circ ( m_{D,C,B} \otimes \id_{\Hc_{A\rightarrow B}})
  \circ \alpha_{\Hc_{C\rightarrow D},\Hc_{B\rightarrow C},\Hc_{A\rightarrow B}} 
\enl
  \text{(unit property)} \etb
m_{B,B,A} \circ (\eta_B \otimes \id_{\Hc_{A \rightarrow B}}) = \lambda_{\Hc_{A \rightarrow B}}
  ~~,~~
  m_{B,A,A} \circ (\id_{\Hc_{A \rightarrow B}} \otimes \eta_A) = \rho_{\Hc_{A \rightarrow B}}
\enl
  \text{(non-degeneracy)} \etb
  \pi_{\Hc_{B \rightarrow A},\Hc_{A \rightarrow B}}^{-1}\big( \eps_A \circ m_{A,B,A} \big) 
  \text{ is an isomorphism.}
\eear\labl{eq:cat-cond-from-VOA}
Here $\alpha$, $\lambda$ and $\rho$ are defined in Section~\ref{sec:tensor}, and
$\pi$ is the isomorphism of Condition C. 
The three conditions above amount to
B1, B2 and half of B3. To account for all of B3 we need to require in addition 
that $\eta_A$ is injective.

\subsection{Internal Homs}\label{sec:iHom-assoc}

Usually the most difficult condition in constructing a consistent boundary theory 
is B1, the associativity of the OPE. In the following we shall describe a 
general construction --- the internal Hom --- which solves this condition 
automatically. In this subsection we shall give a brief overview of some properties 
of internal Homs; for more information the reader could consult {\it e.g.}\ 
\cite[Sect.\,9.3]{Majid-book}.

\begin{Def}\label{def:iHom}
Let $\Cc$ be a tensor category. Given two objects $A,B \in \Cc$, an 
{\em internal Hom} from $A$ to $B$ is an object $[A,B] \in \Cc$ together 
with a natural isomorphism $\phi^{(A,B)} : \Hom(- \otimes A,B) \rightarrow \Hom(-,[A,B])$.
\end{Def}

Naturality of $\phi^{(A,B)}$ is equivalent to the statement that for all 
$X,Y \in \Cc$ and all $g : Y \rightarrow X$, $t : X \otimes A \rightarrow B$,
\be
  \phi^{(A,B)}_X(t) \circ g = \phi^{(A,B)}_Y\big(t \circ (g \otimes \id_A) \big) \ .
\labl{eq:phi-nat}
An internal Hom need not exist, but if it does it is unique up to unique
isomorphism. 
For suppose that $[A,B]$ and $[A,B]'$ are internal Homs from $A$ to $B$ with natural 
isomorphisms $\phi^{(A,B)}$ and $\phi^{(A,B)'}$. Then there exists a unique 
isomorphism $f : [A,B] \rightarrow [A,B]'$ such that the following diagram commutes 
for all $U \in \Cc$,
\be
\begin{xy} 
(25,20)*+{\Hom(U \otimes A,B)};%
(20,20)*+{\phantom{HHH}}="H1a";%
(30,20)*+{\phantom{HHH}}="H2a";%
(0,0)*+{\Hom(U,[A,B])}="H1b";%
(50,0)*+{\Hom(U,[A,B]')}="H2b";%
{\ar "H1a";"H1b"}?*!/^1.0em/{\phi^{(A,B)}_U};%
{\ar "H2a";"H2b"}?*!/_1.4em/{\phi^{(A,B)'}_U};%
{\ar "H1b";"H2b"}?*!/_.6em/{f \circ (-)};
\end{xy}
\labl{eq:iHom-all-iso}
The morphism $f$ is obtained by taking $U =[A,B]$ and using 
${\phi^{(A,B)}_{[A,B]}}^{-1}$ and $\phi^{(A,B)'}_{[A,B]}$ to transport 
$\id_{[A,B]}$ to $\Hom([A,B],[A,B]')$.

As an example of an internal Hom consider the category of finite-dimensional 
complex vector spaces. If $A,B$ are two such vector spaces, then 
$[A,B] = B \otimes_\Cb A^*$, {\it i.e.}\ the space of linear maps from $A$ to $B$. 
Indeed, if $f:U \otimes A \rightarrow B$, is a homomorphism, then 
$\phi_U^{(A,B)}(f)$ is a homomorphism from $U\rightarrow B\otimes_\Cb A^*$. 
Evaluated on $u\in U$, $[\phi^{(A,B)}_U(f)](u)$ is an element of $B\otimes_\Cb A^*$, 
and thus a homomorphism from $A\rightarrow B$, which agrees with $f(u,-)$. 

Internal Homs also provide a different way of stating the second 
part of condition C. It is equivalent to $[V,\one^{\!*}] = V^*$.
\medskip

For Hom-spaces of a category there is an associative composition. For internal 
Hom spaces there is an analogous concept, which we review now 
(see {\it e.g.}\ \cite[Prop.\,9.3.13]{Majid-book} or \cite[Sect.\,3.2]{Ostrik:2001}). Define 
the morphisms (\emph{evaluation}, \emph{multiplication} and \emph{unit} for internal Homs)
\bea
  ev_{A,B} : [A,B] \otimes A \rightarrow B 
  ~~,~~~
  ev_{A,B} = \phi^{(A,B)~-1}_{[A,B]}(\id_{[A,B]})~,
\enl
  m_{C,B,A} : 
  [B,C] \otimes [A,B] \rightarrow [A,C]
  ~,
  \enl \qquad \qquad
  m_{C,B,A} = \phi^{(A,C)}_{[B,C] \otimes [A,B]}\big(
  (ev_{B,C} \circ ( \id_{[B,C]} \otimes ev_{A,B}) )
  \circ \alpha^{~-1}_{[B,C],[A,B],A} \big) ~,
\enl
  \eta_A : \one \rightarrow [A,A] 
  ~~,~~
  \eta_A = \phi^{(A,A)}_{\one}(\lambda_A) ~.
\eear\labl{eq:iHom-morphs}

\begin{Thm}\label{thm:assoc}
The composition of internal Homs is associative, {\it i.e.}\ on 
$[C,D]\otimes ([B,C] \otimes [A,B])$ 
we have
$$
  m_{D,C,A} \circ ( \id_{[C,D]} \otimes m_{C,B,A}) 
  = m_{D,B,A} \circ ( m_{D,C,B} \otimes \id_{[A,B]})
  \circ \alpha_{[C,D],[B,C],[A,B]} \ ,
$$
and it has $\eta$ as unit, {\it i.e.}\ on $\one \otimes [A,B]$ and $[A,B] \otimes \one$ we have
$$
  m_{B,B,A} \circ (\eta_B \otimes \id_{[A,B]}) = \lambda_{[A,B]} \ , \qquad
  m_{B,A,A} \circ (\id_{[A,B]} \otimes \eta_A) = \rho_{[A,B]} \ .
$$
\end{Thm}

The proof is by straightforward calculation. We spell it out for completeness in 
Appendix~\ref{app:iHom-proofs}. The following theorem shows that internal 
Homs exist if the tensor category satisfies condition C.

\begin{Thm}\label{thm:iHom-exist}
Let $\Cc$ be a tensor category satisfying condition $\mathrm{C}$. 
Then $[A,B] = \big( A \otimes B^* \big)^{\!*}$ 
is an internal Hom from $A$ to $B$.
\end{Thm}

\begin{proof}
Consider the sequence of isomorphisms
\bea
  \Hom(U \otimes A,B) \xrightarrow{\delta_B \circ (-)}
  \Hom(U \otimes A,B^{**}) \xrightarrow{\pi_{U\otimes A, B^*}}
  \Hom( (U \otimes A) \otimes  B^*, \one^{\!*}) 
\\[.5em]\displaystyle
  \xrightarrow{ (-) \circ \alpha_{U,A, B^*}}
  \Hom( U \otimes (A \otimes  B^*),\one^{\!*}) \xrightarrow{ \pi_{U,A \otimes  B^*}^{-1} }
  \Hom\big( U , \big( A \otimes B^* \big)^{\!*}\big) ~.
\eear\ee
The above isomorphisms are all natural in $U$, and as a consequence 
so is
$\phi^{(A,B)}_U : \Hom(U \otimes A,B) \rightarrow  \Hom\big( U , \big( A \otimes B^* \big)^{\!*}\big)$,
\be
  \phi^{(A,B)}_U(f) = \pi_{U,A \otimes  B^*}^{-1}
  \big( \pi_{U\otimes A, B^*}(\delta_B \circ f) \circ \alpha_{U,A, B^*} \big) \ .
\labl{eq:cond-C-phi}
This shows that $\big( A \otimes B^* \big)^{\!*}$ is an 
internal Hom from $A$ to $B$.
\end{proof}

For the reasons stated in Section~\ref{sec:tensor} we think it likely that 
$\Rep(\Wc_{2,3})$ is an abelian tensor category satisfying condition C. It therefore 
has internal Homs. This in turn allows to find a boundary theory which satisfies condition 
B1 by setting $\Hc_{A \rightarrow B} = [A,B]$ and choosing the morphisms defined 
in \eqref{eq:iHom-morphs}. Associativity is guaranteed by Theorem~\ref{thm:assoc}. 
However, in general $\eta_A$ need not be injective, nor need there exist a 
non-degenerate two-point correlator. We will address these two problems in the next 
section with the help of dual objects.

\subsection{Dual objects}\label{sec:dual-nondeg}

The notion of a dual object in a tensor category is a generalisation of the properties 
of the dual of a finite dimensional vector space. For a finite dimensional vector space 
$V$ over $\Cb$ (say) there is a linear map $d_V : V^* \otimes_\Cb V \rightarrow \Cb$ 
given by evaluation: $d_V( \varphi \otimes_\Cb v ) = \varphi(v)$. Conversely, if we fix a basis 
$v_i$ of $V$ and denote the dual basis by $v_i^*$ we obtain a linear map 
$b_V : \Cb \rightarrow V \otimes_\Cb V^*$ as 
$\lambda \mapsto \lambda \sum_i v_i \otimes_\Cb v_i^*$. One checks that these 
maps have the properties
\be
 (\id_V \otimes_\Cb d_V) \circ (b_V \otimes_\Cb \id_V) = \id_V \quad \text{and} \quad
 (d_V \otimes_\Cb \id_{V^*}) \circ (\id_{V^*} \otimes_\Cb b_V) = \id_{V^*} \ .
\ee
This notion is generalised to arbitrary tensor categories as follows 
(see {\it e.g.}\ \cite[Def\,2.1.1]{BaKi-book} or \cite[Def.\,9.3.1]{Majid-book}).

\begin{Def}\label{def:right-dual}
Let $\Cc$ be a tensor category. A \emph{right dual} of an object $U$ is an 
object $U^\vee \in \Cc$ together with morphisms 
$b_U : \one \rightarrow U \otimes U^\vee$ and $d_U : U^\vee \otimes U \rightarrow \one$ 
such that
$$
\begin{array}{l}\displaystyle
  \rho_U \circ (\id_U \otimes d_U) \circ \alpha_{U,U^\vee,U}^{-1} \circ (b_U \otimes \id_U) \circ \lambda_U^{-1} = \id_U \quad \text{and} 
  \enl
  \lambda_{U^\vee} \circ (d_U \otimes \id_{U^\vee}) \circ \alpha_{U^\vee,U,U^\vee} \circ  (\id_{U^\vee} \otimes b_U) \circ \rho^{-1}_{U^\vee} = \id_{U^\vee} \ .
\end{array}  
$$
\end{Def}

Just as did condition C, right duals guarantee the existence of internal Homs 
and so allow to solve condition B1 for a boundary theory.

\begin{Lem}\label{lem:dual-iHom}
Let $\Cc$ be a tensor category. If $U \in \Cc$ has a right dual then for all 
$V \in \Cc$ we can choose $[U,V] = V \otimes U^\vee$.
\end{Lem}

\begin{proof} For $f : A \otimes U \rightarrow V$ define 
$\phi^{(U,V)}_A(f) : A \rightarrow V \otimes U^\vee$ as 
\be
  \phi^{(U,V)}_A(f) = (f \otimes \id_{U^\vee}) \circ \alpha_{A,U,U^\vee} \circ 
  (\id_A \otimes b_U) \circ \rho_A^{-1} \ .
\labl{eq:phi-duals}
The properties of $b_U$ and $d_U$ can be used to check that
the map $\tilde\phi^{(U,V)}_A(g) : A \otimes U \rightarrow V$ defined 
for $g : A \rightarrow V \otimes U^\vee$  by 
\be
  \tilde\phi^{(U,V)}_A(g) = \rho_V \circ (\id_V \otimes d_U) \circ 
  \alpha^{-1}_{V,U^\vee,U} \circ (g \otimes \id_U)\ ,
\labl{eq:phi-1-duals}
is a left and right inverse to 
$\phi_A$. Thus $\phi_A$ is an isomorphism. Naturality follows by writing out both 
sides of \eqref{eq:phi-nat}, and using naturality of $\rho$ and $\alpha$ and functoriality 
of the tensor product.
\end{proof}

Substituting the explicit expressions \eqref{eq:phi-duals} and \eqref{eq:phi-1-duals} 
into \eqref{eq:iHom-morphs} gives the following result for the multiplication $m$ 
and unit morphisms $\eta$ (we do not spell out the unit isomorphisms and associators 
of the tensor category $\Cc$)
\bea
  m_{C,B,A} = \id_C \otimes d_B \otimes \id_A : (C \otimes B^\vee) 
  \otimes (B \otimes A^\vee) \rightarrow C \otimes A^\vee ~, \\[.3em]\displaystyle
  \eta_A = b_A  : \one \rightarrow A \otimes A^\vee \ .
\eear\labl{eq:iHom+dual-morphs} 

Similarly to right duals one defines the {\em left dual} of an object $U$ as an 
object $\eev U$ together with morphisms 
$\tilde b_U : \one \rightarrow \eev U \otimes U$ and 
$\tilde d_U : U \otimes \eev U \rightarrow \one$ satisfying analogous conditions to those 
for right duals, see \cite[Sect.\,2.1]{BaKi-book}. The representation categories of
(suitable) 
vertex operator algebras are not only tensor categories, but they also have a braiding 
and a twist. This additional structure ensures that every left dual is also a right dual and 
vice versa \cite[Sect.\,7]{Joyal:1993}. We take this as a motivation to not single out 
right duals and instead treat both on the same footing. 

\begin{Def}\label{def:Cr}
Let $\Cc$ be a tensor category. The {\em category} $\Crr$ 
{\em of rigid objects in}
$\Cc$ is the full subcategory consisting of all objects $U \in \Cc$ that have a right and 
a left dual.
\end{Def}

Not every object in a tensor category need to have a right and/or left dual. However, 
the tensor unit $\one$ always has itself as a right and a left dual, and the objects 
which have right and left duals form a full tensor subcategory. 

\begin{Lem}\label{lem:Cd-tensorcat}
$\one \in \Crr$, and for $U , V \in \Crr$ also $U \otimes V \in  \Crr$.
\end{Lem}

\begin{proof}
It is easy to check that one can choose $\one^\vee = \eev\one = \one$ with 
$b_\one = \tilde b_\one = \lambda_\one^{-1}$ and $d_\one = \tilde d_\one = \lambda_\one$. 
For $U \otimes V$ we set $(U \otimes V)^\vee := V^\vee \otimes U^\vee$ and, omitting the 
associator and unit isomorphisms, 
$b_{U \otimes V} = (\id_U \otimes b_V \otimes \id_{U^\vee}) \circ b_U$ and 
$d_{U \otimes V} = d_V \circ (\id_{V^\vee} \otimes d_U \otimes \id_V)$. The verification 
of the properties in Definition~\ref{def:right-dual} is straightforward. That 
$\eev(U \otimes V) := \eev V \otimes \eev U$ is a left dual can be checked in the same way.
\end{proof}

The category $\Crr$ allows to solve condition B1 in the construction of a boundary theory, 
but does still not guarantee B2 and B3.
For example in $\Rep(\Wc_{2,3})$ the $\Wc$ algebra 
is the tensor unit, and so is its own left and right dual. But as already pointed out in 
Section~\ref{sec:intro-W23-bnd}, $\Wc \not\cong \Wc^*$ and so $\Wc$ does not allow 
for a non-degenerate two-point correlator. Instead we will consider the following 
subcategory of $\Crr$.

\begin{Def}
Let $\Cc$ be a tensor category satisfying condition C. Then $\Cdd$ 
(where $b$ stands for `boundary') denotes the 
sub-category of $\Cc$ consisting of all objects $U$ for which $U^*$ is both a right dual 
and a left dual of $U$, and for which both $b_U : \one \rightarrow U \otimes U^*$ and 
$\tilde b_U : \one \rightarrow U^* \otimes U$ are injective.
\end{Def}

The injectivity requirement will guarantee the injectivity of the unit morphisms in 
\eqref{eq:iHom+dual-morphs}. Note that even if $\Cc$ is abelian, $\Cdd$ is not.
For example it does not contain the zero object ${\bf 0}$ as $b_{\bf 0} = 0$ is not injective. 
The uniqueness of internal Homs, together with Theorem~\ref{thm:iHom-exist} and 
Lemma~\ref{lem:dual-iHom} implies that
\be
  \big( A \otimes B^*\big)^{\!*} \cong B \otimes A^*
  ~~ \text{for}~ A,B \in \Cdd\ .
\labl{eq:AB*-AB*}
The following theorem shows that $\Cdd$ is closed under taking conjugates and 
tensor products. It will be proved in Appendix~\ref{app:some-proofs}.

\begin{Thm} \label{thm:Cdd-prop}
Let $\Cc$ be a tensor category satisfying condition $\mathrm{C}$. \\
{\rm (i)} If $U \in \Cdd$ then also $U^* \in \Cdd$. \\
{\rm (ii)} If $U,V \in \Cdd$, then also $U \otimes V \in \Cdd$.
\end{Thm}

Note that (ii) does not imply that $\Cdd$ is a tensor category, because in 
general $\one \notin \Cdd$. As we have just seen, the $\Wc_{2,3}$ 
model provides an example for this.
On the category $\Cdd$ we can define a boundary theory satisfying B1--B3.
We choose $\Bc$ to consist of the objects of  $\Cdd$. For the 
open string state spaces we again take $\Hc_{A\rightarrow B} = [A,B]$, but 
now we choose the internal Hom defined in Lemma~\ref{lem:dual-iHom}, {\it i.e.}\ 
$[A,B] = B \otimes A^*$. Multiplication and unit morphisms are defined by 
\eqref{eq:iHom+dual-morphs}. Property B1 --- associativity of the multiplication --- holds 
by Theorem~\ref{thm:assoc}, 
but it can also be easily verified directly.
For the one-point correlation function we choose
\be  
  \eps_A = \pi_{A,A^*}(\delta_A) : [A,A] \rightarrow \one^{\!*} \ ,
\labl{eq:epsA-def}
where $\delta_A : A \rightarrow A^{**}$ was defined in \eqref{eq:delta-def}.
Properties B2 and B3
are established in the next theorem, to be proved in 
Appendix~\ref{app:some-proofs}.

\begin{Thm}\label{thm:non-deg}
Let $\Cc$ be an abelian tensor category satisfying condition $\mathrm{C}$ and let 
$A,B \in \Cdd$. Then $[A,B] = B \otimes A^*$ and\\
{\rm (i)} $B \otimes A^*$ is non-zero,\\
{\rm (ii)} the morphism $\eta_A : \one \rightarrow [A,A]$ is injective,\\
{\rm (iii)} the pairing $\eps_A \circ m_{A,B,A} : [B,A] \otimes [A,B] \rightarrow \one^{\!*}$ 
is non-degenerate.
\end{Thm}

\medskip

Altogether we see that, provided $\Rep(\Wc_{2,3})$ is an abelian braided tensor 
category satisfying condition C, we can define a boundary theory satisfying B1--B3
on the set of boundary labels given in \eqref{eq:B-good-choice} (in a braided tensor 
category with twist, $b_A$ is injective iff $\tilde b_A$ is injective). We believe that the 
representations in \eqref{eq:W23-irreps} and \eqref{eq:indec-W-rep} which are not in 
grey boxes are in $\Repbb$. We verify that the $b_A$ are injective in Appendix~\ref{app:ker-bU}.

\subsection{Subgroups of the Grothendieck group}\label{sec:K0-and-K0d}

Let $\Cc$ be an abelian tensor category. The 
Grothendieck group $\mathrm{K}_0(\Cc)$ is defined 
as the free abelian group generated by isomorphism classes of objects in 
$\Cc$, divided by the subgroup generated by the elements $[U]+[W]-[V]$ for each exact 
sequence  $0 \rightarrow U \rightarrow V \rightarrow W \rightarrow 0$ in $\Cc$, 
see {\it e.g.}\ \cite[Def.\,2.1.9]{BaKi-book}.
Note that this definition implies in particular that $[U \oplus V] = [U] + [V]$.

If an object $A \in \Cc$ has the property that for each exact sequence 
$0 \rightarrow U \overset{f}{\rightarrow} V \overset{g}{\rightarrow} W \rightarrow 0$ also 
$0 \rightarrow A \otimes U \xrightarrow{\id_A \otimes f} A \otimes V 
\xrightarrow{\id_A \otimes g} A \otimes W \rightarrow 0$ 
is exact, we say that the functor $A \otimes (-)$ is exact. This is not always the case, 
as we have seen explicitly in Section~\ref{sec:tens-exact}.

However, if $A \otimes (-)$ is exact, then we get a well-defined map 
$[U] \mapsto [A \otimes U]$ on $K_0(\Cc)$. It is proved in \cite[Prop.\,2.1.8]{BaKi-book} 
that if $A$ has both a left and a right dual, then both $A \otimes (-)$ and $(-) \otimes A$ 
are exact. This motivates the definition
\be
  \Krr(\Cc) = (\text{ the subgroup of $K_0(\Cc)$ generated by 
  $[U]$ for all $U \in \Crr$ }) \ .
\ee
By Lemma~\ref{lem:Cd-tensorcat}, the assignment $([U],[V]) \mapsto [U \otimes V]$ 
gives a well-defined map 
$\Krr(\Cc) \times \Krr(\Cc) \rightarrow \Krr(\Cc)$. 
Because the tensor product is associative, this map defines an associative product on 
$\Krr(\Cc)$ with unit element $[\one]$. Thus even if the tensor product does 
not induce a product on $K_0(\Cc)$, we always have a unital ring structure on the abelian 
subgroup $\Krr(\Cc) \subset \mathrm{K}_0(\Cc)$.

In the context of boundary conformal field theory the representation category is
expected to be an abelian 
tensor category  satisfying property C, and we have seen that we can associate a 
boundary theory to the category $\Cdd$. We can then define a corresponding subgroup 
of the Grothendieck group,
\be
  \Kdd(\Cc) = (\text{ the subgroup of $K_0(\Cc)$ generated by 
  $[U]$ for all $U \in \Cdd$ }) \ .
\ee
By definition $\Kdd(\Cc) \subset \Krr(\Cc)$, and by 
Theorem~\ref{thm:Cdd-prop} the product on $\Krr(\Cc)$ restricts to a 
product on $\Kdd(\Cc)$. Because $\Cdd$ does not necessarily 
have a unit, neither does $\Kdd(\Cc)$.

\section{Conclusions and outlook}\label{sec:concl}

In this paper we have studied the $\Wc_{2,3}$ triplet model in some detail. In particular,
we have determined the fusion rules of the theory, {\it i.e.}\ we have determined
the fusion rules of all representations that appear in successive fusions of the irreducible
representations. (The complete list of fusion rules is given in Appendix~\ref{app:fus}.) 
We have also studied some of the unusual properties of these representations and their
fusions. For example, there is a subtle difference between conjugate and dual
representations (see Section~\ref{sec:contra-dual}), and the 
Grothendieck group $\mathrm{K}_0$ that is generated 
by the characters of the $13$ irreducible representations of the $\Wc_{2,3}$ model 
does not admit a straightforward product (see Section~\ref{sec:W23-K0-calc}). 

The second main result concerns
a boundary theory for the $\Wc_{2,3}$ model which is analogous to
the Cardy case in non-logarithmic rational conformal field theory.
We have identified the subset $\Bc$ of representations to which we can assign consistent 
boundary conditions. The resulting boundary conditions have boundary fields whose
OPEs are associative (this is guaranteed by the internal Hom construction, see
Theorem~\ref{thm:assoc}). In addition, the 
boundary two-point correlators are non-degenerate (Theorem~\ref{thm:non-deg}), and 
the spectrum of boundary fields between 
any two such boundary conditions is non-empty (see Theorem~\ref{thm:Cdd-prop}).

The representations in $\Bc$ are characterised by the property that 
the conjugate agrees with the dual representation, and that the intertwiner $b_\Rc$ that is 
needed for duality is an injection, see \eqref{eq:B-good-choice}. If we restrict 
the Grothendieck group $\mathrm{K}_0$ to $\Bc$ --- this defines the group 
$\Kdd$  that is generated by $12$ independent characters, 
see \eqref{eq:K0dd-basis} --- then the fusion rules lead to a well-defined product 
which characterises the cylinder partition functions between these boundary conditions.
\medskip

Our analysis of the boundary theory did not rely on the details of the corresponding bulk 
theory, and indeed, the idea of the approach is to try and reconstruct the bulk theory
starting from our boundary analysis. However, it is a 
priori not clear whether this will be possible, and thus the construction of the corresponding
bulk theory is the main open problem that remains for the $\Wc_{2,3}$ model. 
A good starting point might be the observation in \cite{Quella:2007hr,Gaberdiel:2007jv} 
that, for certain supergroup WZW models and for the $\Wc_{1,p}$ triplet models,
the space of bulk states $\Hc_\text{bulk}$ is a quotient of 
\begin{equation}\label{bulkbig}
\bigoplus_i P_i^{\phantom{*}} \!\otimes_\Cb \bar P_i^*\ , 
\end{equation}
where the sum runs over the indecomposable projective representations, and
the bar refers to right-movers. 
Furthermore, at least in these examples, the character of the quotient was given by
\be 
  Z(q) = \sum_i \chi_{P_i}(q) \cdot \chi_{\bar U_i^*}(\bar q)\ ,
\labl{eq:bulk-ansatz}
where $U_i$ is the irreducible representation of which $P_i$ is the projective cover. 

In analogy with the $\Wc_{1,p}$ models it seems likely to us that the irreducible 
$\Wc_{2,3}$-represen\-tation $\Wc(\tfrac{-1}{24})$ is projective. 
If $\mathcal{P}$ is projective and $\Rc$ has a dual, then 
$\Rc^\vee \otimes \mathcal{P}$ is also projective.
Therefore, if $\Wc(\tfrac{-1}{24})$ is projective the following twelve 
$\Wc_{2,3}$-representations have to be projective as well
\bea
  \Wc(\tfrac{-1}{24}) \,,~
  \Wc(\tfrac{35}{24}) \,,
\\[.3em]\displaystyle 
  \Rc^{(2)}(\tfrac13,\tfrac13)\,,~
  \Rc^{(2)}(\tfrac13,\tfrac{10}{3})\,,~
  \Rc^{(2)}(\tfrac58,\tfrac58)\,,~
  \Rc^{(2)}(\tfrac58,\tfrac{21}{8})\,,~
  \Rc^{(2)}(\tfrac18,\tfrac18)\,,~
  \Rc^{(2)}(\tfrac18,\tfrac{33}{8})\,,
\\[.3em]\displaystyle 
  \Rc^{(3)}(0,0,1,1)\,,~
  \Rc^{(3)}(0,0,2,2)\,,~
  \Rc^{(3)}(0,1,2,5)\,,~
  \Rc^{(3)}(0,1,2,7)\ .
\eear\labl{eq:proj-guess}
The fusion product of such a projective representation with any other 
representation from \eqref{eq:W23-irreps} or \eqref{eq:indec-W-rep} produces 
a direct sum of representations in \eqref{eq:proj-guess}, and we therefore think that 
these are all indecomposable projective representations. In fact, by comparison 
with the embedding diagrams in Appendix~\ref{app:embed} one finds that these are 
the projective covers of the irreducible representations
\bea
  \Wc(\tfrac{-1}{24}) ~,~~
  \Wc(\tfrac{35}{24}) ~,
\\[.3em]\displaystyle 
  \Wc(\tfrac{1}{3})~,~~
  \Wc(\tfrac{10}{3})~,~~
  \Wc(\tfrac{5}{8})~,~~
  \Wc(\tfrac{21}{8})~,~~
  \Wc(\tfrac{1}{8})~,~~ 
  \Wc(\tfrac{33}{8})~,
\\[.3em]\displaystyle 
  \Wc(1)~,~~
  \Wc(2)~,~~
  \Wc(5)~,~~
  \Wc(7)~,~~
\eear \label{eq:irred-for-proj} \ee
in this order.
We do not know if $\Wc(0)$ has a projective cover, but if it has, it is not 
one of the representations we consider in \eqref{eq:W23-irreps} and 
\eqref{eq:indec-W-rep} (see Appendix~\ref{app:embed}).
The characters of the representations in \eqref{eq:proj-guess} 
agree, up to overall factors, with the characters of projective representations proposed 
in \cite[Sect.\,5.2.1]{Feigin:2006iv}, 
and the representations themselves agree with the list proposed in
\cite[Sect.\,3.6]{Rasmussen:2008xi}.
We find by inspection that the ansatz \eqref{eq:bulk-ansatz} by itself is not modular 
invariant, but the following slight modification is,
\be\begin{array}{ll}
  Z_{\Wc_{2,3}}(q) \etb= 
  \big( n \chi_{\Wc(0)}(q){+}2\chi_{\Wc(1)}(q){+}2\chi_{\Wc(2)}(q){+}2\chi_{\Wc(5)}(q){+}
  2\chi_{\Wc(7)}(q) \big) \cdot \chi_{\Wc(0)}(\bar q)
\\[.3em]\displaystyle  
  \etb \hspace{2em} + ~ \sum_i \chi_{P_i}(q) \cdot \chi_{\bar U_i^*}(\bar q)
\\[.3em]\displaystyle  
 \etb = (q \bar q)^{-1/24} + n + 2(q \bar q)^{1/8} + 2(q \bar q)^{1/3} 
 + (q{+}\bar q)\cdot(q \bar q)^{-1/24} + 2(q{+}\bar q) + \cdots \ ,
\eear\label{eq:ansatz} \ee
where the sum runs over the $12$ projectives $P_i$ in \eqref{eq:proj-guess} with the 
corresponding irreducibles $U_i$ given in \eqref{eq:irred-for-proj}. 
The integer $n$ is not constrained by modular invariance, but since there 
is at least one vector of conformal weight zero we have $n \ge 1$.
Even if it is not apparent from the way it is written, the expression  \eqref{eq:ansatz}
is left/right symmetric. 
Furthermore, it agrees with the modular invariant 
combination of characters given in \cite[Sect.\,5.3]{Feigin:2006iv} 
(for $n=1$ and up to an overall factor of 4).
The extra term in \eqref{eq:ansatz} with respect to \eqref{eq:bulk-ansatz} 
could indicate that for the $\Wc_{2,3}$ model, the quotient of 
$\bigoplus_i P_i^{\phantom{*}} \! \otimes_\Cb \bar P_i^*$ needed to 
obtain $\Hc_\text{bulk}$ 
is more complicated. 
It is amusing to note that for $n=1$ we can write (\ref{eq:ansatz}) as 
\be
Z_{\Wc_{2,3}}(q) = \sum_i  
\dim\!\big(\hbox{Hom}(P_i,P_i)\big) ^{-2} \cdot  | \chi_{P_i}(q) |^2 \ ,
\labl{zus1}
where the sum extends over the $12$ projective representations. 
Incidentally, this formula also works for the $\Wc_{1,p}$ models, as well as the 
non-logarithmic rational theories since for these theories (\ref{zus1}) and 
(\ref{eq:bulk-ansatz}) agree. This suggests that (\ref{zus1}) could be more
generally true.

Finally let us remark that the size of the Grothendieck group 
$\Kdd$ suggests that one needs $12$ Ishibashi states to 
construct the boundary states for the boundary conditions in $\Bc$. 
This coincides with the number of 
projective representations in  \eqref{eq:proj-guess}, and thus one could guess that 
one needs one Ishibashi state from each summand in (\ref{bulkbig}). Incidentally,
this is precisely what happened for the $\Wc_{1,p}$ models 
\cite{Gaberdiel:2007jv}. 

It would be very interesting to study the $\Wc_{2,3}$ bulk theory  in more detail
and to see whether these expectations are indeed borne out. We also expect that 
much of the structure we have discovered for the $\Wc_{2,3}$ model
holds more generally for the $\Wc_{p,q}$ models. We hope 
to return to these questions in the near future.

\subsection*{Acknowledgements}
We would like to thank Michael Flohr, J\"urgen Fuchs, Yi-Zhi Huang,
J\o rgen Rasmussen,
David Ridout and Christoph Schweigert 
for helpful discussions and useful comments on a draft of this paper.
IR thanks ETH Z\"urich and the Centre for Theoretical Studies for hospitality
during the final stages of this work.
The research of MRG and SW is supported by the Swiss National Science Foundation. 
IR is partially supported by the EPSRC First Grant EP/E005047/1 and the 
STFC Rolling Grant ST/G000395/1. 

\appendix

\section{More on representations and fusion rules}\label{app:more-rep}

\subsection{Characters}\label{app:W23-chars}

Let us first list the characters of all the irreducible representations; 
these were given in \cite[Sect.\,5.1]{Feigin:2006iv}. We use the formulation 
in \cite[Sect.\,3.2]{Rasmussen:2008ii} where also the characters of the indecomposable
$\Rc^{(\cdot)}(\cdots)$ representations in \eqref{eq:indec-W-rep} can be found.
\begin{align*}
  \chi_{\Wc(0)}&=1\displaybreak[0]\\
  \chi_{\Wc(1)}&=\frac{1}{\eta(q)}\sum_{k\in\mathbb{Z}}k^2\left(q^{(12k-7)^2/24}-q^{(12k+1)^2/24}\right)\\
  &=q\bigl(1+q+2q^2+3q^3+4q^4+6q^5 + \cdots\bigr)\displaybreak[0]\\
  \chi_{\Wc(2)}&=\frac{1}{\eta(q)}\sum_{k\in\mathbb{Z}}k^2\left(q^{(12k-5)^2/24}-q^{(12k-1)^2/24}\right)\\
  &=q^2\bigl(1+q+2q^2+2q^3+4q^4+4q^5 + \cdots\bigr)\displaybreak[0]\\
  \chi_{\Wc(5)}&=\frac{1}{\eta(q)}\sum_{k\in\mathbb{Z}}k(k+1)\left(q^{(12k-1)^2/24}-q^{(12k+7)^2/24}\right)\\
  &=q^5\bigl(2+2q+4q^2+6q^3+10q^4+14q^5 + \cdots\bigr) \displaybreak[0]\\
  \chi_{\Wc(7)}&=\frac{1}{\eta(q)}\sum_{k\in\mathbb{Z}}k(k+1)\left(q^{(12k+1)^2/24}-q^{(12k+5)^2/24}\right)\\
  &=q^7\bigl(2+2q+4q^2+6q^3+10q^4+12q^5 + \cdots\bigr) \displaybreak[0]\\
  \chi_{\Wc\left(\tfrac13\right)}&=\frac{1}{\eta(q)}\sum_{k\in\mathbb{Z}}(2k-1)q^{3(4k-3)^2/8}
=q^{1/3}\bigl(1+q+2q^2+2q^3+4q^4+5q^5+ \cdots\bigr) \displaybreak[0]\\
  \chi_{\Wc\left(\tfrac{10}{3}\right)}&=\frac{1}{\eta(q)}\sum_{k\in\mathbb{Z}}2kq^{3(4k-1)^2/8}
=q^{10/3}\bigl(2+2q+4q^2+6q^3+10q^4+14q^5 + \cdots\bigr) \displaybreak[0]\\
    \chi_{\Wc\left(\tfrac18\right)}&=\frac{1}{\eta(q)}\sum_{k\in\mathbb{Z}}(2k-1)q^{(6k-5)^2/6}
=q^{1/8}\bigl(1+q+2q^2+3q^3+4q^4+6q^5 + \cdots\bigr)\displaybreak[0]\\
  \chi_{\Wc\left(\tfrac58\right)}&=\frac{1}{\eta(q)}\sum_{k\in\mathbb{Z}}(2k-1)q^{(6k-4)^2/6}
=q^{5/8}\bigl(1+q+q^2+2q^3+3q^4+4q^5 + \cdots\bigr) \displaybreak[0]\\
  \chi_{\Wc\left(\tfrac{21}{8}\right)}&=\frac{1}{\eta(q)}\sum_{k\in\mathbb{Z}}2kq^{(6k-2)^2/6}
=q^{21/8}\bigl(2+2q+4q^2+6q^3+10q^4+14q^5 + \cdots\bigr) \displaybreak[0]\\
  \chi_{\Wc\left(\tfrac{33}{8}\right)}&=\frac{1}{\eta(q)}\sum_{k\in\mathbb{Z}}2kq^{(6k-1)^2/6}
=q^{33/8}\bigl(2+2q+4q^2+6q^3+8q^4+12q^5 + \cdots \bigr)\displaybreak[0]\\
  \chi_{\Wc\left(\tfrac{-1}{24}\right)}&=\frac{1}{\eta(q)}\sum_{k\in\mathbb{Z}}(2k-1)q^{(6k-6)^2/6}
=q^{-1/24}\bigl(1+q+2q^2+3q^3+5q^4+7q^5 + \cdots\bigr) \displaybreak[0]\\
  \chi_{\Wc\left(\tfrac{35}{24}\right)}&=\frac{1}{\eta(q)}\sum_{k\in\mathbb{Z}}2kq^{(6k-3)^2/6}
 =q^{35/24}\bigl(2+2q+4q^2+6q^3+10q^4+14q^5 + \cdots \bigr) \ .
\end{align*}
In terms of the irreducible representations the rank 1 representations have the characters
\be\nonumber
  \chi_{\Wc}=\chi_{\Wc^*}=1+\chi_{\Wc(2)}
  \quad , \quad
  \chi_{\Qc}=\chi_{\Qc^*}=1+\chi_{\Wc(1)} \ ,
\ee
while the characters of the rank $2$ representations are 
\begin{align*}
  \chi_{\Rc^{(2)}\left(\tfrac13,\tfrac13\right)}=\chi_{\Rc^{(2)}\left(\tfrac13,\tfrac{10}{3}\right)}&=
  2\chi_{\Wc\left(\tfrac13\right)}+2\chi_{\Wc\left(\tfrac{10}{3}\right)}\\
  \chi_{\Rc^{(2)}\left(\tfrac18,\tfrac18\right)}=\chi_{\Rc^{(2)}\left(\tfrac18,\tfrac{33}{8}\right)}&=
  2\chi_{\Wc\left(\tfrac18\right)}+2\chi_{\Wc\left(\tfrac{33}{8}\right)}\\
  \chi_{\Rc^{(2)}\left(\tfrac58,\tfrac58\right)}=\chi_{\Rc^{(2)}\left(\tfrac58,\tfrac{21}{8}\right)}&=
  2\chi_{\Wc\left(\tfrac58\right)}+2\chi_{\Wc\left(\tfrac{21}{8}\right)}
\end{align*}
\begin{align*}
  \chi_{\Rc^{(2)}(0,2)_7}=1+\chi_{\Rc^{(2)}(2,7)}&=
  1+2\chi_{\Wc(2)}+2\chi_{\Wc(7)}\\
  \chi_{\Rc^{(2)}(0,1)_5}=1+\chi_{\Rc^{(2)}(1,5)}&=
  1+2\chi_{\Wc(1)}+2\chi_{\Wc(5)}\\
  \chi_{\Rc^{(2)}(0,1)_7}=1+\chi_{\Rc^{(2)}(1,7)}&=
  1+2\chi_{\Wc(1)}+2\chi_{\Wc(7)}\\
  \chi_{\Rc^{(2)}(0,2)_5}=1+\chi_{\Rc^{(2)}(2,5)}&=
  1+2\chi_{\Wc(2)}+2\chi_{\Wc(5)}\ .\\
\end{align*}
Finally, all rank 3 representations have the same character
\begin{align}
  \chi_{\Rc^{(3)}(0,k,\ell,m)}=2\chi_{\Wc(0)}+4\chi_{\Wc(1)}
  +4\chi_{\Wc(2)} +4\chi_{\Wc(5)}+4\chi_{\Wc(7)}\ .
\end{align}

\subsection{Dictionary to the notation in other works}\label{app:dict}

\subsubsection*{The notation in \cite{Rasmussen:2008ii,Rasmussen:2008ez}}

It is straightforward to identify the irreducible representations by comparing the
conformal weight of the ground state. We can then successively identify the indecomposable
representations by comparing the fusions of these representations. As a non-trivial consistency 
check we have also compared the embedding 
diagrams\footnote{Note that the embedding diagrams of \cite{Eberle:2006zn} describe
the Virasoro action, while those of \cite{Rasmussen:2008ii,Rasmussen:2008ez} 
refer to the $\Wc$ action.} 
 of \cite[Figure 2--5]{Eberle:2006zn}
with the embedding diagrams of \cite{Rasmussen:2008ii,Rasmussen:2008ez}, see 
in particular the diagram  
\cite[Eq.\,(3.34)]{Rasmussen:2008ii}, relations \cite[Eqs.\,(3.35),\,(3.41)]{Rasmussen:2008ii} 
and  diagram \cite[Eq.\,(4.9)]{Rasmussen:2008ez}.
\begin{center}
\begin{longtable}{c|cc||cc|c}
  our notation & notation in \cite{Rasmussen:2008ii,Rasmussen:2008ez} 
  &&&
  our notation & notation in \cite{Rasmussen:2008ii,Rasmussen:2008ez} \\
\hline
\endfirsthead
  $\dots$ our notation & $\dots$ notation in \cite{Rasmussen:2008ii,Rasmussen:2008ez} &&&
  $\dots$ our notation & $\dots$ notation in \cite{Rasmussen:2008ii,Rasmussen:2008ez} \\
\hline
\endhead
  \(\Wc\)&\((1,1)_\Wc \) &&&
  \(\Rc^{(2)}(0,1)_5\)&\((\Rc^{1,0}_{2,2})_\Wc\) \\
  \(\Qc\)&\((1,2)_\Wc \) &&&
  \(\Rc^{(2)}(1,5)\)&\((\Rc^{1,0}_{4,2})_\Wc\)\\
  \(\Wc(\tfrac13)\)&\((1,3)_\Wc\) &&&
  \(\Rc^{(2)}(0,2)_7\)&\((\Rc^{1,0}_{2,1})_\Wc\)\\
  \(\Wc(\tfrac{10}{3})\)&\((1,6)_\Wc\) &&&
  \(\Rc^{(2)}(2,7)\)&\((\Rc^{1,0}_{4,1})_\Wc\)\\
  \(\Wc(\tfrac58)\)&\((2,1)_\Wc\) &&&
  \(\Rc^{(2)}(\frac13,\frac13)\)&\((\Rc^{1,0}_{2,3})_\Wc\)\\
  \(\Wc(\tfrac{33}{8})\)&\((4,1)_\Wc\) &&&
  \(\Rc^{(2)}(\frac13,\frac{10}{3})\)&\((\Rc^{1,0}_{2,6})_\Wc=(\Rc^{1,0}_{4,3})_\Wc\)\\
  \(\Wc(\tfrac18)\)&\((2,2)_\Wc\) &&&
  \(\Rc^{(2)}(\frac58,\frac58)\)&\((\Rc^{0,2}_{2,3})_\Wc\)\\
  \(\Wc(\tfrac{21}{8})\)&\((4,2)_\Wc\) &&&
  \(\Rc^{(2)}(\frac18,\frac{33}{8})\)&\((\Rc^{0,2}_{2,6})_\Wc\)\\
  \(\Wc(\tfrac{-1}{24})\)&\((2,3)_\Wc\) &&&
  \(\Rc^{(2)}(\frac18,\frac18)\)&\((\Rc^{0,1}_{2,3})_\Wc\)\\
  \(\Wc(\tfrac{35}{24})\)&\((2,6)_\Wc=(4,3)_\Wc\) &&&
  \(\Rc^{(2)}(\frac58,\frac{21}{8})\)&\((\Rc^{0,1}_{2,6})_\Wc\)\\
  \(\Rc^{(2)}(0,1)_7\)&\((\Rc^{0,1}_{1,3})_\Wc\) &&&
  \(\Rc^{(3)}(0,0,1,1)\)&\((\Rc^{1,1}_{2,3})_\Wc\)\\
  \(\Rc^{(2)}(2,5)\)&\((\Rc^{0,1}_{1,6})_\Wc\) &&&
  \(\Rc^{(3)}(0,1,2,5)\)&\((\Rc^{1,1}_{2,6})_\Wc=(\Rc^{1,1}_{4,3})_\Wc\)\\
  \(\Rc^{(2)}(0,2)_5\)&\((\Rc^{0,2}_{1,3})_\Wc\) &&&
  \(\Rc^{(3)}(0,0,2,2)\)&\((\Rc^{1,2}_{2,3})_\Wc\)\\
  \(\Rc^{(2)}(1,7)\)&\((\Rc^{0,2}_{1,6})_\Wc\) &&&
  \(\Rc^{(3)}(0,1,2,7)\)&\((\Rc^{1,2}_{2,6})_\Wc=(\Rc^{1,2}_{4,3})_\Wc\) 
\end{longtable}
\end{center}
The representations $\Wc(0), \Wc(1), \Wc(2), \Wc(5), \Wc(7), \Wc^*, \Qc^*$ 
do not appear in \cite{Rasmussen:2008ii,Rasmussen:2008ez}.
The identifications in the above table are those in 
\cite[Eqs.\,(3.1),\,(3.3)]{Rasmussen:2008ii}.

\subsubsection*{The notation in \cite{Feigin:2006iv}}

The identification can be made by comparing \eqref{eq:W23-irreps} to 
\cite[Table\,1]{Feigin:2006iv} and the sequences \eqref{eq:W-seq} and \eqref{eq:Q-seq} to 
\cite[Sect.\,3.4]{Feigin:2006iv}.

\begin{center}
\begin{tabular}{c|c}
  our notation & notation in \cite{Feigin:2006iv} \\
  \hline
  \(\Wc\)&\( \mathcal{K}_{1,1}^+ \) \\
  \(\Qc\)&\(\mathcal{K}_{1,2}^+ \) \\
  \(\Wc(0)\)& \(\mathcal{X}_{1,1}\) \\
  \(\Wc(2)\)& \(\mathcal{X}_{1,1}^+\) \\
  \(\Wc(7)\)& \(\mathcal{K}_{1,1}^- = \mathcal{X}_{1,1}^-\) \\
  \(\Wc(1)\)& \(\mathcal{X}_{1,2}^+\) \\
  \(\Wc(5)\)& \(\mathcal{K}_{1,2}^- = \mathcal{X}_{1,2}^-\) \\
  \phantom{.}
\end{tabular}
\hspace{3em}
\begin{tabular}{c|c}
  our notation & notation in \cite{Feigin:2006iv} \\
  \hline
  \(\Wc(\tfrac13)\)& \(\mathcal{K}_{1,3}^+ = \mathcal{X}_{1,3}^+\)\\
  \(\Wc(\tfrac{10}{3})\)&\(\mathcal{K}_{1,3}^- = \mathcal{X}_{1,3}^-\) \\
  \(\Wc(\tfrac58)\)&\(\mathcal{K}_{2,1}^+ = \mathcal{X}_{2,1}^+\) \\
  \(\Wc(\tfrac{33}{8})\)&\(\mathcal{K}_{2,1}^- = \mathcal{X}_{2,1}^-\) \\
  \(\Wc(\tfrac18)\)&\(\mathcal{K}_{2,2}^+ = \mathcal{X}_{2,2}^+\) \\
  \(\Wc(\tfrac{21}{8})\)&\(\mathcal{K}_{2,2}^- = \mathcal{X}_{2,2}^-\) \\
  \(\Wc(\tfrac{-1}{24})\)&\(\mathcal{K}_{2,3}^+ = \mathcal{X}_{2,3}^+\) \\
  \(\Wc(\tfrac{35}{24})\)&\(\mathcal{K}_{2,3}^- = \mathcal{X}_{2,3}^-\) \\
\end{tabular}
\end{center}

\noindent
The representations $\Wc^*$, $\Qc^*$ and those of the form $\Rc^{(\cdot)}(\cdots)$ 
are not considered in \cite{Feigin:2006iv}.

\subsection{Embedding structure of the $\Wc$-representations}\label{app:embed}

For the convenience of the reader we transcribe the 
embedding diagrams of \cite[(3.34)]{Rasmussen:2008ii} 
in our notation.

\subsubsection*{Rank 2 Representations}

The rank 2 representations are indecomposable combinations of the irreducible 
representations. For $\ell>h>0$, they
are given in \cite{Rasmussen:2008ii} as
\begin{align}
  \Rc^{(2)}(0,h)_\ell: \qquad 
  &\xymatrix{\Wc(\ell)\ar[d]&&\Wc(\ell)\ar[dll]\\
  \Wc(h)&&\ar[u]\ar[ll]\ar[dl]\ar[ull]\Wc(h)\\
&\Wc(0)\ar[ul]}
\end{align}
while for $h$ fractional the diagrams of \cite{Rasmussen:2008ii}  are
\begin{align}
  \Rc^{(2)}(h,h) : \qquad 
  &\xymatrix{\ar[d]\Wc(h+n)&&\ar[dll]\Wc(h+n)\\
  \Wc(h)&&\ar[u]\ar[ll]\ar[ull]\Wc(h)} 
\end{align}
where $n=2,3,4$ for $h=\tfrac58, \tfrac13, \tfrac18$, respectively. Finally,
\begin{align}
  \Rc^{(2)}(h,h+n): \qquad 
  &\xymatrix{\Wc(h+n)&&\ar[ll]\ar[dll]\ar[d]\Wc(h+n)\\
  \ar[u]\Wc(h)&&\ar[ull]\Wc(h)}
\end{align}
where $h$ and $h+n$ as in the previous case, and additionally
$h=1,2$ and $h+n=5,7$ (all four combinations).

\subsubsection*{Rank 3 Representations}

Similarly the rank 3 representations are indecomposable combinations of the rank 2 
representations.
 For each representation two equivalent embedding diagrams are given in 
 \cite{Rasmussen:2008ii}.
\begin{align*}
  \Rc^{(3)}(0,0,1,1):\quad 
  &\xymatrix{\ar[d]\Rc^{(2)}(2,7)&&\ar[dll]\Rc^{(2)}(2,7)\\
  \Rc^{(2)}(0,1)_5&&\ar[u]\ar[ll]\ar[ull]\Rc^{(2)}(0,1)_5}
  &\xymatrix{\ar[d]\Rc^{(2)}(2,5)&&\ar[dll]\Rc^{(2)}(2,5)\\
  \Rc^{(2)}(0,1)_7&&\ar[u]\ar[ll]\ar[ull]\Rc^{(2)}(0,1)_7}
\end{align*}

\begin{align*}
  \Rc^{(3)}(0,1,2,5):\quad 
  &\xymatrix{\ar[d]\Rc^{(2)}(0,2)_7&&\ar[dll]\Rc^{(2)}(0,2)_7\\
  \Rc^{(2)}(1,5)&&\ar[u]\ar[ll]\ar[ull]\Rc^{(2)}(1,5)}
  &\xymatrix{\ar[d]\Rc^{(2)}(0,1)_7&&\ar[dll]\Rc^{(2)}(0,1)_7\\
  \Rc^{(2)}(2,5)&&\ar[u]\ar[ll]\ar[ull]\Rc^{(2)}(2,5)}
\end{align*}

\begin{align*}
  \Rc^{(3)}(0,0,2,2):\quad
  &\xymatrix{\ar[d]\Rc^{(2)}(1,5)&&\ar[dll]\Rc^{(2)}(1,5)\\
  \Rc^{(2)}(0,2)_7&&\ar[u]\ar[ll]\ar[ull]\Rc^{(2)}(0,2)_7}
  &\xymatrix{\ar[d]\Rc^{(2)}(1,7)&&\ar[dll]\Rc^{(2)}(1,7)\\
  \Rc^{(2)}(0,2)_5&&\ar[u]\ar[ll]\ar[ull]\Rc^{(2)}(0,2)_5}
\end{align*}

\begin{align*}
  \Rc^{(3)}(0,1,2,7): \quad
  &\xymatrix{\ar[d]\Rc^{(2)}(0,1)_5&&\ar[dll]\Rc^{(2)}(0,1)_5\\
  \Rc^{(2)}(2,7)&&\ar[u]\ar[ll]\ar[ull]\Rc^{(2)}(2,7)}
  &\xymatrix{\ar[d]\Rc^{(2)}(0,2)_5&&\ar[dll]\Rc^{(2)}(0,2)_5\\
  \Rc^{(2)}(1,7)&&\ar[u]\ar[ll]\ar[ull]\Rc^{(2)}(1,7)}
\end{align*}
The embedding diagram of the conjugate representation is obtained from these
diagrams by reversing all arrows. It is easy to see that at least the embedding 
diagrams of all rank 2 and rank 3 representations are self-conjugate.

\subsubsection*{Homomorphisms}

Using the relation
\be
 \Hom(U,V) \cong \Hom(U\otimes V^*,\Wc^*) 
\ee
one can determine the dimension of $\Hom(U,V)$ from $\Hom(-,\Wc^*)$.
One finds that
\be
  \dim \Hom(U,\Wc^*) = 
  \begin{cases} 1 & : ~ U \in \big\{ \, \Wc(0),\, \Wc,\, \Wc^*,\, \Qc, \\
  & \hspace{4em} \Rc^{(2)}(0,2)_5,\, \Rc^{(2)}(0,2)_7,\, \Rc^{(3)}(0,0,2,2)\, \big\} \\
  0 & : ~ \text{else.}
  \end{cases} 
\ee  
For example $\dim\Hom(U,\Wc(0)) = 1$ for $U \in\{ \Wc(0), \Wc, \Qc \}$ and 
$\dim\Hom(U,\Wc(0)) = 0$ for all other representations in \eqref{eq:W23-irreps} 
and \eqref{eq:indec-W-rep}.

As an application, let us show that if $\Wc(0)$ has a projective cover at all, it is not one of the 
representations listed in \eqref{eq:W23-irreps} and \eqref{eq:indec-W-rep}.
Recall that a representation $P$ is projective if, given an intertwiner $f : P \rightarrow V$, 
for any surjective $\pi : U \rightarrow V$ we can find a (typically non-unique) intertwiner 
$g : P \rightarrow U$ such that $f = \pi \circ g$. We can now check that none of 
$\Wc(0)$, $\Wc$, $\Qc$ are projective. To see this note that 
\be
  \dim\Hom(\Wc(0),\Wc) = 0 ~~,~~~
  \dim\Hom(\Qc,\Wc) = 0 ~~,~~~
  \dim\Hom(\Wc,\Qc) = 0\ .
\ee
If $\Qc$ were projective, then for the
non-zero morphism $f : \Qc \rightarrow \Wc(0)$ and the 
surjection $\pi : \Wc \rightarrow \Wc(0)$ we would have to find a 
$g : \Qc \rightarrow \Wc$ such that $f = \pi \circ g$. But there is no 
non-zero intertwiner $g : \Qc \rightarrow \Wc$, so this is not possible. 
Replacing $\Qc$ by $\Wc(0)$ shows that $\Wc(0)$ is not projective. 
For $\Wc$ one can consider $f : \Wc \rightarrow \Wc(0)$ and 
$\pi : \Qc \rightarrow \Wc(0)$.

\subsection{The complete list of fusion rules}\label{app:fus}

\subsubsection*{The action of $\Wc(0)$}
The fusion product of \(\Wc(0)\) with everything is zero with the
exception of
\be
  \Wc(0)\otimes\Wc(0) = \Wc(0) ~~,\quad
  \Wc(0)\otimes\Wc =\Wc(0) ~~,\quad
  \Wc(0)\otimes{\cal Q}=\Wc(0)\ .
\ee

\subsubsection*{The action of $\Wc$, $\Wc^*$, $\Wc(2)$ and $\Qc$, $\Qc^*$, $\Wc(1)$}

The representation $\Wc$ is the vertex operator algebra and acts as the identity in
all fusion products. The fusion with the representations $\Wc^*$ and $\Wc(2)$ acts as 
the identity on all representations in \eqref{eq:W23-irreps} and \eqref{eq:indec-W-rep}
that are not in grey boxes; on the representations in grey boxes, the fusion is explicitly given 
as 
\begin{center}
\begin{tabular}{rcl|l||rcl|l}
  &&Factors&Fusion product&&&Factors&Fusion product\\
  \hline
  \(\Wc^*\)&\(\otimes\)&\(\Wc^*\)&\(\Wc^*\)&\(\Wc(2)\)&\(\otimes\)&\(\Wc^*\)&\(\Wc^*\)\\
  &\(\otimes\)&\({\Qc}\)&\(\Qc^*\)&&\(\otimes\)&\({\Qc}\)&\(\Wc(1)\)\\
  &\(\otimes\)&\({\Qc^*}\)&\(\Qc^*\)&&\(\otimes\)&\(\Qc^*\)&\(\Qc^*\)\\
  &\(\otimes\)&\(\Wc(0)\)&0&&\(\otimes\)&\(\Wc(0)\)&0\\
  &\(\otimes\)&\(\Wc(1)\)&\(\Qc^*\)&&\(\otimes\)&\(\Wc(1)\)&\(\Qc^*\)\\
  &\(\otimes\)&\(\Wc(2)\)&\(\Wc^*\)&&\(\otimes\)&\(\Wc(2)\)&\(\Wc^*\)\\
  &\(\otimes\)&\(\Wc(5)\)&\(\Wc(5)\)&&\(\otimes\)&\(\Wc(5)\)&\(\Wc(5)\)\\
  &\(\otimes\)&\(\Wc(7)\)&\(\Wc(7)\)&&\(\otimes\)&\(\Wc(7)\)&\(\Wc(7)\)
\end{tabular}
\end{center}
Similarly ${\cal Q}$, $\Qc^*$ and $\Wc(1)$ have the same fusion rules with all
representations that are not in grey boxes, and the fusion rules of $\Wc(1)$ are 
explicitly given below; on the representations in grey boxes the fusion rules of 
$\Qc$ and $\Qc^*$ are 
\begin{center}
\begin{tabular}{rcl|l||rcl|l}
  &&Factors&Fusion product&&&Factors&Fusion product\\
  \hline
  \({\cal Q}\)&\(\otimes\)&\({\cal Q}\)&\(\Wc\oplus\Wc(\tfrac13)\)&\(\Qc^*\)&\(\otimes\)&\({\cal Q}\)
  &\(\Wc^*\oplus\Wc(\tfrac13)\)\\
  &\(\otimes\)&\(\Qc^*\)&\(\Wc^* \oplus\Wc(\tfrac13)\)&&\(\otimes\)&\(\Qc^*\)&
  \(\Wc^*\oplus\Wc(\tfrac13)\)\\
  &\(\otimes\)&\(\Wc(0)\)&\(\Wc(0)\)&&\(\otimes\)&\(\Wc(0)\)&0\\
  &\(\otimes\)&\(\Wc(1)\)&\(\Wc(2)\oplus\Wc(\tfrac13)\)&&\(\otimes\)&\(\Wc(1)\)&
  \(\Wc^*\oplus\Wc(\tfrac13)\)\\
  &\(\otimes\)&\(\Wc(5)\)&\(\Wc(7)\oplus\Wc(\tfrac{10}{3})\)&&\(\otimes\)&\(\Wc(5)\)&
  \(\Wc(7)\oplus\Wc(\tfrac{10}{3})\)\\
  &\(\otimes\)&\(\Wc(7)\)&\(\Wc(5)\)&&\(\otimes\)&\(\Wc(7)\)&\(\Wc(5)\)
\end{tabular}
\end{center}

\subsubsection*{The action of $\Wc(7)$}
The simple current $\Wc(7)$ squares to $\Wc^*$, and with the exception
of $\Wc(0)$, $\Wc(1)$, $\Wc(2)$, $\Wc$ and ${\cal Q}$, the fusion rules 
organise themselves into $\Wc(7)$-pairs. The fusion of $\Wc(7)$ with these special
representations is 
\begin{equation}
\begin{array}{rclrcl}
\Wc(7) \otimes \Wc(0) & = & 0 \qquad \qquad
& \Wc(7) \otimes \Wc(1) & = & \Wc(5) \\
\Wc(7) \otimes \Wc(2) & = & \Wc(7)  \qquad \qquad
& \Wc(7) \otimes \Wc & = & \Wc(7) \\
 \Wc(7) \otimes {\cal Q} & = & \Wc(5) \ , & & & 
\end{array}
\end{equation}
while on the remaining representations we have
\begin{center}
  \begin{longtable}{rcl||rcl}
   \(\Wc^*\)&\(\stackrel{\Wc(7)}{\longleftrightarrow}\)&\(\Wc(7)\)&
   \(\Qc^*\)&\(\stackrel{\Wc(7)}{\longleftrightarrow}\)&\(\Wc(5)\) \\
  \(\Wc(\tfrac13)\)&\(\stackrel{\Wc(7)}{\longleftrightarrow}\)&\(\Wc(\tfrac{10}{3})\) &
\(\Wc(\tfrac58)\)&\(\stackrel{\Wc(7)}{\longleftrightarrow}\)&\(\Wc(\tfrac{33}{8})\) \\
 \(\Wc(\tfrac18)\)&\(\stackrel{\Wc(7)}{\longleftrightarrow}\)&\(\Wc(\tfrac{21}{8})\) &
\(\Wc(\tfrac{-1}{24})\)&\(\stackrel{\Wc(7)}{\longleftrightarrow}\)&\(\Wc(\tfrac{35}{24})\) \\
\({\cal R}^{(2)}(0,2)_5\)&\(\stackrel{\Wc(7)}{\longleftrightarrow}\)&\({\cal R}^{(2)}(1,7)\) & 
\({\cal R}^{(2)}(0,1)_7\)&\(\stackrel{\Wc(7)}{\longleftrightarrow}\)&\({\cal R}^{(2)}(2,5)\) \\
\({\cal R}^{(2)}(0,2)_7\)&\(\stackrel{\Wc(7)}{\longleftrightarrow}\)&\({\cal R}^{(2)}(2,7)\) &
\({\cal R}^{(2)}(0,1)_5\)&\(\stackrel{\Wc(7)}{\longleftrightarrow}\)&\({\cal R}^{(2)}(1,5)\) \\
\({\cal R}^{(2)}(\tfrac58,\tfrac58)\)&\(\stackrel{\Wc(7)}{\longleftrightarrow}\)&
                    \({\cal R}^{(2)}(\tfrac18,\tfrac{33}{8})\) &
\({\cal R}^{(2)}(\tfrac18,\tfrac18)\)&\(\stackrel{\Wc(7)}{\longleftrightarrow}\) &
                    \({\cal R}^{(2)}(\tfrac58,\tfrac{21}{8})\) \\
\(\Rc^{(2)}(\tfrac13,\tfrac13)\)&\(\stackrel{\Wc(7)}{\longleftrightarrow}\)&
                   \(\Rc^{(2)}(\tfrac13,\tfrac{10}{3})\) &
 \(\Rc^{(3)}(0,0,1,1)\)&\(\stackrel{\Wc(7)}{\longleftrightarrow}\)&\(\Rc^{(3)}(0,1,2,5)\) \\
 \(\Rc^{(3)}(0,0,2,2)\)&\(\stackrel{\Wc(7)}{\longleftrightarrow}\)&\(\Rc^{(3)}(0,1,2,7)\)  & & & 
\end{longtable}
\end{center}
We only list the fusion products for the first representative of each $\Wc(7)$ pair. 
To obtain the fusion of for example \(\Wc(\tfrac58)\) and \(\Wc(\tfrac{21}{8})\) one
computes
\be
  \Wc(\tfrac58)\otimes\Wc(\tfrac{21}{8}) = \Wc(7)\otimes\Wc(\tfrac58)\otimes\Wc(\tfrac18)
  =\Wc(7)\otimes\Rc^{(2)}(0,1)_5
  = \Rc^{(2)}(1,5)\ .
\ee

\subsubsection*{The remaining products}

\begin{center}
\begin{longtable}{rcl|l}
 &&Factors&Fusion product\\
\hline
\endfirsthead
 &&$\dots$ factors&$\dots$ fusion product\\
\hline
\endhead
  \(\Wc(1)\)&\(\otimes\)&\(\Wc(1)\)&\(\Wc^*\oplus\Wc(\tfrac13)\)\\
  &\(\otimes\)&\(\Wc(\tfrac13)\)&\({\cal
    R}^{(2)}(0,1)_7\)\\
  &\(\otimes\)&\(\Wc(\tfrac58)\)&\(\Wc(\tfrac18)\)\\
  &\(\otimes\)&\(\Wc(\tfrac18)\)&\(\Wc(\tfrac58)\oplus\Wc(\tfrac{-1}{24})\)\\
  &\(\otimes\)&\(\Wc(\tfrac{-1}{24})\)&\({\cal
    R}^{(2)}(\tfrac18,\tfrac18)\)\\
  &\(\otimes\)&\(\Rc^{(2)}(0,1)_7\)&\(2\,\Wc(\tfrac13)\oplus\Rc^{(2)}(0,2)_5\)\\
  &\(\otimes\)&\(\Rc^{(2)}(0,2)_5\)&\(2\,\Wc(\tfrac{10}{3})\oplus\Rc^{(2)}(0,1)_7\)\\
  &\(\otimes\)&\(\Rc^{(2)}(0,2)_7\)&\(\Rc^{(2)}(0,1)_5\)\\
  &\(\otimes\)&\(\Rc^{(2)}(0,1)_5\)&\(\Rc^{(2)}(0,2)_7\oplus{\cal
    R}^{(2)}(\tfrac13,\tfrac13)\)\\
  &\(\otimes\)&\(\Rc^{(2)}(\tfrac58,\tfrac58)\)
  &\(2\,\Wc(\tfrac{35}{24})\oplus\Rc^{(2)}(\tfrac18,\tfrac18)\)\\
  &\(\otimes\)&\(\Rc^{(2)}(\tfrac18,\tfrac18)\)&\(2\,\Wc(\tfrac{-1}{24})\oplus\Rc^{(2)}(\tfrac58,\tfrac58)\)\\
  &\(\otimes\)&\(\Rc^{(2)}(\tfrac13,\tfrac13)\)&\(\Rc^{(3)}(0,0,1,1)\)\\
  &\(\otimes\)&\(\Rc^{(3)}(0,0,1,1)\)
  &\(2\,\Rc^{(2)}(\tfrac13,\tfrac13)\oplus\Rc^{(3)}(0,0,2,2)\)\\
  &\(\otimes\)&\(\Rc^{(3)}(0,0,2,2)\)
  &\(2\,\Rc^{(2)}(\tfrac13,\tfrac{10}{3})\oplus\Rc^{(3)}(0,0,1,1)\)\\[.3em]
  \hline
  \(\Wc(\tfrac13)\)&\(\otimes\)&\(\Wc(\tfrac13)\)&\(\Wc(\tfrac13)\oplus{\cal
    R}^{(2)}(0,2)_5\)\\
  &\(\otimes\)&\(\Wc(\tfrac58)\)&\(\Wc(\tfrac{-1}{24})\)\\
  &\(\otimes\)&\(\Wc(\tfrac18)\)&\({\cal
    R}^{(2)}(\tfrac18,\tfrac18)\)\\
  &\(\otimes\)&\(\Wc(\tfrac{-1}{24})\)&\(\Wc(\tfrac{-1}{24})\oplus\Rc^{(2)}(\tfrac58,\tfrac58)\)\\
  &\(\otimes\)&\(\Rc^{(2)}(0,1)_7\)&\(2\,\Wc(\tfrac{10}{3})\oplus2\,\Rc^{(2)}(0,1)_7\)\\
  &\(\otimes\)&\(\Rc^{(2)}(0,2)_5\)&\(2\,\Wc(\tfrac13)\oplus2\,\Rc^{(2)}(2,5)\)\\
  &\(\otimes\)&\(\Rc^{(2)}(0,2)_7\)&\(\Rc^{(2)}(\tfrac13,\tfrac13)\)\\
  &\(\otimes\)&\(\Rc^{(2)}(0,1)_5\)&\(\Rc^{(3)}(0,0,1,1)\)\\
  &\(\otimes\)&\(\Rc^{(2)}(\tfrac58,\tfrac58)\)&\(2\,\Wc(\tfrac{-1}{24})\oplus2\,\Rc^{(2)}(\tfrac58,\tfrac{21}{8})\)\\
  &\(\otimes\)&\(\Rc^{(2)}(\tfrac18,\tfrac18)\)&\(2\,\Wc(\tfrac{35}{24})\oplus2\,\Rc^{(2)}(\tfrac18,\tfrac18)\)\\
  &\(\otimes\)&\(\Rc^{(2)}(\tfrac13,\tfrac13)\)&\(\Rc^{(2)}(\tfrac13,\tfrac13)\oplus\Rc^{(3)}(0,0,2,2)\)\\
  &\(\otimes\)&\(\Rc^{(3)}(0,0,1,1)\)&\(2\,\Rc^{(2)}(\tfrac13,\tfrac{10}{3})\oplus2\,\Rc^{(3)}(0,0,1,1)\)\\
  &\(\otimes\)&\(\Rc^{(3)}(0,0,2,2)\)&\(2\,\Rc^{(2)}(\tfrac13,\tfrac13)
  \oplus2\,\Rc^{(3)}(0,1,2,5)\)\\[.3em]
  \hline
  \(\Wc(\tfrac58)\)&\(\otimes\)&\(\Wc(\tfrac58)\)&\(\Rc^{(2)}(0,2)_7\)\\
  &\(\otimes\)&\(\Wc(\tfrac18)\)&\(\Rc^{(2)}(0,1)_5\)\\
  &\(\otimes\)&\(\Wc(\tfrac{-1}{24})\)&\(\Rc^{(2)}(\tfrac13,\tfrac13)\)\\
  &\(\otimes\)&\(\Rc^{(2)}(0,1)_7\)&\(\Rc^{(2)}(\tfrac18,\tfrac18)\)\\
  &\(\otimes\)&\(\Rc^{(2)}(0,2)_5\)&\(\Rc^{(2)}(\tfrac58,\tfrac58)\)\\
  &\(\otimes\)&\(\Rc^{(2)}(0,2)_7\)&\(2\,\Wc(\tfrac58)\oplus2\,\Wc(\tfrac{33}{8})\)\\
  &\(\otimes\)&\(\Rc^{(2)}(0,1)_5\)&\(2\,\Wc(\tfrac18)\oplus2\,\Wc(\tfrac{21}{8})\)\\
  &\(\otimes\)&\(\Rc^{(2)}(\tfrac58,\tfrac58)\)&\(\Rc^{(3)}(0,0,2,2)\)\\
  &\(\otimes\)&\(\Rc^{(2)}(\tfrac18,\tfrac18)\)&\(\Rc^{(3)}(0,0,1,1)\)\\
  &\(\otimes\)&\({\cal
    R}^{(2)}(\tfrac13,\tfrac13)\)&\(2\,\Wc(\tfrac{-1}{24})\oplus2\,\Wc(\tfrac{35}{24})\)\\
  &\(\otimes\)&\(\Rc^{(3)}(0,0,1,1)\)&\(2\,\Rc^{(2)}(\tfrac18,\tfrac18)
  \oplus2\,\Rc^{(2)}(\tfrac58,\tfrac{21}{8})\)\\
  &\(\otimes\)&\(\Rc^{(3)}(0,0,2,2)\)&\(2\,\Rc^{(2)}(\tfrac58,\tfrac58)
  \oplus2\,\Rc^{(2)}(\tfrac18,\tfrac{33}{8})\)\\[.3em]
  \hline
  \(\Wc(\tfrac18)\)&\(\otimes\)&\(\Wc(\tfrac18)\)&\(\Rc^{(2)}(0,2)_7\oplus{\cal
    R}^{(2)}(\tfrac13,\tfrac13)\)\\
  &\(\otimes\)&\(\Wc(\tfrac{-1}{24})\)&\(\Rc^{(3)}(0,0,1,1)\)\\
  &\(\otimes\)&\(\Rc^{(2)}(0,1)_7\)&\(2\,\Wc(\tfrac{-1}{24})\oplus\Rc^{(2)}(\tfrac58,\tfrac58)\)\\
  &\(\otimes\)&\(\Rc^{(2)}(0,2)_5\)&\(2\,\Wc(\tfrac{35}{24})\oplus\Rc^{(2)}(\tfrac18,\tfrac18)\)\\
  &\(\otimes\)&\(\Rc^{(2)}(0,2)_7\)&\(2\,\Wc(\tfrac18)\oplus2\,\Wc(\tfrac{21}{8})\)\\
  &\(\otimes\)&\(\Rc^{(2)}(0,1)_5\)&\(2\,\Wc(\tfrac58)\oplus
    2\,\Wc(\tfrac{33}{8})\oplus2\,\Wc(\tfrac{-1}{24})\oplus2\,\Wc(\tfrac{35}{24})\)\\
  &\(\otimes\)&\(\Rc^{(2)}(\tfrac58,\tfrac58)\)
  &\(2\,\Rc^{(2)}(\tfrac13,\tfrac{10}{3})\oplus\Rc^{(3)}(0,0,1,1)\)\\
  &\(\otimes\)&\(\Rc^{(2)}(\tfrac18,\tfrac18)\)
  &\(2\,\Rc^{(2)}(\tfrac13,\tfrac13)\oplus\Rc^{(3)}(0,0,2,2)\)\\
  &\(\otimes\)&\(\Rc^{(2)}(\tfrac13,\tfrac13)\)
  &\(2\,\Rc^{(2)}(\tfrac18,\tfrac18)\oplus2\,\Rc^{(2)}(\tfrac58,\tfrac{21}{8})\)\\
  &\(\otimes\)&\(\Rc^{(3)}(0,0,1,1)\)&\(4\,\Wc(\tfrac{-1}{24})\oplus
     4\,\Wc(\tfrac{35}{24})\oplus2\,\Rc^{(2)}(\tfrac58,\tfrac58)\oplus2\,\Rc^{(2)}(\tfrac18,\tfrac{33}{8})\)\\
  &\(\otimes\)&\(\Rc^{(3)}(0,0,2,2)\)&\(4\,\Wc(\tfrac{-1}{24})\oplus
     4\,\Wc(\tfrac{35}{24})\oplus2\,\Rc^{(2)}(\tfrac18,\tfrac18)\oplus2\,\Rc^{(2)}(\tfrac58,\tfrac{21}{8})\)\\[.3em]
  \hline
  \(\Wc(\tfrac{-1}{24})\)&\(\otimes\)&\(\Wc(\tfrac{-1}{24})\)&\({\cal
    R}^{(2)}(\tfrac13,\tfrac13)\oplus\Rc^{(3)}(0,0,2,2)\)\\
  &\(\otimes\)&\(\Rc^{(2)}(0,1)_7\)&\(2\,\Wc(\tfrac{35}{24})\oplus2\,\Rc^{(2)}(\tfrac18,\tfrac18)\)\\
  &\(\otimes\)&\(\Rc^{(2)}(0,2)_5\)&\(2\,\Wc(\tfrac{-1}{24})\oplus2\,\Rc^{(2)}(\tfrac58,\tfrac{21}{8})\)\\
  &\(\otimes\)&\(\Rc^{(2)}(0,2)_7\)&\(2\,\Wc(\tfrac{-1}{24})\oplus2\,\Wc(\tfrac{35}{24})\)\\
  &\(\otimes\)&\(\Rc^{(2)}(0,1)_5\)
  &\(2\,\Rc^{(2)}(\tfrac18,\tfrac18)\oplus2\,\Rc^{(2)}(\tfrac58,\tfrac{21}{8})\)\\
  &\(\otimes\)&\(\Rc^{(2)}(\tfrac58,\tfrac58)\)
  &\(2\,\Rc^{(2)}(\tfrac13,\tfrac13)\oplus2\,\Rc^{(3)}(0,1,2,5)\)\\
  &\(\otimes\)&\(\Rc^{(2)}(\tfrac18,\tfrac18)\)
  &\(2\,\Rc^{(2)}(\tfrac13,\tfrac{10}{3})\oplus2\,\Rc^{(3)}(0,0,1,1)\)\\
  &\(\otimes\)&\(\Rc^{(2)}(\tfrac13,\tfrac13)\)
  &\(2\,\Wc(\tfrac{-1}{24})\oplus2\,\Wc(\tfrac{35}{24})\oplus2\,{\cal
    R}^{(2)}(\tfrac58,\tfrac58)\oplus2\,{\cal
    R}^{(2)}(\tfrac18,\tfrac{33}{8})\)\\
  &\(\otimes\)&\(\Rc^{(3)}(0,0,1,1)\)&\(4\,\Wc(\tfrac{-1}{24})\oplus4\,\Wc(\tfrac{35}{24})
    \oplus4\,\Rc^{(2)}(\tfrac18,\tfrac18)\oplus4\,{\cal
    R}^{(2)}(\tfrac58,\tfrac{21}{8})\)\\
  &\(\otimes\)&\(\Rc^{(3)}(0,0,2,2)\)&\(4\,\Wc(\tfrac{-1}{24})\oplus4\,\Wc(\tfrac{35}{24})
    \oplus4\,\Rc^{(2)}(\tfrac18,\tfrac18)\oplus4\,{\cal
    R}^{(2)}(\tfrac58,\tfrac{21}{8})\)\\[.3em]
  \hline
  \(\Rc^{(2)}(0,1)_7\)&\(\otimes\)&\(\Rc^{(2)}(0,1)_7\)
  &\(4\,\Wc(\tfrac13)\oplus2\,\Rc^{(2)}(2,5)
    \oplus2\,\Rc^{(2)}(0,2)_5\)\\
  &\(\otimes\)&\(\Rc^{(2)}(0,2)_5\)
  &\(4\,\Wc(\tfrac{10}{3})\oplus2\,\Rc^{(2)}(0,1)_7\oplus2\,\Rc^{(2)}(1,7)\)\\
  &\(\otimes\)&\(\Rc^{(2)}(0,2)_7\)&\(\Rc^{(3)}(0,0,1,1)\)\\
  &\(\otimes\)&\(\Rc^{(2)}(0,1)_5\)&\(2\,{\cal
    R}^{(2)}(\tfrac13,\tfrac13)
   \oplus\Rc^{(3)}(0,0,2,2)\)\\
  &\(\otimes\)&\(\Rc^{(2)}(\tfrac58,\tfrac58)\)
  &\(4\,\Wc(\tfrac{35}{24})\oplus2\,\Rc^{(2)}(\tfrac18,\tfrac{33}{8})
    \oplus2\,\Rc^{(2)}(\tfrac18,\tfrac18)\)\\
  &\(\otimes\)&\(\Rc^{(2)}(\tfrac18,\tfrac18)\)
  &\(4\,\Wc(\tfrac{-1}{24})\oplus2\,\Rc^{(2)}(\tfrac58,\tfrac58)
    \oplus2\,\Rc^{(2)}(\tfrac58,\tfrac{21}{8})\)\\
  &\(\otimes\)&\(\Rc^{(2)}(\tfrac13,\tfrac13)\)
  &\(2\,\Rc^{(2)}(\tfrac13,\tfrac{10}{3})\oplus2\,\Rc^{(3)}(0,0,1,1)\)\\
  &\(\otimes\)&\(\Rc^{(3)}(0,0,1,1)\)&\(4\,\Rc^{(2)}(\tfrac13,\tfrac13)
  \oplus2\,\Rc^{(3)}(0,1,2,5)
    \oplus2\,\Rc^{(2)}(0,0,2,2)\)\\
  &\(\otimes\)&\(\Rc^{(3)}(0,0,2,2)\)
  &\(4\,\Rc^{(2)}(\tfrac13,\tfrac{10}{3})\oplus2\,{\cal
    R}^{(3)}(0,0,1,1)\oplus2\,\Rc^{(2)}(0,1,2,7)\)\\[.3em]
  \hline
  \(\Rc^{(2)}(0,2)_5\)&\(\otimes\)&\(\Rc^{(2)}(0,2)_5\)
  &\(4\,\Wc(\tfrac13)\oplus2\,\Rc^{(2)}(2,5)\oplus2\,\Rc^{(2)}(0,2)_5\)\\
  &\(\otimes\)&\(\Rc^{(2)}(0,2)_7\)&\(\Rc^{(3)}(0,0,2,2)\)\\
  &\(\otimes\)&\(\Rc^{(2)}(0,1)_5\)&\(2\,\Rc^{(2)}(\tfrac13,\tfrac{10}{3})\oplus\Rc^{(3)}(0,0,1,1)\)\\
  &\(\otimes\)&\(\Rc^{(2)}(\tfrac58,\tfrac58)\)
  &\(4\,\Wc(\tfrac{-1}{24})\oplus2\,\Rc^{(2)}(\tfrac58,\tfrac58)
    \oplus2\,\Rc^{(2)}(\tfrac58,\tfrac{21}{8})\)\\
  &\(\otimes\)&\(\Rc^{(2)}(\tfrac18,\tfrac18)\)
  &\(4\,\Wc(\tfrac{35}{24})\oplus2\,\Rc^{(2)}(\tfrac18,\tfrac{33}{8})
    \oplus2\,\Rc^{(2)}(\tfrac18,\tfrac18)\)\\
  &\(\otimes\)&\(\Rc^{(2)}(\tfrac13,\tfrac13)\)
  &\(2\,\Rc^{(2)}(\tfrac13,\tfrac13)\oplus2\,\Rc^{(3)}(0,1,2,5)\)\\
  &\(\otimes\)&\(\Rc^{(3)}(0,0,1,1)\)&\(4\,\Rc^{(2)}(\tfrac13,\tfrac{10}{3})
  \oplus2\,\Rc^{(3)}(0,0,1,1)\oplus2\,\Rc^{(2)}(0,1,2,7)\)\\
  &\(\otimes\)&\(\Rc^{(3)}(0,0,2,2)\)&\(4\,\Rc^{(2)}(\tfrac13,\tfrac13)
  \oplus2\,\Rc^{(3)}(0,1,2,5)\oplus2\,\Rc^{(2)}(0,0,2,2)\)\\[.3em]
  \hline
  \(\Rc^{(2)}(0,2)_7\)&\(\otimes\)&\(\Rc^{(2)}(0,2)_7\)
  &\(2\,\Rc^{(2)}(0,2)_7\oplus2\,\Rc^{(2)}(2,7)\)\\
  &\(\otimes\)&\(\Rc^{(2)}(0,1)_5\)&\(2\,\Rc^{(2)}(0,1)_5\oplus2\,\Rc^{(2)}(1,5)\)\\
  &\(\otimes\)&\(\Rc^{(2)}(\tfrac58,\tfrac58)\)
  &\(2\,\Rc^{(2)}(\tfrac58,\tfrac58)\oplus2\,\Rc^{(2)}(\tfrac18,\tfrac{33}{8})\)\\
  &\(\otimes\)&\(\Rc^{(2)}(\tfrac18,\tfrac18)\)
  &\(2\,\Rc^{(2)}(\tfrac18,\tfrac18)\oplus 2\,\Rc^{(2)}(\tfrac58,\tfrac{21}{8})\)\\
  &\(\otimes\)&\(\Rc^{(2)}(\tfrac13,\tfrac13)\)
  &\(2\,\Rc^{(2)}(\tfrac13,\tfrac13)\oplus2\,\Rc^{(2)}(\tfrac13,\tfrac{10}{3})\)\\
  &\(\otimes\)&\(\Rc^{(3)}(0,0,1,1)\)&\(2\,\Rc^{(3)}(0,0,1,1)
  \oplus2\,\Rc^{(3)}(0,1,2,5)\)\\
  &\(\otimes\)&\(\Rc^{(3)}(0,0,2,2)\)&\(2\,\Rc^{(3)}(0,0,2,2)
  \oplus2\,\Rc^{(3)}(0,1,2,7)\)\\[.3em]
  \hline
  \(\Rc^{(2)}(0,1)_5\)&\(\otimes\)&\(\Rc^{(2)}(0,1)_5\)
  &\(2\,\Rc^{(2)}(0,2)_7\oplus2\,{\cal
    R}^{(2)}(2,7)\oplus2\,\Rc^{(2)}(\tfrac13,\tfrac13)\oplus2\,\Rc^{(2)}(\tfrac13,\tfrac{10}{3})\)\\
  &\(\otimes\)&\(\Rc^{(2)}(\tfrac58,\tfrac58)\)
  &\(4\,\Wc(\tfrac{-1}{24})\oplus4\,\Wc(\tfrac{35}{24})\oplus2\,{\cal
    R}^{(2)}(\tfrac18,\tfrac18)\oplus2\,\Rc^{(2)}(\tfrac58,\tfrac{21}{8})\)\\
  &\(\otimes\)&\(\Rc^{(2)}(\tfrac18,\tfrac18)\)
  &\(4\,\Wc(\tfrac{-1}{24})\oplus4\,\Wc(\tfrac{35}{24})\oplus2\,{\cal
    R}^{(2)}(\tfrac58,\tfrac58)\oplus2\,\Rc^{(2)}(\tfrac18,\tfrac{33}{8})\)\\
  &\(\otimes\)&\(\Rc^{(2)}(\tfrac13,\tfrac13)\)
  &\(2\,\Rc^{(3)}(0,0,1,1)\oplus2\,\Rc^{(3)}(0,1,2,5)\)\\
  &\(\otimes\)&\(\Rc^{(3)}(0,0,1,1)\)&\(4\,{\cal
    R}^{(2)}(\tfrac13,\tfrac13)\oplus4\,{\cal
    R}^{(2)}(\tfrac13,\tfrac{10}{3})\)\\
  &&&\(\qquad \oplus 2\,\Rc^{(3)}(0,0,2,2)\oplus2\,\Rc^{(3)}(0,1,2,7)\)\\  
  &\(\otimes\)&\(\Rc^{(3)}(0,0,2,2)\)&\(4\,{\cal
    R}^{(2)}(\tfrac13,\tfrac13)\oplus4\,{\cal
    R}^{(2)}(\tfrac13,\tfrac{10}{3})\)\\
  &&&\(\qquad \oplus 2\,\Rc^{(3)}(0,0,1,1)\oplus2\,\Rc^{(3)}(0,1,2,5)\)\\[.3em]
  \hline
  \(\Rc^{(2)}(\tfrac58,\tfrac58)\)&\(\otimes\)&\({\cal
    R}^{(2)}(\tfrac58,\tfrac58)\)
  &\(4\,\Rc^{(2)}(\tfrac13,\tfrac13)\oplus2\,{\cal
    R}^{(3)}(0,1,2,5)\oplus2\,\Rc^{(3)}(0,0,2,2)\)\\
  &\(\otimes\)&\(\Rc^{(2)}(\tfrac18,\tfrac18)\)
  &\(4\,\Rc^{(2)}(\tfrac13,\tfrac{10}{3})\oplus2\,{\cal
    R}^{(3)}(0,0,1,1)\oplus2\,\Rc^{(3)}(0,1,2,7)\)\\
  &\(\otimes\)&\(\Rc^{(2)}(\tfrac13,\tfrac13)\)
  &\(4\,\Wc(\tfrac{-1}{24})\oplus4\,\Wc(\tfrac{35}{24})\oplus4\,{\cal
    R}^{(2)}(\tfrac18,\tfrac18)\oplus4\,\Rc^{(2)}(\tfrac58,\tfrac{21}{8})\)\\
  &\(\otimes\)&\(\Rc^{(3)}(0,0,1,1)\)&\(8\,\Wc(\tfrac{-1}{24})\oplus8\,\Wc(\tfrac{35}{24})\oplus4\,{\cal
    R}^{(2)}(\tfrac58,\tfrac58)\)\\
  &&&\(\qquad \oplus 4\,\Rc^{(2)}(\tfrac18,\tfrac{33}{8})\oplus4\,{\cal
    R}^{(2)}(\tfrac18,\tfrac18)\oplus4\,\Rc^{(2)}(\tfrac58,\tfrac{21}{8})\)\\
  &\(\otimes\)&\(\Rc^{(3)}(0,0,2,2)\)&\(8\,\Wc(\tfrac{-1}{24})\oplus8\,\Wc(\tfrac{35}{24})\oplus4\,{\cal
    R}^{(2)}(\tfrac58,\tfrac58)\)\\
  &&&\(\qquad \oplus 4\,\Rc^{(2)}(\tfrac18,\tfrac{33}{8})\oplus4\,{\cal
    R}^{(2)}(\tfrac18,\tfrac18)\oplus4\,\Rc^{(2)}(\tfrac58,\tfrac{21}{8})\)\\[.3em]
  \hline
  \(\Rc^{(2)}(\tfrac18,\tfrac18)\)&\(\otimes\)&\({\cal
    R}^{(2)}(\tfrac18,\tfrac18)\)
  &\(4\,\Rc^{(2)}(\tfrac13,\tfrac13)\oplus2\,{\cal
    R}^{(3)}(0,1,2,5)\oplus2\,\Rc^{(3)}(0,0,2,2)\)\\
  &\(\otimes\)&\(\Rc^{(2)}(\tfrac13,\tfrac13)\)
  &\(4\,\Wc(\tfrac{-1}{24})\oplus4\,\Wc(\tfrac{35}{24})\oplus4\,\Rc^{(2)}(\tfrac18,\tfrac18)
  \oplus4\,\Rc^{(2)}(\tfrac58,\tfrac{21}{8})\)\\
  &\(\otimes\)&\(\Rc^{(3)}(0,0,1,1)\)&\(8\,\Wc(\tfrac{-1}{24})\oplus8\,\Wc(\tfrac{35}{24})\oplus4\,{\cal
    R}^{(2)}(\tfrac58,\tfrac58)\)\\
  &&&\(\qquad \oplus 4\,\Rc^{(2)}(\tfrac18,\tfrac{33}{8})\oplus4\,{\cal
    R}^{(2)}(\tfrac18,\tfrac18)\oplus4\,\Rc^{(2)}(\tfrac58,\tfrac{21}{8})\)\\
  &\(\otimes\)&\(\Rc^{(3)}(0,0,2,2)\)&\(8\,\Wc(\tfrac{-1}{24})\oplus8\,\Wc(\tfrac{35}{24})\oplus4\,{\cal
    R}^{(2)}(\tfrac58,\tfrac58)\)\\
  &&&\(\qquad \oplus 4\,\Rc^{(2)}(\tfrac18,\tfrac{33}{8})\oplus4\,{\cal
    R}^{(2)}(\tfrac18,\tfrac18)\oplus4\,\Rc^{(2)}(\tfrac58,\tfrac{21}{8})\)\\[.3em]
  \hline
  \(\Rc^{(2)}(\tfrac13,\tfrac13)\)&\(\otimes\)&\({\cal
    R}^{(2)}(\tfrac13,\tfrac13)\)&\(2\,\Rc^{(2)}(\tfrac13,\tfrac13)
  \oplus2\,\Rc^{(2)}(\tfrac13,\tfrac{10}{3})\)\\
  &&&\(\qquad \oplus2\,{\cal
   R}^{(3)}(0,0,2,2)\oplus2\,\Rc^{(3)}(0,1,2,7)\)\\
  &\(\otimes\)&\(\Rc^{(3)}(0,0,1,1)\)&\(4\,{\cal
    R}^{(2)}(\tfrac13,\tfrac13)\oplus4\,{\cal
    R}^{(2)}(\tfrac13,\tfrac{10}{3})\)\\
  &&&\(\qquad \oplus 4\,\Rc^{(3)}(0,0,1,1)\oplus4\,\Rc^{(3)}(0,1,2,5)\)\\
  &\(\otimes\)&\(\Rc^{(3)}(0,0,2,2)\)&\(4\,{\cal
    R}^{(2)}(\tfrac13,\tfrac13)\oplus4\,{\cal
    R}^{(2)}(\tfrac13,\tfrac{10}{3})\)\\
  &&&\(\qquad \oplus 4\,\Rc^{(3)}(0,0,1,1)\oplus4\,{\cal
    R}^{(3)}(0,1,2,5)\)\\[.3em]
\hline
  \(\Rc^{(3)}(0,0,1,1)\hspace{-.5em}\)&\(\otimes\)&\({\cal
    R}^{(3)}(0,0,1,1)\)&\(8\,\Rc^{(2)}(\tfrac13,\tfrac13)\oplus8\,\Rc^{(2)}(\tfrac13,\tfrac{10}{3})
      \oplus 4\,\Rc^{(3)}(0,0,1,1)\)\\
  &&&\(\oplus4\,\Rc^{(3)}(0,1,2,5) \oplus 4\,\Rc^{(3)}(0,0,2,2)\oplus4\,\Rc^{(3)}(0,1,2,7)\)\\
  &\(\otimes\)&\(\Rc^{(3)}(0,0,2,2)\)&\(8\,{\cal
    R}^{(2)}(\tfrac13,\tfrac13)\oplus8\,\Rc^{(2)}(\tfrac13,\tfrac{10}{3})
      \oplus 4\,\Rc^{(3)}(0,0,1,1)\)\\
  &&&\(\oplus4\,\Rc^{(3)}(0,1,2,5) \oplus 4\,\Rc^{(3)}(0,0,2,2)\oplus4\,\Rc^{(3)}(0,1,2,7)\)\\[.3em]
  \hline
  \(\Rc^{(3)}(0,0,2,2)\hspace{-.5em}\)&\(\otimes\)&\({\cal
    R}^{(3)}(0,0,2,2)\)&\(8\,\Rc^{(2)}(\tfrac13,\tfrac13)\oplus8\,\Rc^{(2)}(\tfrac13,\tfrac{10}{3})
      \oplus 4\,\Rc^{(3)}(0,0,1,1)\)\\
  &&&\(\oplus4\,\Rc^{(3)}(0,1,2,5)\oplus 4\,\Rc^{(3)}(0,0,2,2)\oplus4\,{\cal
    R}^{(3)}(0,1,2,7)\)
\end{longtable}
\end{center}

\section{Some technical lemmas and proofs}\label{app:cat-stuff}

\subsection{Associativity for composition of internal Homs}\label{app:iHom-proofs}

\begin{Lem}\label{lem:phi-nat-2}
For $g : B \rightarrow [U,V]$ we have 
$g = \phi_B^{(U,V)}\big(ev_{U,V} \circ (g \otimes \id_U)\big)$.
\end{Lem}

\begin{proof} This is a consequence of applying 
\eqref{eq:phi-nat} for $X = [U,V]$, $Y=B$ and $t = ev_{U,V}$.
\end{proof}

\begin{Lem}\label{lem:c-via-ev}
$ev_{V,W} \circ (\id_{[V,W]} \otimes ev_{U,V}) \circ 
\alpha^{~-1}_{[V,W],[U,V],U} = ev_{U,W} \circ ( m_{W,V,U} \otimes \id_U )$.
\end{Lem}

\begin{proof} First apply $\phi^{(U,W)}_{[V,W]\otimes[U,V]}$ to both sides. 
The next step is to show that the resulting morphisms are equal. By definition in 
\eqref{eq:iHom-morphs}, the left hand side is equal to $m_{W,V,U}$. On the right 
hand side one uses Lemma~\ref{lem:phi-nat-2} with $g = m_{W,V,U}$. 
\end{proof}

\begin{Lem}\label{lem:eta-and-ev}
$ev_{U,U} \circ (\eta_U \otimes \id_U) = \lambda_U$.
\end{Lem}

\begin{proof} From Lemma~\ref{lem:phi-nat-2} with $g=\eta_U$ we get 
$\eta_U = \phi_\one^{(U,U)} \big(ev_{U,U} \circ (\eta_U \otimes \id_U)$. 
By definition, $\eta_U =  \phi_\one^{(U,U)} (\lambda_U)$, and the statement follows.
\end{proof}

\medskip\noindent
{\em Proof of Theorem~\ref{thm:assoc}:}\\[.3em]
\emph{Associativity:} Let $g$ be the left hand side of the associativity equation, 
and $g'$ the right hand side. By Lemma~\ref{lem:phi-nat-2} it is enough to show that 
$ev_{A,D} \circ (g \otimes \id_A) = ev_{A,D} \circ (g' \otimes \id_A)$. For the two sides 
one finds (omitting the indices (exercise: add the indices) and using 
Lemma~\ref{lem:c-via-ev})
\bea
  ev \circ (g \otimes \id)
  = ev \circ (m \otimes \id) \circ ( (\id \otimes m) \otimes \id)
  = ev \circ (\id \otimes ev) \circ \alpha^{-1} \circ ( (\id \otimes m) \otimes \id)
\enl  
  = ev \circ \big[ \id \otimes \big( ev \circ (m \otimes id) \big) \big] \circ \alpha^{-1}
  = ev \circ \big[ \id \otimes \big( ev \circ (\id \otimes ev) \circ \alpha^{-1} \big) \big] \circ 
      \alpha^{-1}
\enl  
  = ev \circ (\id \otimes ev) \circ (\id \otimes (\id \otimes ev))
  \circ (\id \otimes \alpha^{-1}) \circ \alpha^{-1}
\eear\ee
and
\bea
  ev \circ (g' \otimes \id)
  = ev \circ (m \otimes \id) \circ ( (m \otimes \id) \otimes \id) \circ (\alpha \otimes \id)
\enl  
  = ev \circ (\id \otimes ev) \circ \alpha^{-1} \circ ( (m \otimes \id) \otimes \id) \circ 
  (\alpha \otimes \id)
\enl  
  = ev \circ (m \otimes \id) \circ (\id \otimes ev) \circ \alpha^{-1} \circ (\alpha \otimes \id)
\enl  
  = ev \circ (\id \otimes ev) \circ (\id \otimes (\id \otimes ev))  \circ \alpha^{-1}  \circ \alpha^{-1} 
  \circ (\alpha \otimes \id)   \ .
\eear\ee
The two expressions are equal if (putting the indices back)
\bea
  (\id_{[C,D]} \otimes \alpha^{-1}_{[B,C],[A,B],A}) \circ \alpha^{-1}_{[C,D],[B,C]
  \otimes[A,B],A}
\enl
  = \alpha^{-1}_{[C,D],[B,C],[A,B]\otimes A}  \circ \alpha^{-1}_{[C,D]\otimes[B,C],[A,B],A} 
  \circ (\alpha_{[C,D],[B,C],[A,B]} \otimes \id_{A}) \ .
\eear\ee
This equality holds because any two ways of rebracketing are equal in a tensor category. 
Concretely, it follows from the pentagon equation satisfied by the associator,
\be
   \alpha_{T\otimes U,V,W} \circ \alpha_{T,U,V \otimes W}  =
  (\alpha_{T,U,V} \otimes \id_W) \circ \alpha_{T , U\otimes V, W} 
  \circ (\id_T \otimes \alpha_{U,V,W})  \ .
\ee
\emph{Unit:} For the first of the two unit conditions set 
$g = m_{B,B,A} \circ (\eta_B \otimes \id_{[A,B]})$ and 
$g' = \lambda_{[A,B]}$. By Lemma~\ref{lem:phi-nat-2} it is enough to show 
that $ev \circ (g \otimes \id) = ev \circ (g' \otimes id)$. 
Using also Lemma~\ref{lem:eta-and-ev} we get
\bea
  ev \circ (g \otimes \id) 
  = ev \circ (m \otimes \id) \circ ((\eta \otimes \id) \otimes \id)
  = ev \circ (\id \otimes ev) \circ \alpha^{-1} \circ ((\eta \otimes \id) \otimes \id)
\enl
  = ev \circ (\eta \otimes \id) \circ (\id \otimes ev) \circ \alpha^{-1} 
  = \lambda \circ (\id \otimes ev) \circ \alpha^{-1} 
  = ev \circ \lambda \circ \alpha^{-1} \ .
\eear\ee
This is equal to $ev \circ (g' \otimes \id)$ if $\lambda_{[A,B]\otimes A} \circ 
\alpha^{-1}_{\one,[A,B],A} = \lambda_{[A,B]} \otimes \id_A$. The last identity 
follows from the axioms of a tensor category, see \cite[Prop.\,1.1]{Joyal:1993}.
For the second unit condition set $g=m_{B,B,A} \circ (\id_{[A,B]} \otimes \eta_A)$ and 
$g' = \rho_{[A,B]}$. We get
\be
  ev \circ (g \otimes \id) 
  = ev \circ 
    (m \otimes \id) 
  \circ ((\id \otimes \eta) \otimes \id)
  = ev \circ (\id \otimes \lambda) \circ \alpha^{-1}\ .
\ee
For this to be equal to $ev \circ (g' \otimes \id)$ we need 
$(\id_{[A,B]} \otimes \lambda_A) \circ \alpha^{-1}_{[A,B],\one,A} = \rho_{[A,B]} \otimes \id_A$, 
which is an instance of the triangle condition,
\be
  \id_U \otimes \lambda_V = (\rho_U \otimes \id_V) \circ \alpha_{U,\one,V} \ .
\ee
\qed

\subsection{Proof of Theorems \ref{thm:Cdd-prop} and \ref{thm:non-deg}}\label{app:some-proofs}

\begin{Lem}\label{lem:Cdbar-dualclosed}
Let $\Cc$ be a tensor category satisfying condition $\mathrm{C}$. 
If $U \in \Cdd$ then so is $U^*$.
\end{Lem}

\begin{proof}
The duality morphisms for $U^*$ are constructed from $\delta_U$ 
and the duality morphisms of $U$ as
\be\begin{array}{ll}
b_{U^*} = \one \xrightarrow{\tilde b_U} U^* \otimes U \xrightarrow{\id_{U^*} 
\otimes \delta_U} U^* \otimes  U^{**} ~,~~
\etb
d_{U^*} = U^{**} \otimes U^* \xrightarrow{\delta_U^{-1} \otimes \id_{U^*}} 
U \otimes U^* \xrightarrow{\tilde d_U} \one \ ,
\enl
\tilde b_{U^*} = \one \xrightarrow{b_U} U \otimes U^* \xrightarrow{\delta_U 
\otimes \id_{U^*}} U^{**} \otimes U^*~,~~
\etb
\tilde d_{U^*} = U^* \otimes  U^{**}  \xrightarrow{\id_{U^*} \otimes \delta_U^{-1}}   
U^* \otimes U  \xrightarrow{d_U} \one \ .
\end{array}
\ee
The check that these satisfy the duality properties is a straightforward calculation 
using the duality properties of $b_{U}$, $d_{U}$, $\tilde b_{U}$, $\tilde d_{U}$.
It is also clear that $b_{U^*}$, $\tilde b_{U^*}$ are injective because they are the 
composition of an injective map and a bijection.
\end{proof}

\begin{Lem}\label{lem:b-inject}
Let $\Cc$ be a tensor category.\\[.3em]
{\rm (i)} Let $U \in \Cc$ have a right dual. Then $b_U : \one \rightarrow U \otimes U^\vee$ 
is injective if and only if the map 
$f \mapsto f \otimes \id_U : \Hom(X,\one) \rightarrow (X\otimes U,\one \otimes U)$ is 
injective for all $X \in \Cc$.\\[.3em]
{\rm (ii)} Let $U \in \Cc$ have a left dual. Then $\tilde b_U : \one \rightarrow \eev U \otimes U$ 
is injective if and only if 
the map $f \mapsto \id_U \otimes f : \Hom(X,\one) \rightarrow (U \otimes X,U \otimes \one)$ 
is injective for all $X \in \Cc$.
\end{Lem}

\begin{proof}
The proof is straightforward. For example, for (i) one shows with the help of the 
duality morphisms that $b_U \circ f = b_U \circ g$ is equivalent to 
$f \otimes \id_U = g \otimes \id_U$.
\end{proof}

\begin{Lem}\label{lem:Cdbar-tensorclosed}
Let $\Cc$ be a tensor category satisfying condition $\mathrm{C}$. If $U,V \in \Cdd$ 
then so is $U \otimes V$.
\end{Lem}

\begin{proof}
The duality morphisms for $U\otimes V$ are constructed as in the proof of 
Lemma~\ref{lem:Cd-tensorcat}, 
together with the observation \eqref{eq:AB*-AB*} that for $U,V \in \Cdd$ we have 
$(U \otimes V)^* \cong V^* \otimes U^*$.
It remains to check that $b_{U \otimes V}$ and 
$\tilde b_{U \otimes V}$ are injective. This follows from Lemma~\ref{lem:b-inject}; 
let us go through the argument for 
$b_{U \otimes V} : \one \rightarrow (U \otimes V) \otimes (U \otimes V)^*$. This 
morphism is injective if and only if the map $f \mapsto f \otimes \id_{U \otimes V}$ is 
injective. But by assumption $b_U$ and $b_V$ are injective so that, again by 
Lemma~\ref{lem:b-inject},
\be
  f \otimes  \id_{U \otimes V} =0
  ~\Rightarrow~
  (f \otimes  \id_U) \otimes \id_V =0
  ~\Rightarrow~
  f \otimes  \id_U =0
  ~\Rightarrow~
  f =0 \ .
\ee  
Thus $b_{U \otimes V}$ is injective.
\end{proof}

\begin{proof}[Proof of Theorem~\ref{thm:Cdd-prop}.]
Part (i) amounts to Lemma~\ref{lem:Cdbar-dualclosed} and part (ii) to 
Lemma \ref{lem:Cdbar-tensorclosed}.
\end{proof}

As in \eqref{eq:cat-cond-from-VOA} we call a morphism 
$p : U \otimes V \rightarrow \one^{\!*}$ {\em non-degenerate} if $\pi^{-1}_{U,V}(p) : U \rightarrow V^*$ 
is an isomorphism.

\begin{Lem}\label{lem:pair-nondeg}
Let $\Cc$ be an abelian tensor category satisfying condition $\mathrm{C}$. For a morphism 
$p :  U \otimes V \rightarrow \one^{\!*}$ the following are equivalent:\\[.1em]
{\rm (i)} $p$ is non-degenerate.
\\[.3em]
{\rm (ii)} For all $X,Y \in \Cc$ and all $f : X \rightarrow U$, $g : Y \rightarrow V$ we 
have that $p \circ (f \otimes \id_V) = 0$ implies $f=0$ and $p \circ (\id_U \otimes g)=0$ 
implies $g=0$.
\end{Lem}

\begin{proof}
Let $f$ and $g$ be as in part (ii). Since $\pi_{U,V}$ is natural in $U$ and $V$, so is $\pi_{U,V}^{-1}$. This in turn means that 
$\pi_{X,V}^{-1}(p \circ (f \otimes \id_V)) = \pi_{U,V}^{-1}(p) \circ f$ and
$\pi_{U,Y}^{-1}(p \circ (\id_U \otimes g)) = g^* \circ \pi_{U,V}^{-1}(p)$.
\\[.3em]
(i)$\Rightarrow$(ii): Suppose $p \circ (f \otimes \id_V) = 0$. Then $0 = \pi_{X,V}^{-1}(p \circ (f \otimes \id_V)) = \pi_{U,V}^{-1}(p) \circ f$. Since  $\pi_{U,V}^{-1}(p)$ 
is an isomorphism, this implies $f=0$. That $p \circ (\id_U \otimes g)=0$ implies $g=0$ follows in the same way.
\\[.3em]
(ii)$\Rightarrow$(i): 
Suppose $\pi_{U,V}^{-1}(p) \circ f = 0$. Then $\pi_{X,V}^{-1}(p \circ (f \otimes \id_V)) = 0$ and consequently $p \circ (f \otimes \id_V)=0$. By assumption this implies $f=0$. Thus $\pi_{U,V}^{-1}(p)$ is injective. Similarly, $g^* \circ \pi_{U,V}^{-1}(p) = 0$ implies $g=0$ (and so $g^*=0$). Thus $\pi_{U,V}^{-1}(p)$ is surjective. Since $\Cc$ is abelian this implies that $\pi_{U,V}^{-1}(p)$ is an isomorphism.
\end{proof}

\begin{proof}[Proof of Theorem~\ref{thm:non-deg}.]
Part (i) follows from Lemmas \ref{lem:Cdbar-dualclosed} and 
\ref{lem:Cdbar-tensorclosed} because they show that if $A,B \in \Cdd$, so is 
$B \otimes A^*$, and objects in $\Cdd$ are necessarily non-zero (otherwise the 
duality morphism $b_{B \otimes A^*}$ cannot be injective). Part (ii) holds by definition of 
$\Cdd$ because by \eqref{eq:iHom+dual-morphs} the unit morphism is just 
$\eta_A = b_A$. Part (iii) can be proved as follows.
By \eqref{eq:iHom+dual-morphs} and \eqref{eq:epsA-def} we can write 
$\eps_A \circ m_{A,B,A} = \pi_{A,A^*}(\delta_A) \circ 
(\id_A \otimes d_B \otimes \id_{A^*}) =: p$. 
We will show that $p$ satisfies condition (ii) of Lemma~\ref{lem:pair-nondeg}. Let 
$f : X \rightarrow A \otimes B^*$ and suppose that $p \circ (f \otimes \id_{B \otimes A^*}) = 0$. 
We can write
\be
  0 = p \circ (f \otimes \id_{B \otimes A^*}) = \eps_A \circ (\tilde f \otimes \id_{A^*})
  \quad \text{with}~~
  \tilde f = (\id_A \otimes d_B) \circ (f \otimes \id_B) \ .
\ee
By definition $\pi_{A,A^*}^{-1}(\eps_A) = \delta_A$, so that $\eps_A$ is non-degenerate. 
By Lemma~\ref{lem:pair-nondeg} the above equation implies $\tilde f=0$. Applying the 
duality morphism $b_B$ to remove $d_B$ shows that then also $f=0$. 
The argument that 
$p \circ (\id_{A\otimes B^*} \otimes g)=0$ implies $g=0$ is similar. Thus $p$ is non-degenerate.
\end{proof}

\subsection[The kernel of $b_U$ and $\tilde b_U$]{The kernel of $\boldsymbol{b_U}$ and 
$\boldsymbol{\tilde b_U}$}\label{app:ker-bU}

The following lemma provides a method to deduce the kernel of $b_U$ and $\tilde b_U$ 
from the action of the tensor product on objects.

\begin{Lem}\label{lem:KU=0}
Let $\Cc$ be an abelian tensor category and suppose that $U$ has a right and a left dual. 
Let $K$ be the kernel of $b_U$ and $\tilde K$ the kernel of $\tilde b_U$. \\
{\rm (i)} $K \otimes U = 0$ and $U^\vee \otimes K = 0$.
\\
{\rm (ii)} If $S$ is a subobject of $\one$ such that $S \otimes U = 0$ or 
$U^\vee \otimes S = 0$ then $S$ is a subobject of $K$.
\\
{\rm (iii)} $\tilde K \otimes \eev U = 0$ and $U \otimes \tilde K = 0$.
\\
{\rm (iv)} If $\tilde S$ is a subobject of $\one$ such that $\tilde S \otimes \eev U = 0$ or 
$U \otimes \tilde S = 0$ then $\tilde S$ is a subobject of $\tilde K$.
\end{Lem}

\begin{proof}
Let us prove (i) and (ii) in detail, parts (iii) and (iv) work similarly. We will not write 
out unit isomorphisms and associators.
\\[.3em]
(i) 
Let $k : K \rightarrow \one$ be the embedding of the kernel.
As in the proof of Lemma~\ref{lem:b-inject}, applying the duality morphisms to $b_U \circ k = 0$ 
gives $k \otimes \id_U = 0$. From this we conclude
\be
  0 = \Big( K \otimes U \otimes \eev U \xrightarrow{k \otimes \id_U \otimes \id_{\eev U}} 
  U\otimes\eev  U \xrightarrow{\,\tilde d_U\,} \one
  \Big) = \Big( K \otimes U \otimes \eev U \xrightarrow{\id_K \otimes \tilde d_U} 
  K \xrightarrow{\,k\,} \one \Big) \ .
\ee
Since $k$ is injective it follows that $\id_K \otimes \tilde d_U = 0$. Using the left duality 
morphisms, this in turn implies $\id_{K \otimes U} = 0$, i.e.\ $K \otimes U=0$.
That $U^\vee \otimes K=0$ can be seen similarly.
\\[.3em]
(ii) Let $s : S \rightarrow \one$ be the subobject embedding. If $S \otimes U = 0$ then 
also $s \otimes \id_U = 0$. Again as in the proof of Lemma~\ref{lem:b-inject}, this implies 
$b_U \circ s = 0$. Thus $s : S \rightarrow \one$ will factor through $K$ via an injective 
morphism. The argument starting from $U^\vee \otimes S = 0$ is similar.
\end{proof}

Note that the statement of the lemma cannot be split into two independent 
statements about right and left duals, because the proof of (i), which is a statement 
about the right dual, did require the left dual, and vice versa for part (iii). The lemma 
tells us that if $U$ has a right and a left dual, then the kernel of $b_U$ is the maximal 
subobject $S$ of $\one$ for which $S \otimes U=0$, and the kernel of $\tilde b_U$ is the 
maximal subobject $\tilde S$ of $\one$ for which $U \otimes \tilde S = 0$.
\medskip

Let us now turn again to the $\Wc_{2,3}$ model. As already mentioned a number of times, 
we believe that the representations listed in \eqref{eq:W23-irreps} and \eqref{eq:indec-W-rep} 
which are not in grey boxes have the property that $U^*$ is a right and left dual of $U$. 
To check which of these are in $\Repbb$ it remains to select those $U$ for 
which $b_U$ and $\tilde b_U$ are injective. We will do that with the help of 
Lemma~\ref{lem:KU=0}. The only non-trivial subobject of $\Wc$ is $\Wc(2)$. From the 
fusion rules in Appendix~\ref{app:fus} we see that $\Wc(2) \otimes \Rc \cong \Rc \neq 0$ 
for all representations in \eqref{eq:W23-irreps} and \eqref{eq:indec-W-rep} not in grey boxes. 
Therefore, $\Wc(2)$ cannot be in the kernel of $b_{\Rc}$ or $\tilde b_{\Rc}$ for any of these 
$\Rc$, and so the kernels of $b_{\Rc}$ and $\tilde b_{\Rc}$ are trivial.

\small

\newcommand\arxiv[2]      {\href{http://arXiv.org/abs/#1}{\tt #2}}
\newcommand\doi[2]        {\href{http://dx.doi.org/#1}{#2}}
\newcommand\httpurl[2]    {\href{http://#1}{#2}}

\end{document}